\documentclass{aa}

\usepackage{graphicx}
\usepackage{txfonts}
\usepackage{hyperref}
\usepackage{upgreek}
\usepackage{color}
\usepackage{comment}
\usepackage{mathtools}
\usepackage{ulem}
\usepackage{xspace}
\usepackage{siunitx}
\usepackage{supertabular}
\makeatletter
\let\MYcaption\@makecaption
\makeatother

\usepackage[font={footnotesize}]{subcaption}
\makeatletter
\let\@makecaption\MYcaption
\makeatother

\newcommand{\sectionname}{Sect.}
\newcommand{\dd}{\textrm{d}}

\newcommand{\derivp} [2] {\frac {\partial #1 } {\partial #2} }

\newcommand{\algn} [1] {
\begin{align} #1
\end{align}}

\let\originaleqref\eqref
\renewcommand{\eqref}{Eq.~\originaleqref}
\newcommand{\eq}[1] {Eq.\,(\ref{#1})}
\newcommand{\eqss}[2]{Eqs.~(\ref{#1})-(\ref{#2})}
\makeatletter

\newcommand{\checknextarg}{\@ifnextchar\bgroup{\gobblenextarg}{}}
\newcommand{\gobblenextarg}[1]{\@ifnextchar\bgroup{, (\ref{#1})\gobblenextarg}{ and (\ref{#1})}}
\makeatother

\begin{document}

   \title{Coriolis darkening in late-type stars}

   \subtitle{II. Effect of self-sustained magnetic fields in stratified convective envelopes}

 \authorrunning{C. Pinçon et al.}
   \titlerunning{Coriolis darkening in late-type stars II}

   \author{C. Pinçon\inst{1,2}, L. Petitdemange\inst{2}, R. Raynaud\inst{3}, L. J. Garcia\inst{4}, A. Guseva\inst{2}, M. Rieutord\inst{5} \and E. Alecian\inst{6}
   }

   \institute{\inst{1}~Université Paris-Saclay, CNRS, Institut d’astrophysique spatiale, 91405, Orsay, France\\
   ~~~\email{charly.pincon@universite-paris-saclay.fr}\\
   \inst{2}~LERMA, Observatoire de Paris, Sorbonne Université, Université PSL, CNRS, 75014 Paris, France\\
   \inst{3}~Université Paris Cité, Université Paris-Saclay, CEA, CNRS, AIM, F-91191 Gif-sur-Yvette, France\\
   \inst{4} {Center for Computational Astrophysics, Flatiron Institute, New York, NY, USA}\\
   \inst{5}~IRAP, Université de Toulouse, CNRS, UPS, CNES, 14 avenue Édouard Belin, 31400 Toulouse, France\\
  \inst{6}~Univ.  Grenoble Alpes, CNRS, IPAG, F-38000 Grenoble, France
             }

   \date{\today}


  \abstract
   {Modeling the surface brightness distribution of stars is of prime importance to interpret the large amount of available interferometric, spectropolarimetric, or photometric observations. Beyond stellar physics, this is also a prerequisite to characterize exoplanets or our Galaxy. Nevertheless, this remains quite challenging for cool stars as it requires one to model the magnetohydrodynamic turbulence that develops in their convective envelope.
   }
   {In Paper I, the effect of the Coriolis acceleration on the surface heat flux has been studied by means of hydrodynamic simulations. In this paper, we aim to investigate the additional effect of dynamo magnetic fields that can be generated in the thick convective envelopes of cool stars. We focus on an envelope thickness that is representative of either a $\sim0.35~M_\odot$ M dwarf, a young red giant star or a pre-main sequence star.}
   {We performed a parametric study using numerical magnetohydrodynamic simulations of anelastic convection in thick rotating spherical shells. The stratification in density ranges from a few tens to a few hundreds. The setup assumes a constant entropy jump between the inner and outer layers to force convection, with stress-free boundary conditions for the velocity field. The magnetic Prandtl number was systematically varied in order to vary the magnetic field intensity. For each model, we computed the azimuthally and temporally averaged surface distribution of the heat flux, and examined the leading-order effect of the magnetic field on the obtained latitudinal luminosity profile.}
   {We identify three different regimes. Close to the onset of convection, while the first unstable modes tend to convey heat more efficiently near the equator, magnetic fields are shown to generally enhance the mean heat flux close to the polar regions (and the tangent cylinder). By progressively increasing the Rayleigh number, the development of a prograde equatorial jet was previously shown to make the equator darker when no magnetic field is taken into account. For moderate Rayleigh numbers, magnetic fields can instead inverse the mean pole-equator brightness contrast (which means going from a darker to a brighter equator when a dynamo sets in) and finally induce a similar regime to that found close to the onset of convection. For more turbulent models with larger Rayleigh numbers, magnetic fields alternatively tend to smooth out the brightness contrast. This general behavior is shown to be related to the quenching of the surface differential rotation by magnetic fields and remains valid regardless of the magnetic morphology.}
   {Mean global trends regarding the impact of rotation and magnetic fields on the surface brightness distribution of cool stars are theoretically depicted and need to be tested by future observations. This work opens the door to more detailed theoretical studies including the effect of nonaxisymmetric and time-variable surface features associated with magnetic activity.}

   \keywords{Convection -- Dynamo -- Magnetohydrodynamics (MHD) -- Methods : numerical -- Stars: interiors, magnetic field
               }

   \maketitle
%

\section{Introduction}


Understanding the spatial distribution of the light emission at the surface of stars is of prime importance to characterize their properties. This is not only necessary to convert the observed magnitude of a star into its actual luminosity, but also essential to help analyze the detailed mapping of surface features obtained by interferometry or Zeeman-Doppler reconstruction techniques \citep[e.g.,][]{Rottenbacher2016,Finociety2021,Petit2022}, as well as interpret the photometric light curves of eclipsing binaries or transiting exoplanets \citep[e.g.,][]{Valio2017,Ozavci2018}.

In addition to the limb-darkening effect, which refers to the decrease in the light flux emitted onto the line of sight from the center of the stellar disk to its limb
\citep[e.g.,][]{Morello2017,Maxted2023}, internal dynamics itself can affect the brightness distribution on the surface of stars. Stellar rotation is known to be a key ingredient in this regard. In rapidly rotating, hot (early-type) stars with a stably stratified radiative envelope, the centrifugal structural flattening along the rotation axis makes the poles hotter and brighter than the equator; this is the so-called gravity-darkening effect, which was first formulated by \cite{Zeipel1924} and well constrained by interferometric data and advanced 2D stellar models \citep[e.g.,][]{ELR11,Domiciano2014,Bouchaud2020}. In contrast, in cooler (late-type) stars with a convective envelope, the picture remains unclear. Modeling the outgoing heat flux through turbulent rotating envelopes is indeed a complex task and requires one to account for the impact of three main physical ingredients on the convective flows. The first one is naturally the effect of the Coriolis acceleration that tends to organize the convection in a columnar shape; the second one is the effect of magnetic fields; and the third one is the effect of the centrifugal acceleration. The latter is expected to make the star oblate as well as decrease the convective buoyancy driving in the regions close to the equator. Understanding the impact of the centrifugal acceleration on the surface brightness distribution of stars is difficult and is deferred to future works.

Regarding the Coriolis force, its effect has already been investigated in previous work \citep[][hereafter Paper I]{Raynaud2018}. Using numerical simulations of highly stratified rotating convective envelopes, we have demonstrated that in turbulent regimes (i.e., far from the onset of convection), the latitudinal distribution of brightness is determined by the competition between the inertia of convective eddies and the Coriolis acceleration. When the Coriolis acceleration dominates inertial forces at the surface (i.e., low surface Rossby number), a strong prograde azimuthal flow is sustained close to the equator owing to Reynolds stresses induced by Rossby-like eddies. This tends to inhibit the outgoing heat flux in the equatorial region, making the poles brighter. On the contrary, when inertia tends to dominate the Coriolis acceleration (i.e., large surface Rossby number),
the heat flux tends to become uniform in latitude and decouple from the surface differential rotation profile.
To avoid any misleading confusion with the original concept of gravity darkening, we decide in this second paper of the series to term Coriolis darkening this modulation of the surface brightness. In this work, we aim to extend the results of Paper I by taking magnetic effects generated by dynamo action into account.

As shown by spectropolarimetric observations, magnetic fields are ubiquitous in low-mass stars and exhibit diverse topologies, ranging from small-scale fields to large-scale dipolar fields \citep{Donati2009,Morin2010,Folsom2018,Kochukhov2021,Donati2023}. Such magnetic fields result from dynamo mechanisms  that are ruled by the turbulent motions in the convective envelopes of these stars. These mechanisms permit to transfer a part of the kinetic energy into magnetic energy through nonlinear amplification processes of electrical currents, allowing magnetic fields to be sustained against Ohmic dissipation.
Since the pioneer works of \cite{Gilman1983,Glatzmaier1985,Glatzmaiers1995}, numerical modeling of this multiscale mechanism has been developing considerably. However, direct numerical simulations still cannot meet realistic stellar conditions and have to rely on some convective approximations to simplify the problem, such as the Boussinesq or the anelastic approximations~\citep[e.g.,][]{Ogura1962,Gough1969,Braginsky1995,Lantz1999}. Both are sound-proof approximations of the magnetohydrodynamic (MHD) equations that are well adapted to model low-Mach turbulence. 
Either in the incompressible Boussinesq paradigm \citep[e.g.,][]{Christensen2006, Schrinner2012,Yadav2013,Petitdemange2018,Menu2020} or in the anelastic framework that allows for variations in the medium density \citep[e.g.,][]{Gastine2012,Duarte2013,Schrinner2014,Raynaud2015,Schwaiger2021,Zaire2022}, magnetic fields obtained in numerical simulations can be roughly classified as either dominated by the large-scale poloidal dipolar component or not, even though a more sophisticated description in terms of force balance and temporal evolution is clearly suggested by the numerous results on the subject. In this framework, anelastic dynamos were shown to exhibit some specific behaviors compared to Boussinesq models. For instance, an increase in the density stratification seems to disadvantage dipolar configurations \citep{Gastine2012,Jones2014}, unless the magnetic Prandtl number, $Pm$, is large enough \citep{Schrinner2014}. Such differences between Boussinesq and anelastic configurations result from the difference of mass distribution and gravity profile in both cases, which directly affect the properties of the convective cells with depth \citep{Raynaud2014}. For instance, \citet{Raynaud2015} showed that a large density drop toward the outer regions is associated with an increase in the convective inertia, which can destabilize at some point a dipolar magnetic configuration. Besides, when the Lorentz force plays a major role at large scales, that is, above the dynamo threshold and with large values of the magnetic Prandtl number, strong dipole magnetic fields can be maintained even in highly stratified fluids. This regime (or strong branch) is characterized by a balance between the Coriolis acceleration and the Lorentz force although the convective inertia remains significant close to the surface \citep{Dormy2016,Dormy2018,Menu2020,Schwaiger2021, Zaire2022}.

These previous studies constitute an appropriate reference basis to investigate the combined effects of the Coriolis acceleration and self-sustained magnetic fields on the brightness distribution at the surface of convective stellar envelopes. Actually, \cite{Yadav2016} have already partially tackled the question. In the regime where the Coriolis acceleration dominates turbulent inertia at the surface, they have shown that self-sustained magnetic fields can globally quench the azimuthal mean flow close to the equator, relaxing its blocking action on the outgoing heat flux in this region. However, this last study is restricted to Boussinesq models with a constant magnetic Prandtl number (namely, equal to unity). In this work, we aim to question this result exploring a wider parameter space (e.g., large density stratification, varying magnetic Prandtl numbers), and to extend at the same time the results of Paper I to the magnetic case. To do so, we use about 100 three-dimensional, self-consistent dynamo models computed by direct numerical simulations. Our goal is to understand general physical trends as a function of the main control parameters. Hereafter, we specifically focus on stars cooler than the Sun with thick convective envelopes. For the sake of simplicity, the aspect ratio of the simulated convective shell is set to $\chi=0.35$, where $\chi=r_{\rm i}/r_{\rm o}$ with $r_{\rm i}$ and $r_{\rm o}$ the inner and outer radii of the shell, respectively. We recall that the Sun is characterized by a relatively thin convective envelope, with $\chi_\odot \approx 0.7$. In comparison, our setup can be regarded as representative of different evolutionary stages:
M-dwarf stars of about $0.35~M_\odot$ \citep[e.g.,][]{Chabrier1997}; pre-main sequence stars leaving the Hayashi phase while developing a radiative core \citep[e.g.,][]{Siess2000,Scott2016,Antona2017}; low- and intermediate-mass red giant stars at the beginning of the red giant branch \citep[e.g.,][]{Pincon2020}. On the one hand, cool M dwarfs are currently one of the most interesting targets to characterize rocky Earth-like exoplanets by the method of the transits owing to the accessible large signal-to-noise ratio \citep[e.g.,][]{Garcia2022}.
On the other hand, pre-main sequence evolution is still subject to huge uncertainties related for instance to disk accretion phenomena and stellar magnetism \citep[e.g.,][]{Alecian2020}. Finally, red giant stars, with their wide mass range (and thus age range) and their large luminosity, stand for a mine of information for galactic archaeology research \citep[e.g.,][]{Mosser2016,Miglio2021}. All these types of stars are therefore important targets and deserve special attention, motivating our choice. The numerical simulations used in this work are introduced in \sectionname{}~\ref{models}. The different magnetic morphologies encountered in such simulations are briefly recalled in \sectionname{}~\ref{morpho}. The effect of the magnetic fields on the mean latitudinal luminosity distribution is then addressed in \sectionname{}~\ref{brightness}. The results are discussed in \sectionname{}~\ref{discussion} and the conclusions are formulated in \sectionname{}~\ref{conclu}.


\section{Anelastic simulations of convective envelopes}

\label{models}

\subsection{Physical setup}

Our setup is similar to \cite{Raynaud2015}.  We consider a convective spherical shell of width $d$ that rotates about the $z$-axis at angular velocity $\Omega$; it is filled with a perfect, electrically conducting gas of kinematic viscosity~$\nu$, thermal diffusivity~$\kappa$, specific heat~$c_p$, and magnetic diffusivity~$\eta$ (all supposed to be constant). We assume that the mass is concentrated inside the inner sphere and neglect the centrifugal force. The gravitational acceleration is thus approximated by
$\vec{g} = - (GM/r^2)~ \vec{e}_r$, where $G$ is the gravitational constant, $M$ is the central mass, and $\vec{e}_r$ is the radial unit vector. For the reference state, we assume an adiabatic hydrostatic equilibrium (i.e., at marginal stability with respect to convection). The reference state is characterized by a constant polytropic index~$n$ (hereafter, $n=2$ by hypothesis\footnote{We choose the same value as in the dynamo benchmarks of \cite{Jones2011} for comparison, i.e., an adiabatic index of $\gamma = 1.5$. This practical choice is close to the value expected in the convective envelope of stars, i.e., $\gamma = 1.66$, and is not expected to affect the main trends studied in this work.}) and a given number of density scale heights between the bottom boundary and the surface, that is,
\begin{equation}
    N_\rho = \ln \left[\frac{\rho(r_{\rm i})}{\rho(r_{\rm o})}\right]
    \,,
\end{equation}
where $\rho(r)$ is the reference state density. Convection is then driven by imposing a fixed entropy drop~$\Delta S$ between the inner and outer spheres.
We apply stress-free boundary conditions for the velocity field, while the magnetic field matches a potential field inside and outside the fluid shell.
 
The governing  equations rely on the LBR formulation of the anelastic approximation \citep{Jones2011}
and are recalled in Appendix~\ref{app:eq} for completeness.
The magnetohydrodynamic system in \eqss{moment}{flux m} involves four other control parameters, namely the Rayleigh number, the Ekman number, and the thermal and magnetic Prandtl numbers, which are defined respectively as
\begin{equation}
    Ra = \frac{GM d \Delta S}{ \nu\kappa c_p},~~E= \frac{\nu}{\Omega d^2},~~Pr= \frac{\nu}{\kappa}~~{\rm and}~~ Pm= \frac{\nu}{\eta} \,.
\end{equation}
In the convective envelope of stars, these dimensionless parameters reach extreme values owing to the very low values of the viscosity and diffusivity; for instance, in the solar convective zone, we expect $Ra \sim 10^{20}$, $E\sim 10^{-15}$, $Pr \sim 10^{-7}$ and $Pm\sim 10^{-5}$. Considering such regimes in direct numerical simulations of extended stellar magnetized convective zones is currently impossible, as it will demand a much too long computing time to resolve the smallest dissipation length scale \citep[e.g,][]{Kupka2017}. Given these numerical limitations, we can instead rely on systematic parameter studies in more moderate and accessible regimes; the aim is to obtain valuable insights on the global dynamics, applicable to some extents to the stellar case (e.g., see the discussion in \sectionname{}~\ref{applicability}). The unrealistic large values of the viscosity and diffusivity can therefore be regarded as effective transport coefficients mimicking the effect of turbulent small-scale motions on the large-scale flow. Keeping this in mind for our set of simulations,
 the Ekman number, $E$, is set to low accessible values between $3\times 10^{-5}$ and $3\times 10^{-4}$, and the thermal Prandtl number, $Pr$, remains close to unity. The parameter $N_\rho$ covers values between 3 and 6 (i.e., a density contrast between the inner and outer boundaries going from about 20 to 400). We notice that even setting $N_\rho=6$ does not permit to model the thin surface layer of stars where the density is expected to drop by several orders of magnitude. In this very narrow shell, the anelastic approximation fails as the flow becomes supersonic and is dominated by small-scale eddies. Nevertheless, this setup is expected to grasp the general large-scale dynamics in a substantial part of the convective bulk, independently of these very near-surface layers. We also explore increasing levels of turbulence by varying the ratio $Ra/Ra_{\rm c}$ from unity up to about 30, with $Ra_{\rm c}$ the critical Rayleigh number at the linear onset of convection. This latter has been calculated solving the boundary value problem of the linearized hydrodynamic equations as in \cite{Jones2009}. They are reported in \tablename{}~\ref{table:Rac} for the control parameters considered in this study, with typical values on the order of $10^6$. The magnetic Prandtl number, $Pm$, is then systematically varied between 0.1 and 5, which enables us to explore different magnetic field intensities and morphologies. The covered parameter space is briefly summarized in \tablename{}~\ref{param}.
Finally, as initial conditions, both strong dipoles or random seeds have generally been used to account as much as possible for the two dynamo branches at low Rossby numbers \citep[e.g.,][]{Raynaud2015}.

\subsection{Numerical setup and output diagnostics}
For all the considered models, the equations are integrated at least over about one magnetic diffusion timescale $\tau_\eta = d^2/\eta$ with the 3D-MHD codes \textsc{PaRoDy} \citep{Dormy1998D,Schrinner2014} or \textsc{MagIC} \citep{Gastine2012,Schaeffer2013}. In both codes, the vector fields are transformed into scalars using a poloidal-toroidal decomposition. The equations are discretized in the radial direction with a finite-difference scheme for \textsc{PaRoDy} and an expansion onto the Chebyshev polynomials for \textsc{MagIC}; on each concentric sphere, variables are expanded onto the spherical harmonic basis, with angular degrees $\ell$ and azimuthal orders~$m$. Typical resolutions in the radial direction are up to 240 mesh points and the horizontal spectral decomposition is truncated at $\ell_{\rm max} \sim m_{\rm max} \le 400$ for the more turbulent and stratified models. Both solvers give comparable results for the set of models considered here. For each model, the total kinetic energy spectrum has been checked to drop monotonically by at least two orders of magnitude from the scale at maximum power spectrum toward the smallest considered scale. This represents our empirical criterion to ensure numerical convergence.

By construction, the entropy is set to a constant value and the radial velocity vanishes at the surface, so that the surface heat flux results only from the radial entropy gradient. It is commonly represented by the surface Nusselt number $Nu(r_{\rm o},\theta,\varphi)=(\vec{e}_r\cdot \vec{\nabla} S )/(\vec{e}_r\cdot \vec{\nabla} S_{\rm c}) $, where $S_{\rm c}$ is the conductive entropy profile, that is, the profile of entropy that we would get if the energy was transported by thermal diffusion only. The Nusselt number hence measures the efficiency of the heat transport by convection. In this work, we are interested in studying the combined effects of the magnetic fields and the Coriolis acceleration on the global surface brightness distribution. For the sake of convenience, we consider the mean surface Nusselt number averaged over the duration of the simulation $T_{\rm run}$ (i.e., the time spent after reaching a stationary state and larger at least than about $\tau_\eta$) and over the azimuthal direction, that is (see Paper I for details),
\algn{
\overline{Nu}_{r_{\rm o}}(\theta) = - \frac{(1-e^{-N_\rho}) r_{\rm o}^2 \varw_{\rm o}}{nc_1} \frac{1}{2\pi T_{\rm run} } \iint_{0}^{2\pi,T_{\rm run}} \left.\derivp{S}{r}\right|_{r_{\rm o}} \dd \varphi \dd t \; .
\label{Nu}
}
This diagnostic is sufficient in a first step to investigate the effect of the magnetic field on the mean heat flux distribution observed in the hydrodynamic case (Paper I). It will nevertheless not permit to account for the rapid temporal variations generated by nonaxisymmetric features (e.g., nonpolar stellar spots), which deserve a dedicated study out of the scope of this paper.


\section{Magnetic morphologies in stratified spherical dynamo models}

\label{morpho}

\begin{figure*}

\begin{subfigure}[c]{0.5\hsize}
\includegraphics[width = \hsize]{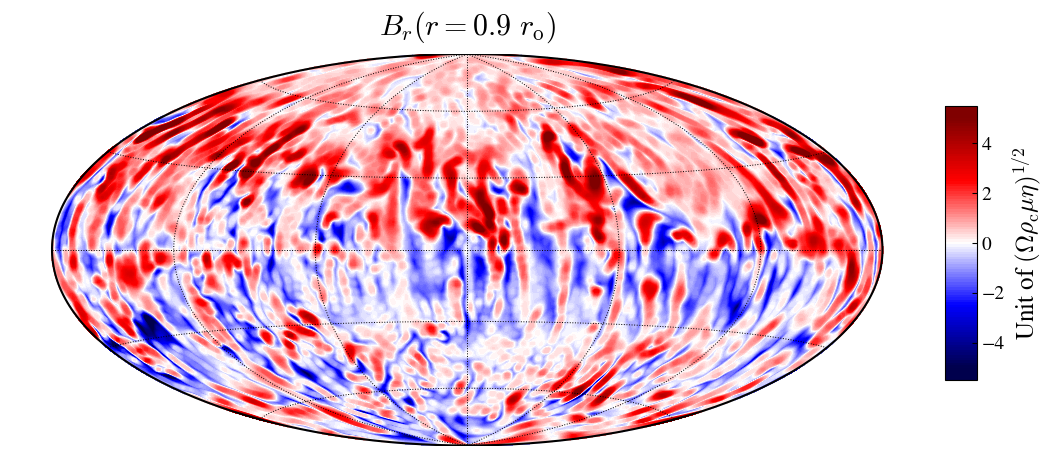}
\caption{Axial dipole}
\label{fig:axial}
\end{subfigure}
\begin{subfigure}[c]{0.5\hsize}
\centering
\includegraphics[width=\hsize]{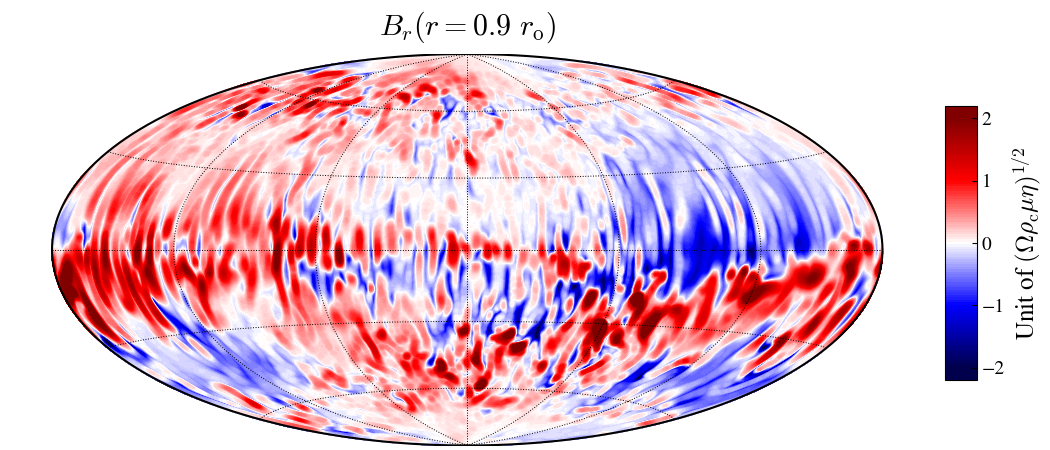}\\
\caption{Equatorial dipole}
\label{fig:equat}
\end{subfigure}

\vspace{0.5cm}
\begin{subfigure}[c]{0.5\hsize}
\centering
\includegraphics[width=\hsize]{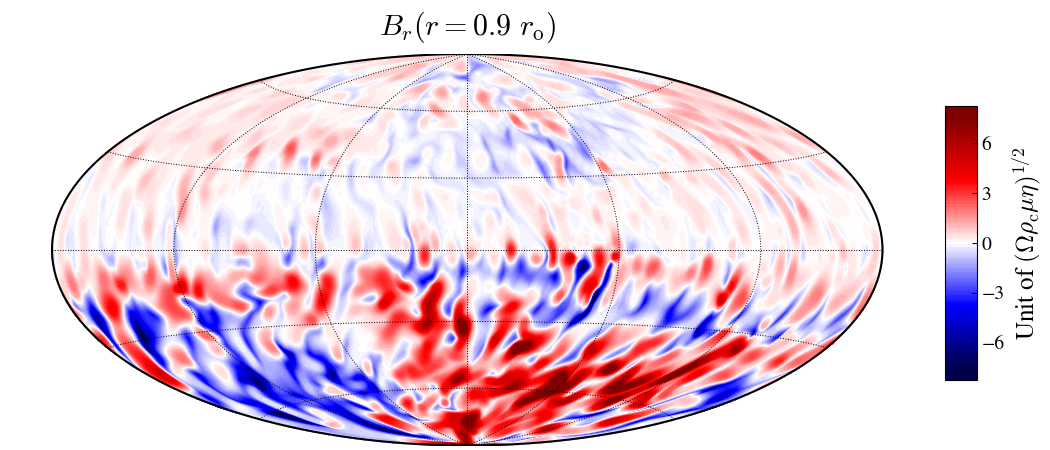}
\caption{Hemispherical dynamo}
\label{fig:hemi}
\end{subfigure}
\begin{subfigure}[c]{0.5\hsize}
\centering
\includegraphics[width=1\hsize]{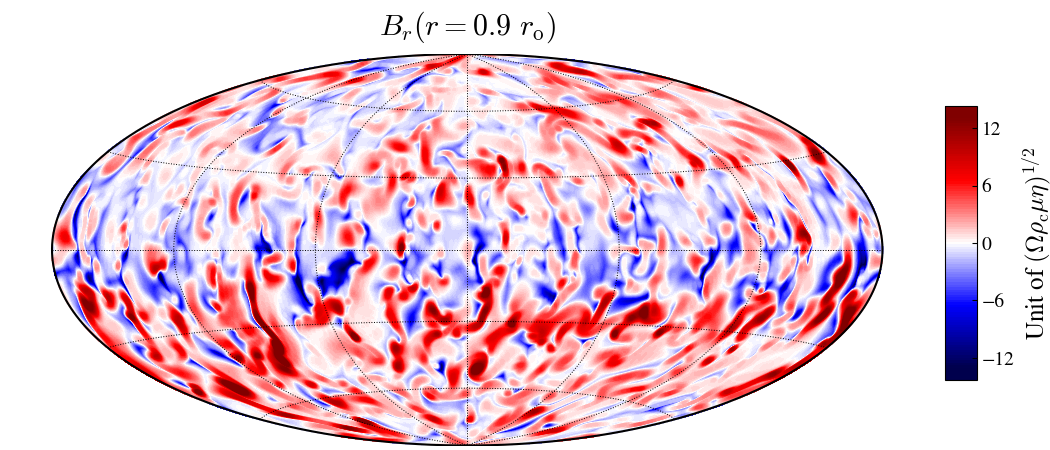}\\
\caption{Coherently oscillating dynamo}
\label{fig:multi}
\end{subfigure}

\caption{Hammer projection of the radial component of the magnetic field at $r=0.9r_{\rm o}$ and at a given time, for four typical dynamo models observed in direct numerical simulations. (a) Axial dipole with $E=10^{-4}$, $Pr = 1$, $N_\rho= 3.5$, $Ra = 2.5~Ra_{\rm c}$ and $Pm = 5$. (b) Equatorial dipole with $E=3\times 10^{-5}$, $Pr = 1$, $N_\rho= 3$, $Ra = 2.48~Ra_{\rm c}$, and $Pm = 2$. (c) Hemispherical dynamo with $E=3\times 10^{-4}$, $Pr = 1$, $N_\rho= 6$, $Ra = 4~Ra_{\rm c}$, and $Pm = 2$. (d) Coherently oscillating dynamo with $E=3\times 10^{-4}$, $Pr = 1$, $N_\rho= 6$, $Ra = 8~Ra_{\rm c}$, and $Pm = 2$.}
\label{topologies}
\end{figure*}

\begin{figure*}

\begin{subfigure}[c]{0.5\hsize}
\includegraphics[width=\hsize]{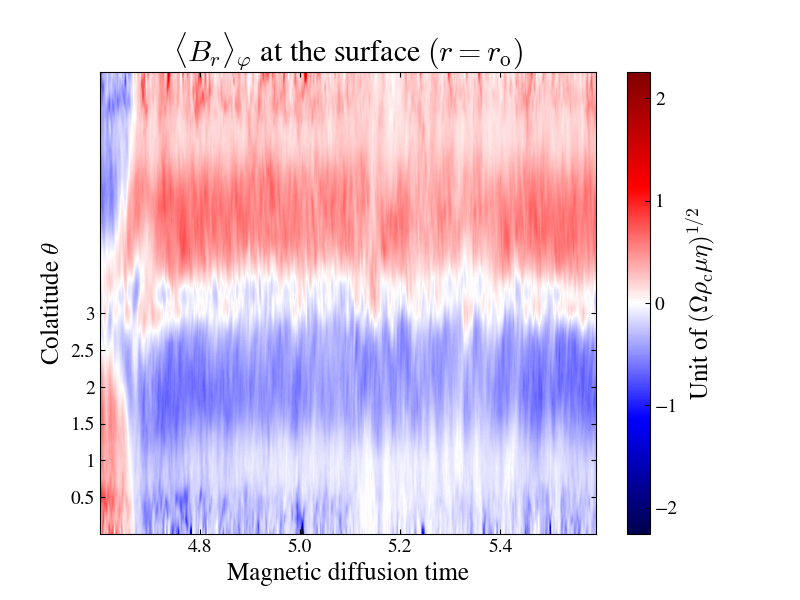}
\caption{Axial dipole}
\label{fig:but_dip}
\end{subfigure}
\begin{subfigure}[c]{0.5\hsize}
\centering
\includegraphics[width=\hsize]{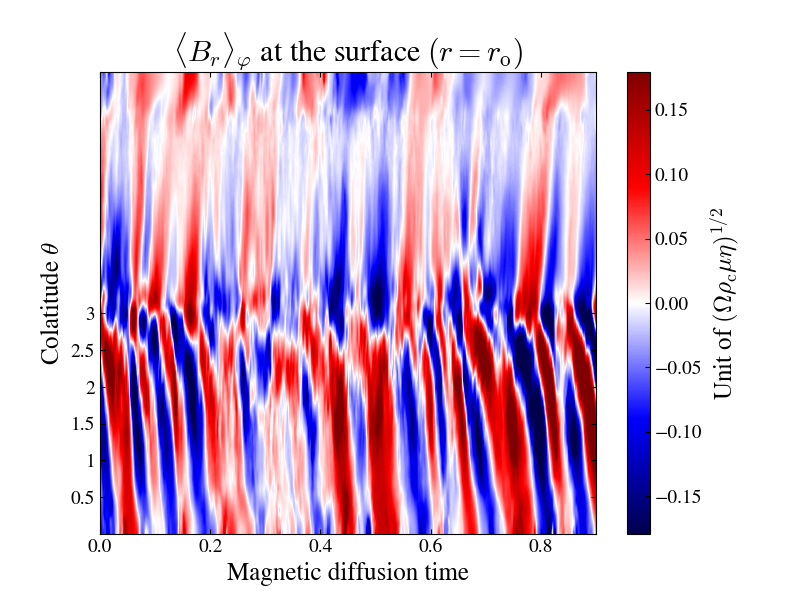}
\caption{Equatorial dipole}
\label{fig:but_equat}
\end{subfigure}

\vspace{0.5cm}
\begin{subfigure}[c]{0.5\hsize}
\centering
\includegraphics[width=1\hsize]{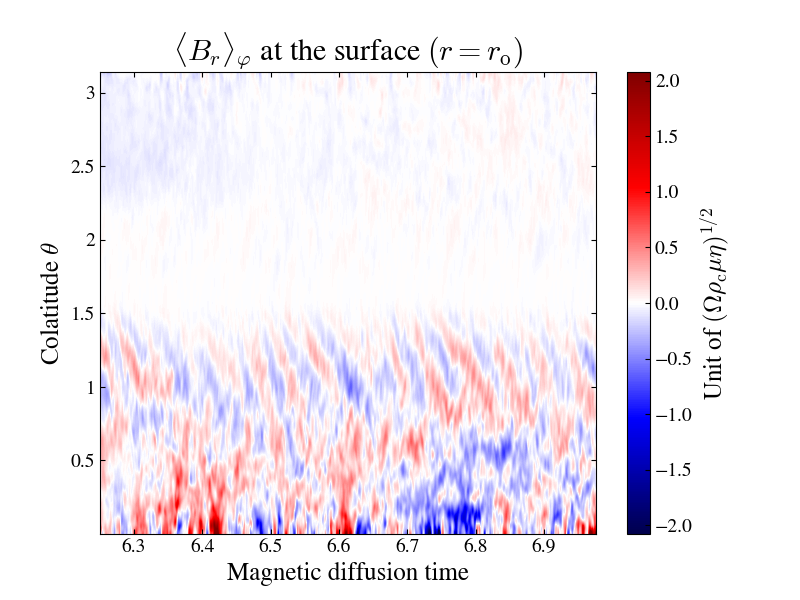}
\caption{Hemispherical dynamo}
\label{fig:but_hemi}
\end{subfigure}
\begin{subfigure}[c]{0.5\hsize}
\centering
\includegraphics[width=1\hsize]{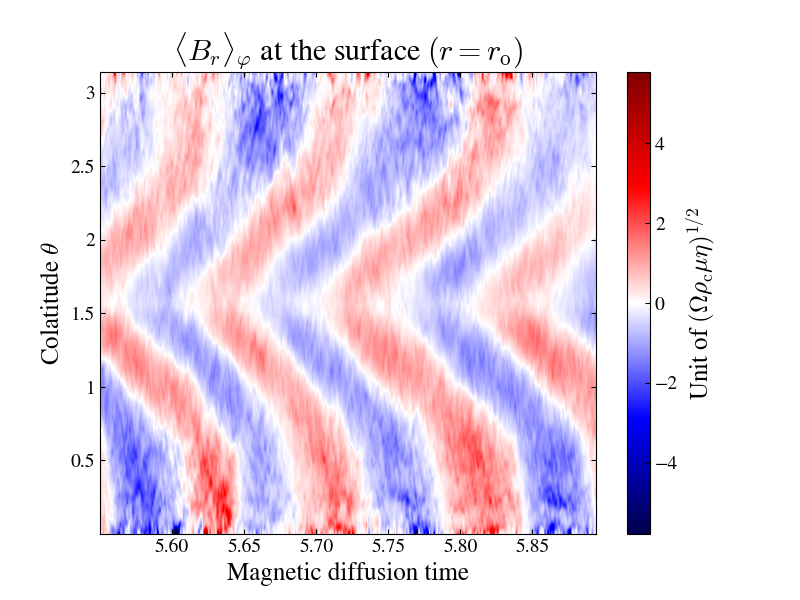}
\caption{Coherently oscillating dynamo}
\label{fig:but_multi}
\end{subfigure}

\caption{Surface butterfly diagram of the four typical examples shown in \figurename{}~\ref{topologies}, i.e., the azimuthal average of the magnetic field radial component as a function of time and colatitude at the surface. (a) Axial dipole with $E=10^{-4}$, $Pr = 1$, $N_\rho= 3.5$, $Ra = 2.5~Ra_{\rm c}$, and $Pm = 5$. (b) Equatorial dipole with $E=3\times 10^{-5}$, $Pr = 1$, $N_\rho= 3$, $Ra = 2.48~Ra_{\rm c}$, and $Pm = 2$. (c) Hemispherical dynamo with $E=3\times 10^{-4}$, $Pr = 1$, $N_\rho= 6$, $Ra = 4~Ra_{\rm c}$, and $Pm = 2$. (d) Coherently oscillating dynamo with $E=3\times 10^{-4}$, $Pr = 1$, $N_\rho= 6$, $Ra = 8~Ra_{\rm c}$, and $Pm = 2$.}
\label{butterflies}
\end{figure*}

In this section, we briefly recall the general magnetic morphologies that are encountered in numerical simulations of thick convective envelopes and that have already been observed in previous studies. For later purpose, we also emphasize the global impact of magnetic fields on the large-scale differential rotation.

\subsection{A diversity of dynamos}

Based on the previous studies, we choose to classify the magnetic fields observed in spherical dynamo simulations into three main categories: axial dipoles, equatorial dipoles, and multipolar configurations, among which we also distinguish hemispherical and coherently oscillating dynamos.

\begin{figure}
\includegraphics[width=\hsize,trim=0cm 0cm 1cm 1cm,clip]{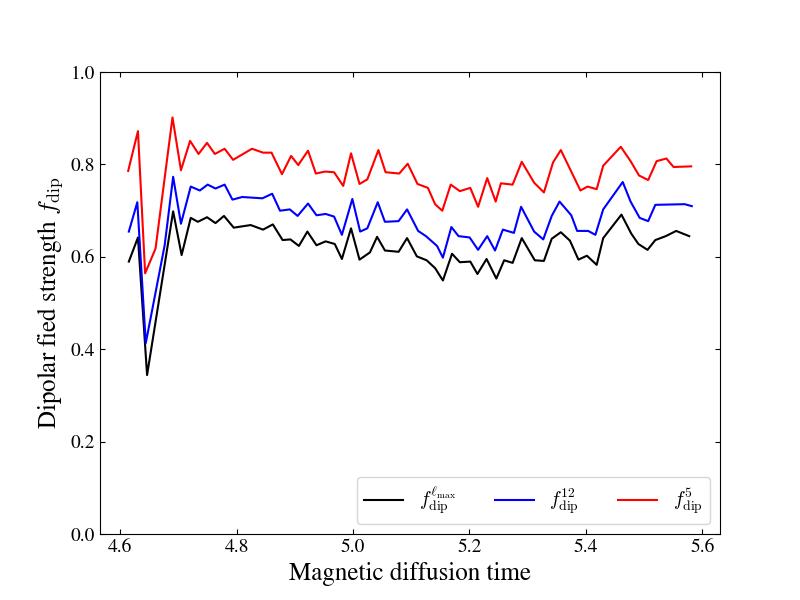}
\caption{Dipolarity indicator as defined in \eq{f_dip} for the axial dipole considered in the panel \figurename{}~\ref{fig:axial}, with $E=10^{-4}$, $Pr = 1$, $N_\rho= 3.5$, $Ra = 2.5~Ra_{\rm c}$, and $Pm = 5$. The relative contribution of the dipole is computed for the cut-off degrees $\ell_{\rm cut}=\ell_{\rm max},12,$ and $5$ (black, blue, and red lines, respectively).}
\label{f_dip plot}
\end{figure}
\begin{figure}
\includegraphics[width=\hsize,trim=0cm 0cm 1cm 1cm,clip]{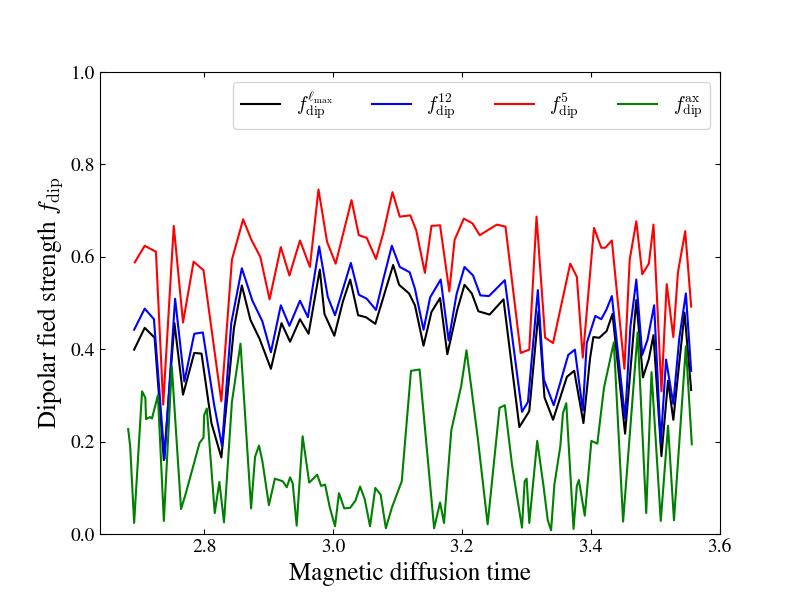}
\caption{Same as in \figurename{}~\ref{f_dip plot}, but for the equatorial dipole considered in \figurename{}~\ref{fig:equat}, with $E=3\times 10^{-5}$, $Pr = 1$, $N_\rho= 3$, $Ra = 2.48~Ra_{\rm c}$, and $Pm = 2$. The relative contribution of the axial dipole alone ($m=0$) for the cut-off degrees $\ell_{\rm cut}=\ell_{\rm max}$, denoted with $f_{\rm dip}^{\rm ax}$, is also represented in green.}
\label{f_dipax}
\end{figure}

\paragraph{Axial dipoles:}

Dipolar magnetic fields almost aligned with the rotation axis were shown to be stable, firstly, when the differential rotation between the poles and the equator is moderate and do not enter in conflict with the growth of large-scale axisymmetric components; and secondly, when the Coriolis force dominates the influence of turbulent inertia in a subsequent part of the convective bulk \cite[e.g.,][and references there in]{Menu2020,Zaire2022}. \cite{Raynaud2015} showed that this second criterion is met when $Ro_{\rm surf} \lesssim 0.1$, with the surface (convective) Rossby number
 \begin{equation}
    Ro_{\rm surf}=\frac{\varv_{\rm rms}}{ \ell_\varv \Omega}
    \,,
    \label{Ro_l}
 \end{equation}
where $\varv_{\rm rms}$ and $\ell_\varv$ are the root mean square and the characteristic length scale of the nonaxisymmetric component of the surface convective velocity, respectively. In the covered parameter space, 
$Ro_{\rm surf} < 0.1$ is achieved for weakly-magnetized models with moderate stratification (typically, up to $N_\rho \approx 3.5$ and $Pm \approx 3$).
In the strong branch (i.e., $Pm\gtrsim 10$), \cite{Menu2020} and \cite{Zaire2022} showed that the dipolar branch can survive at higher values of $Ro_{\rm surf}$ since the Lorentz force starts playing an important role and competing turbulent inertia. \cite{Raynaud2015} also demonstrated that the stability domain of the dipolar branch is associated with a narrower and narrower range of values of $Ra/Ra_{\rm c}$ as the density stratification increases, so that it seems more and more difficult to generate magnetic dipoles. We show the surface map of the radial component of the magnetic field for such a dipolar model in \figurename{}~\ref{fig:axial}, with $N_\rho=3.5$ and $Ra/Ra_{\rm c} \approx 2.5$. This snapshot exhibits relatively small-scale structures close to the surface, even if the dipole character remains obvious. The butterfly diagram in \figurename{}~\ref{fig:but_dip} clearly confirms the mean dipole behavior of the model over a magnetic diffusion timescale. At the very beginning of the time series, we can see the occurrence of a polarity reversal, which is common in such a model \citep[e.g.,][]{Menu2020}. As another diagnosis, we can also consider the time series of the $f_{\rm dip}^{\ell_{\rm cut}}$ indicator, which is defined as the surface ratio of the dipole magnetic energy to the magnetic energy of the angular degrees between $\ell=1$ and a certain cut-off degree $\ell_{\rm cut}$, that is,
\algn{
f_{\rm dip}^{\ell_{\rm cut}} =\left(\dfrac{\sum_{m=-1}^{1} \int \vec{B}^2_{\ell=1,m}(r_{\rm o},\theta,\varphi)  \sin \theta \dd \theta \dd \varphi}{\sum_{\ell=1}^{\ell_{\rm cut}} \sum_{m=-\ell}^{m=+\ell} \int \vec{B}^2_{\ell,m} (r_{\rm o} \theta, \varphi) \sin \theta \dd \theta \dd \varphi}\right)^{1/2} \; .
\label{f_dip}
}
This quantity is plotted for the present axial dipole in \figurename{}~\ref{f_dip plot}. Aside from the short reversal event at the very beginning of the time series, we can see that $f_{\rm dip}^{\ell_{\rm cut}}$ remains higher than $0.6$ for any value of $\ell_{\rm cut}$, as expected. A value of $\ell_{\rm cut}=12$ is generally considered when tackling the question of the Earth's core magnetism since it corresponds to the maximal spatial resolution accessible by observations \citep[e.g.,][]{Christensen2006}. Using the same value, we can see that $f_{\rm dip}^{12}$ drops on average around $0.5$ during the reversal event, showing the partial transfer of energy from the $\ell=1$ components toward higher degrees. It is also worth considering smaller values for $\ell_{\rm cut}$. For instance, by choosing $\ell_{\rm cut}=5$, we can see that $f_{\rm dip}^5$ remains above $0.5$ even during the reversal event. The choice of $\ell_{\rm cut}$ naturally appears important if one wants to define relevant diagnostics of the dipolar field strength in stars. This is true in particular when confronting theoretical models to observations, the latter being limited regarding the spatial resolution (see also discussion in \sectionname{}~\ref{disc:morph}).

\paragraph{Equatorial dipoles:}

In some cases, the generated magnetic field can also be structured as a large-scale dipole orthogonal to the rotation axis, that is, with the energy mainly contained in the $\ell=1$ and $|m|=1$ components; this is commonly called an equatorial dipole \citep[e.g.,][]{Schrinner2014,Raynaud2014}. The surface radial component of the magnetic field for a model exhibiting such a topology is plotted in \figurename{}~\ref{fig:equat}, with $Ra=2.48~Ra_{\rm c}$ and $N_\rho=3$. This configuration is generally observed for models close to the onset of convection and dynamo threshold, and at low values of the Rossby numbers, where it is favored by the columnar shape of the convective flow. This columnar structure can be discerned near the equator in \figurename{}~\ref{fig:equat}. While the magnetic axis can remain near the equator over time at low values of $Ra/Ra_{\rm c}$, it may start oscillating around the equator over time as the value of $Ra/Ra_{\rm c}$ is progressively increased. This is actually the case for this model, as can be seen in the butterfly diagram plotted in \figurename{}~\ref{fig:but_equat}. The mean polarity alternates between positive and negative values and are always opposite in each hemisphere. To better reveal this oscillation, it is instructive to show the $f_{\rm dip}^{\ell_{\rm cut}}$ indicator and the contribution of the axial component alone (i.e., $m=0$), denoted with $f_{\rm dip}^{\rm ax}$; this is plotted in \figurename{}~\ref{f_dipax}. As we can see, the value of $f_{\rm dip}^{\ell_{\rm cut}}$ remains on average around $0.5$. In contrast, the axial contribution $f_{\rm dip}^{\rm ax}$ oscillates between $0$ and a maximum of about $0.4$, the maximum being reached when the dipole axis tends to align with the rotation axis. Otherwise, $f_{\rm dip}^{\rm ax}$ is very low. Indeed, as $Ra/Ra_{\rm c}$ is increased from the onset of convection, an equatorial prograde jet progressively forms and enters at some point in conflict with the nonaxisymmetric part of the magnetic field. The system then leads to the generation of an axisymmetric dipole, which in contrast tends to kill the differential rotation. A nonaxisymmetric equatorial dipole is then rebuilt, whence the oscillations observed in the butterfly diagram. Increasing slightly again the value of $Ra/Ra_{\rm c}$, the (axial) dipole branch discussed in the previous paragraph may be retrieved with adequate initial conditions, especially for large values of $Pm$. Otherwise, the reinforcement of the differential rotation level as $Ra/Ra_{\rm c}$ increases tends to disrupt the emergence of equatorial dipoles and favors multipolar morphology, which we now discuss \citep[e.g.,][]{Raynaud2015}.

\paragraph{Multipolar dynamos:}

When the conditions to sustain a dominant large-scale dipolar magnetic field are not met, the possible magnetic topologies are generally encapsulated into the complementary branch, that is, the so-called multipolar branch \cite[e.g.,][and references there in]{Kutzner2002,Christensen2006,Goudard2008,Gastine2012,Schrinner2014}. We distinguish here two sub-classes in this branch. First, we have hemispherical dynamos for which a given hemisphere is significantly more magnetized than the other one. This kind of models is generally observed for moderate values of $Ra/Ra_{\rm c}$ \citep{Schrinner2014,Raynaud2014,Raynaud2016}.
As an illustration, we plot in \figurename{}~\ref{fig:hemi} the surface radial component of the magnetic field for such an hemispherical model, with $Ra/Ra_{\rm c}=4$ and $N_\rho=6$. As this figure suggests, this configuration is the result of the superposition of a symmetric and an antisymmetric dynamo modes with respect to the equator that share similar amplitudes, leading to the global canceling of the magnetic field in one given hemisphere.
We can see in the butterfly diagram plotted in \figurename{}~\ref{fig:but_hemi} that this cancellation is maintained over about one magnetic diffusion timescale. In the magnetized hemisphere, the dynamics seems noisy close to the pole but oscillating features tend to emerge and survive very close to the equator, somehow overshadowed by the antisymmetric mode. Such equatorial dynamo waves remarkably look like the structures observed in the second classes of the multipolar branch, that is, the so-called coherently oscillating dynamos. An example of models exhibiting such a behavior is plotted in \figurename{}~\ref{fig:multi}, for $N_\rho=6$ and $Ra/Ra_{\rm c}=8$. This kind of dynamo is generally obtained for large values of $Ra/Ra_{\rm c}$. In this case, the convective velocity fluctuations are large enough and the inertia dominates the Coriolis force in a substantial part of the convective bulk so that any dipolar component would collapse. As shown by \cite{Raynaud2015}, this seems to be met at lower and lower values of $Ra/Ra_{\rm c}$ as $N_\rho$ increases. As seen in \figurename{}~\ref{fig:multi}, the dynamics is dominated by small-scale structures. Nevertheless, the butterfly diagram in \figurename{}~\ref{fig:but_multi} shows that the magnetic structures migrate coherently on average from the equator to the poles in both hemispheres. We note that the migration tends to be quasi symmetrical with respect to the equator, even if some asymmetries persist. These oscillatory dynamo modes have been shown to be well explained by the Parker's wave formalism to a good approximation \citep{Parker1955,Yoshimura1975,Gastine2012,Warnecke2014,Duarte2016,Warnecke2018}. Nevertheless, even if this attractive scenario has long been proposed to explain the origin of the solar cycle, it may not explain all the complexity of stellar magnetic cycles \cite[e.g.,][]{Baudoin2016,Raynaud2016,Strugarek2017,Strugarek2018,Brandenburg2017,Beer2018,Brun2022,Ortiz2023}.

\subsection{Magnetic quenching of the differential rotation}
\label{quenching}

\begin{figure*}
\captionsetup[subfigure]{justification=centering}
\begin{subfigure}[c]{0.24\hsize}
\centering
\includegraphics[width=0.95\hsize]{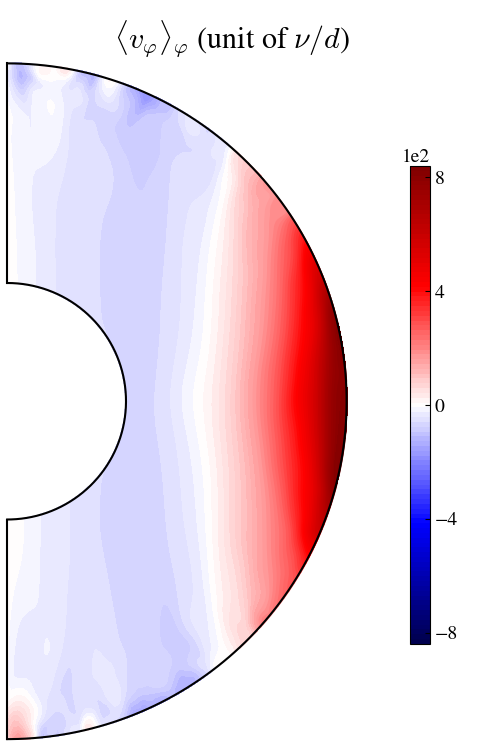}
\caption{Hydro}
\label{fig:zonala}
\end{subfigure}
\begin{subfigure}[c]{0.24\hsize}
\centering
\includegraphics[width=0.95\hsize]{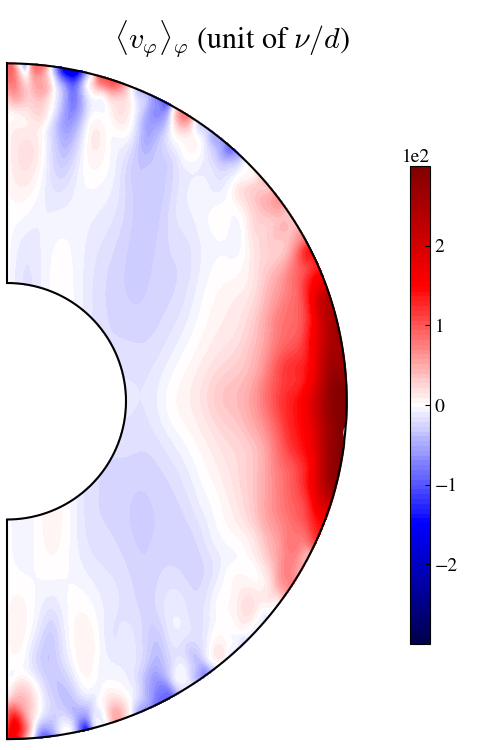}
\caption{$Pm=1$}
\label{fig:zonalb}
\end{subfigure}
\begin{subfigure}[c]{0.24\hsize}
\centering
\includegraphics[width=0.95\hsize]{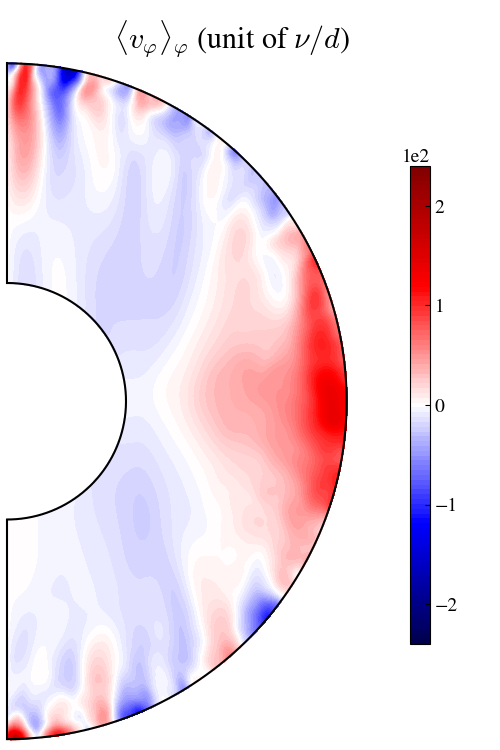}
\caption{$Pm=2$}
\label{fig:zonalc}
\end{subfigure}
\begin{subfigure}[c]{0.24\hsize}
\centering
\includegraphics[width=0.95\hsize]{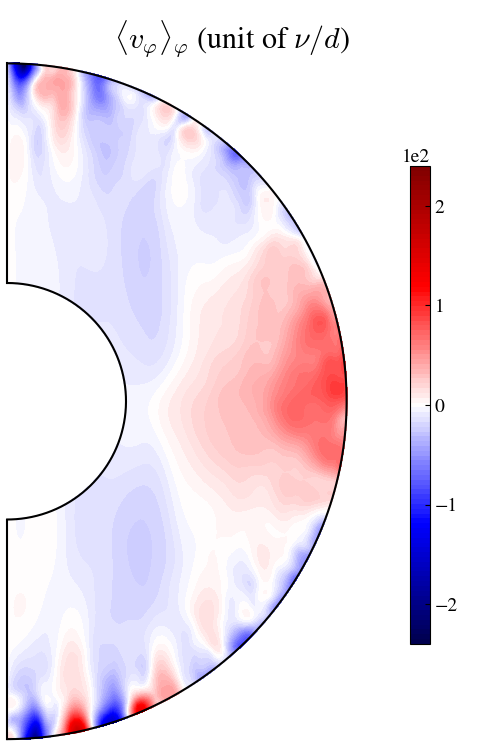}
\caption{$Pm=3$}
\label{fig:zonald}
\end{subfigure}
\caption{Meridional slices of  the azimuthally averaged zonal velocity, for different values of the magnetic Prandtl number $Pm$, in unit of the viscous velocity scale $\nu/d$. Models with $Pr=1$, $E=3\times 10^{-4}$, $N_\rho = 6$, and $Ra = 8~Ra_{\rm c}$.}
\label{fig:zonal}
\end{figure*}

One of the most important properties of thermal convection in rotating shells is the generation of a large-scale differential rotation \citep{Christensen2002,Aurnou2007,Gastine2012,Gastine2013,Guerrero2013,Brun2017a} which can play an important role in the dynamo process \citep[e.g.,][]{Schrinner2011,Schrinner2012,Brun2017b,Kapyla2023}. In Paper I, the outgoing surface heat flux has been shown to remain uniform when the influence of turbulent inertia on the convective motions dominates the Coriolis acceleration close to the surface, that is, when $Ro_{\rm surf} \gtrsim 1$, which is usually met for large values of $Ra$. In contrast, when $Ro_{\rm surf} \lesssim 1$, the Coriolis acceleration can significantly affect the surface turbulent convective rolls and couple the heat flux with the surface differential rotation. In this case, the presence of a strong prograde equatorial jet at the surface is associated with a darker equatorial band.

Before addressing the impact of the magnetic fields on the surface heat flux distribution, it is thus worth questioning its impact on the differential rotation.
To illustrate this point, we consider models with $N_\rho=6$, $Ra/Ra_{\rm c}=8$, $Pr=1$, and $E=3\times 10^{-4}$, and plot their azimuthally averaged zonal velocity in \figurename{}~\ref{fig:zonal}. First, in \figurename{}~\ref{fig:zonala}, we consider the hydrodynamic case without magnetic field. We see that the mean zonal flow exhibits a cylindrical geometry, as expected from the Taylor-Proudman constraint. The outer column is prograde and the inner cylinder embedding the inner core is retrograde. Indeed, the convective Rossby number remains smaller than unity in a substantial part of the convective bulk and the turbulent-mixed superficial layers are not able to evince the resulting positive angular momentum deposition induced by Reynolds stress in the external layers \citep{Aurnou2007,Gastine2013}. In turn, the inner part has to rotate in the opposite sense in order to conserve the total angular momentum. Then, accounting for dynamo magnetic fields, we are left with a new control parameter, which is the magnetic Prandtl number $Pm$. In \figurename{}~\ref{fig:zonalb}, we show the result obtained with $Pm=1$. This model exhibits a coherently oscillating multipolar topology, similar to that illustrated in \figurename{}~\ref{fig:multi}. We see that the magnetic Lorentz force has two main effects on the mean zonal flow. First, it breaks the cylindrical geometry of the flow: the prograde region extends in deeper layers close to the equator while its latitudinal extension is reduced compared to the case without magnetic fields, and the retrograde region takes a conical shape. At the surface, the development of small-scale surface zonal jets at high latitudes is favored; these are very similar to the alternation of retrograde and prograde bands observed in Jupiter or Saturn, which have already been reproduced in numerical simulations \citep[e.g.,][]{Gastine2014.icarus,Heimpel2016,Christensen2020,Heimpel2022,Yadav2022}. Second, the magnetic field clearly reduces the intensity of the zonal mean flow (by a factor of about 3 for $Pm=1$, see the colorbar). In Figs.~\ref{fig:zonalc} and \ref{fig:zonald}, we plot the results for two other models computed by $Pm=2$ and $3$. We note that the model with $Pm=2$ is the same as the one considered in Figs.~\ref{fig:multi} and \ref{fig:but_multi}. When increasing the value of $Pm$, we actually increase the intensity of the magnetic field. Indeed, at a given value of $Ra/Ra_{\rm c}$, we expect the Reynolds number $Re=\varv_{\rm rms} d /\nu$, with $\varv_{\rm rms}$ the root mean square of the convective fluctuations of velocity, to remain about constant when varying $Pm$. In turn, when increasing $Pm$, the magnetic Reynolds number, $Rm=\varv_{\rm rms} d /\eta=Pm Re$, increases. In other words, the effect of the Ohmic dissipation decreases and the dynamo-induced magnetic fields have larger magnitudes. We can see in Figs.~\ref{fig:zonalc} and \ref{fig:zonald} that increasing the magnetic intensity leads to a decrease in the differential rotation. This clearly indicates that magnetic fields globally quench the differential rotation compared to the case without magnetic fields. We check that this trend holds for the set of models we have considered in this work, as we subsequently show in detail in \sectionname{}~\ref{brightness}. This result extends the conclusions of \cite{Yadav2016} and \cite{Kapyla2017} obtained for Boussinesq and moderately stratified models, respectively, to highly-stratified models and a wider parameter space: whatever their morphology, the self-sustained magnetic fields tend to globally quench the large-scale differential rotation in typical thick convective envelopes of cool stars. Since the dynamo is triggered at higher values of $Pm$ when $N_\rho$ is increased \citep[][]{Raynaud2015}, we also expect that the higher the density stratification, the larger the value of $Pm$ for magnetic fields to start quenching the differential rotation.


\section{Effect of self-sustained magnetic fields on the global surface heat flux distribution}

\label{brightness}

    \begin{figure*}
    \centering
    
    \begin{subfigure}[c]{\hsize}
    \includegraphics[trim=0cm 0cm 0cm 0cm, clip, width=0.49\hsize]{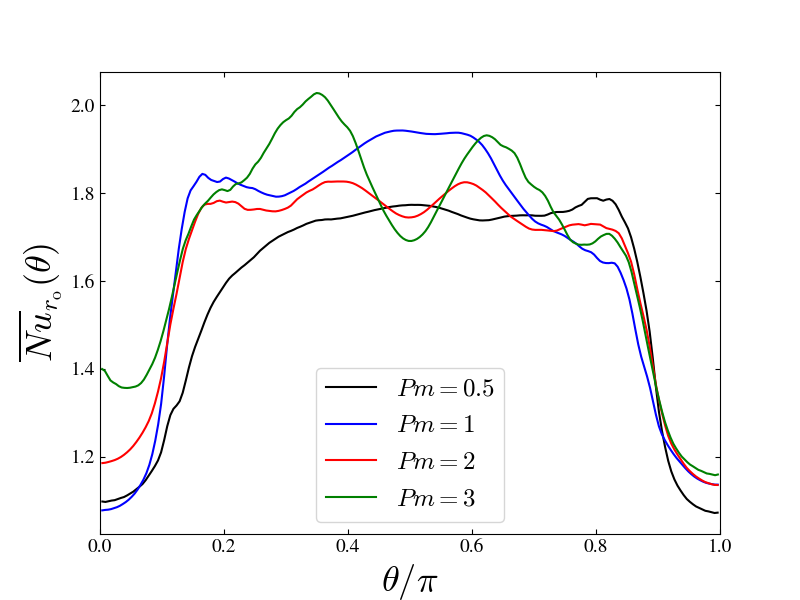}
    \includegraphics[trim=0cm 0cm 0cm 0cm, clip, width=0.49\hsize]{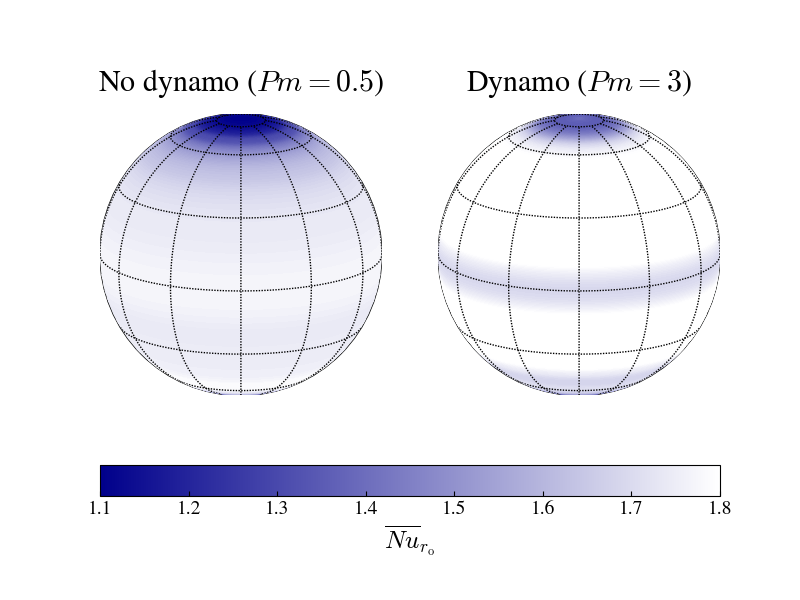}
    \caption{{\bf Onset of convection ($Ra \approx 1.7~Ra_{\rm c}$)}}
    \label{fig:Nu_onset}
    \end{subfigure}
    \par \medskip
    
    \begin{subfigure}[c]{\hsize}
    \includegraphics[trim=0cm 0cm 0cm 0cm, clip, width=0.49\hsize]{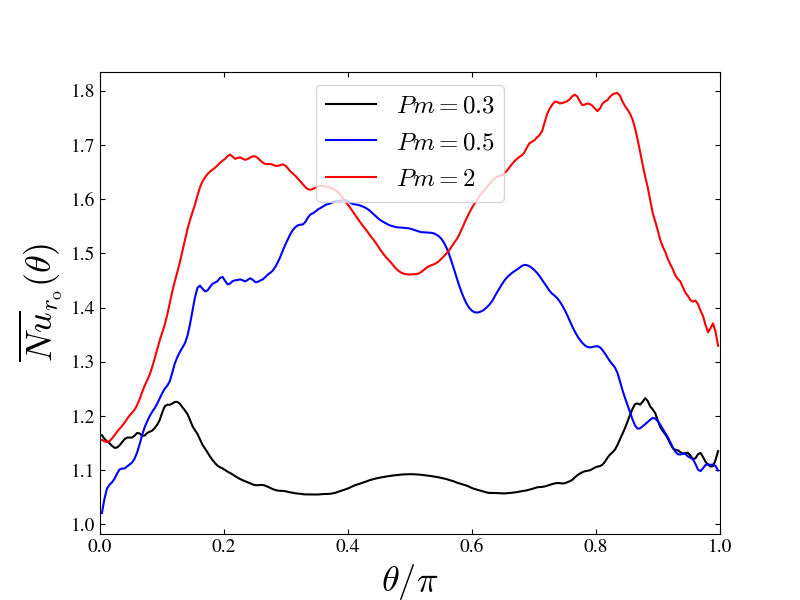}
    \includegraphics[trim=0cm 0cm 0cm 0cm, clip, width=0.49\hsize]{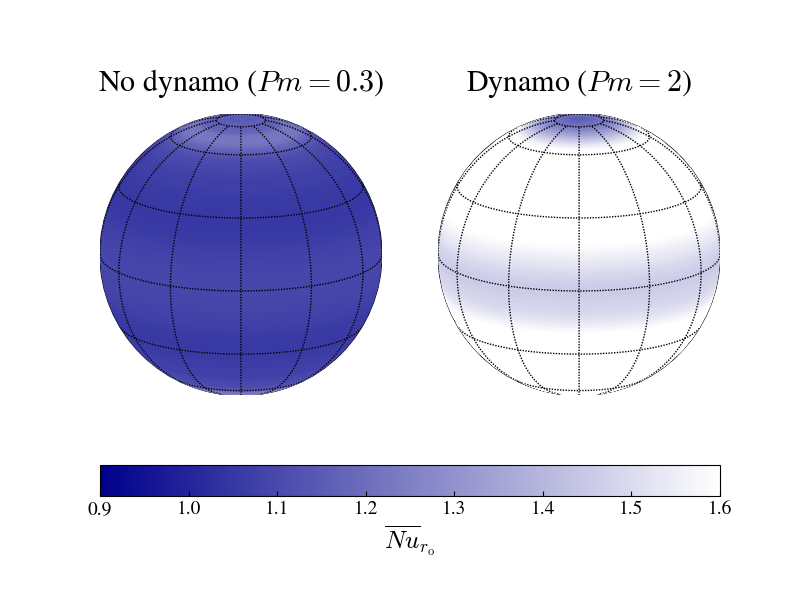}
    \caption{{\bf Transition toward turbulent flows ($Ra \approx 3~Ra_{\rm c}$)}}
    \label{fig:Nu_transition}
    \end{subfigure}
    \par \medskip
    
    \begin{subfigure}[c]{\hsize}
    \includegraphics[trim=0cm 0cm 0cm 0cm, clip, width=0.49\hsize]{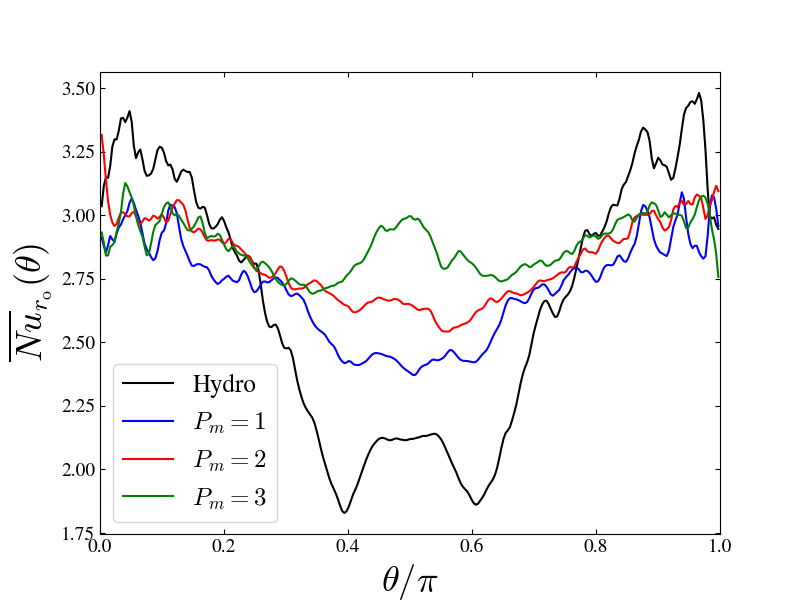}
    \includegraphics[trim=0cm 0cm 0cm 0cm, clip, width=0.49\hsize]{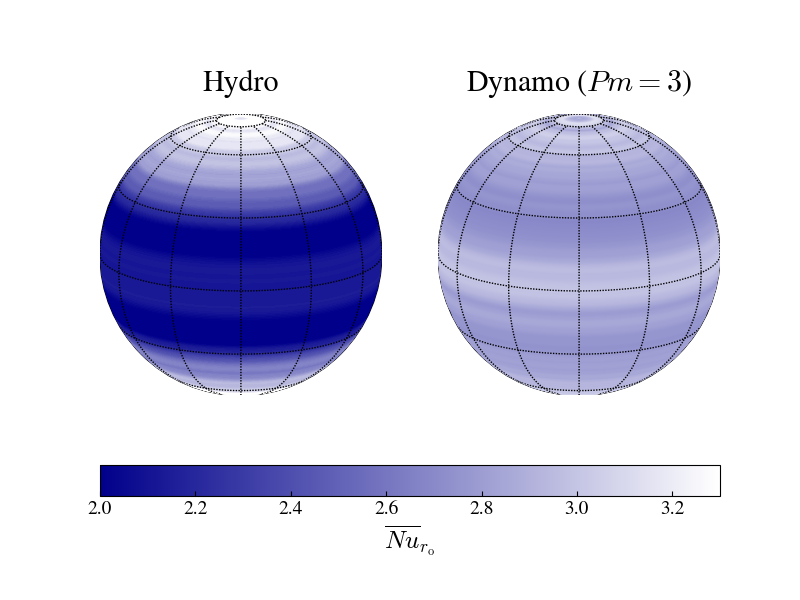}
    \caption{\textbf{Turbulent models ($Ra \approx 8~Ra_{\rm c}$)}}
    \label{fig:Nu_turb}
    \end{subfigure}
    \caption{Mean latitudinal profile of the surface Nusselt number defined in \eq{Nu}, for three typical models. (a) $E=3\times10^{-5}$, $Pr=1$, $N_\rho=3$, and $Ra\approx 1.7~Ra_{\rm c}$ close to the onset of convection (b) $E=3\times10^{-5}$, $Pr=0.3$, $N_\rho=4$, and $Ra\approx 3~Ra_{\rm c}$ at the transition toward turbulent flows (c) $E=3\times10^{-4}$, $Pr=1$, $N_\rho=6$, and $Ra\approx 8~Ra_{\rm c}$. {\bf Left:} Variations with the magnetic Prandtl number $Pm$. Solid black lines represent either pure hydrodynamic models or MHD models without magnetic dynamo (i.e., too low values of $Pm$). {\bf Right}: Spherical representation for the nonmagnetized model (i.e., the lowest value of $Pm$) and the most magnetized dynamo model (i.e., the highest value of $Pm$) in each case. Note that we use different colorbars in each row to enhance the visualization contrast.}
    \end{figure*}

After having reviewed the different magnetic topologies expected in stratified dynamo models, we now investigate the global effect of magnetic fields on the mean latitudinal distribution of luminosity at the surface of cool stars, with thick convective envelopes. To that end, we compute the mean latitudinal Nusselt number in \eq{Nu} for each simulation. We go step-by-step and first tackle the case of models close to the onset of convection before dealing with more turbulent models, which are more representative of stellar conditions. The results obtained from our set of simulations are summarized through the figures presented in \appendixname{}~\ref{app:Nu} and \ref{app:V}, showing the mean Nusselt and large-scale zonal flow profiles, respectively. We note that for most of our models, the flow is rather turbulent close to the surface especially for $N_\rho > 3$, so that the mean Nusselt profile may fluctuate over time and appear noisy in case the time average is performed on only few snapshots (e.g., see time variations in Fig.~2 of Paper I). We checked that averaging on few dozens of snapshots taken at fixed intervals over a magnetic diffusion timescale is sufficient to grasp the mean brightness and velocity profiles. This is systematically performed for each model. Typical standard deviations around the mean value of the Nusselt profile depends on the latitude but we expect through a visual inspection in few models that it is generally lower than 10-15\%; the mean brightness distribution as computed in this work thus appears relevant for a first step investigation.

\subsection{Close to the onset of convection}
\label{close to onset}

At the onset of convection, the first unstable modes take the form of columns aligned with the rotation axis. In the hydrodynamic case, it was shown in Paper I that the surface heat flux is preferentially conveyed close the equator. As the value of $N_\rho$ increases at a given value of $Ra/Ra_{\rm c}$, the convective cells get more and more concentrated near the surface layers, and the heat flux tends to be maximum near the tangent cylinder\footnote{The cylinder that is tangent to the inner spherical boundary and whose revolution axis is colinear with the rotation axis. For thick convective envelopes with $\chi=0.35$, the tangent cylinder location corresponds to the colatitudes $\theta/\pi\sim 0.1$ and $0.9$ at the surface.} while keeping an equator brighter than the poles. A typical example of the surface Nusselt profile in this case is shown by the solid black line in \figurename{}~\ref{fig:Nu_onset}, for a model with $N_\rho=3$, $E=3\times 10^{-5}$, $Pr=1$, and $Ra=1.7~Ra_{\rm c}$; the value of $Pm$ is equal to $0.5$, which is too low to trigger a magnetic dynamo (see \sectionname{}~\ref{morpho}).

We now investigate the effect of magnetic fields on the mean surface heat flux distribution, remaining first in a regime close to the onset of convection. In this case, the magnetic Prandtl number, $Pm$, is varied above the minimal critical value needed to trigger a self-sustained dynamo while attempting to cover both the dipolar (axial and equatorial) and multipolar branches (for given $E$, $Ra/Ra_{\rm c}$, $N_{\rho}$, and $Pr$). As explained in \sectionname{}~\ref{quenching}, this in addition allows us to quantify the impact of the intensity of the magnetic field on the surface Nusselt profile; the larger $Pm$ for given parameters, the larger the magnetic intensity.
For the previous illustrative model in \figurename{}~\ref{fig:Nu_onset}, we can clearly see that close to the onset of convection, magnetic fields generally tend to increase, on average, the integrated surface Nusselt number. In particular, the poles and the regions near the tangent cylinder can become brighter than in the hydrodynamic case as $Pm$ increases. This is a systematic trend observed in several models close to the onset of convection, as we can see in Figs.~\ref{fig:Np3a} and \ref{fig:Np3b}, Figs.~\ref{fig:onset_Np3a}-\ref{fig:onset_Np4a}, and Fig.~\ref{fig:onset_Np35}, for values of $N_\rho$ between $3$ and $4$ and $Ra/Ra_{\rm c}$ varying up to a value of $4$. In all these figures and in the subsequent sections of the paper, we note that the models represented by solid black lines are either purely hydrodynamic or without dynamo (i.e., with a value of $Pm$ too low to sustain a magnetic field). In this regime, a surface equatorial prograde jet starts developing but it is inhibited by magnetic fields, as expected from \sectionname{}~\ref{quenching}. For example, this can be seen in Figs.~\ref{fig:V_onset_Np4a} and \ref{fig:V_onset_Np35} for two models with $N_\rho=3.5$ and $4$, $Ra \sim 2.5~Ra_{\rm c}$. Similarly to the hydrodynamic case (see Paper I), the surface heat flux close to the onset of convection thus appears not to be anticorrelated with the surface differential rotation, but to be a property of the first unstable convective modes, in contrast with more turbulent models as we see in the next sections.

\subsection{Transition toward turbulent flows}

\label{transition}

\begin{figure}
    \centering
    \includegraphics[trim=0cm 0cm 0cm 0cm, clip, width=\hsize]{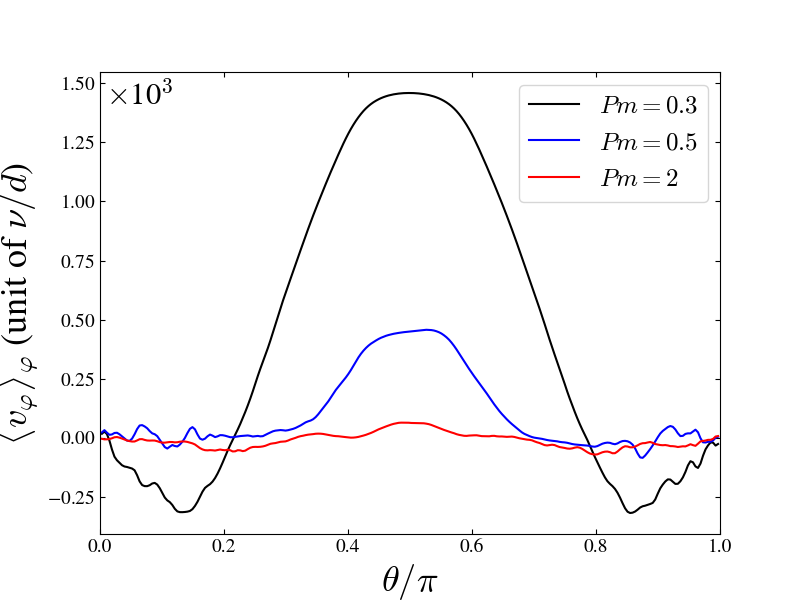}
    \caption{
    Mean surface azimuthal velocity as a function of the colatitude, for the models in Fig.~\ref{fig:Nu_transition}, i.e., with $E=3\times 10^{-5}$, $Pr=0.3$, $N_\rho=4$, and $Ra=3~Ra_{\rm c}$. The magnetic Prandtl number $Pm$ is varied and the solid black line represents the model with no dynamo (i.e., too low values of $Pm$).}
    \label{fig:V_transition_Np4b_main}
\end{figure}

Continuously increasing the Rayleigh number from the onset of convection progressively reinforces the equatorial prograde jet. As already explained, this jet results from the Reynolds stress induced by coherent convective rolls, and that can redistribute positive angular momentum to the mean flow in the outer layers. This typical enhancement of the surface differential rotation as the Rayleigh number is increased from the onset of convection toward turbulent flows is observed in most of our models with no dynamo (e.g., see Figs.~\ref{fig:V_Np4}, \ref{fig:V_Np35} and \ref{fig:V_Np6_3}-\ref{fig:V_Np6_8}, solid black lines). When the equatorial jet is strong enough, it progressively inhibits the outgoing heat flux close to the equator, making at some point the equator as bright as or even darker than the poles, as expected from Paper I. A typical example of the surface Nusselt profile in this situation is shown in \figurename{}~\ref{fig:Nu_transition} (solid black line) for a model with $E=3\times10^{-5}$, $Pr=0.3$, $N_\rho=4$, and $Ra \approx 3~Ra_{\rm c}$. When adding magnetic fields, we expect according to \sectionname{}~\ref{quenching} the surface differential rotation to be quenched by the Lorentz force.  For this example model, Fig.~\ref{fig:V_transition_Np4b_main} indeed shows that the prograde equatorial jet slows down as the magnitude of the magnetic field (i.e., the value of $Pm$) increases. As seen in Fig.~\ref{fig:Nu_transition}, the equator concomitantly becomes brighter than the poles when going from $Pm=0.3$ to $Pm=0.5$ (black and blue lines). Increasing $Pm$ to a value of two in a second step (red line), we can see that the heat flux tends to increase in the region inside the tangent cylinder, producing a slight dip in the vicinity of the equator. The change in the Nusselt profile resulting from this second increase in $Pm$ is very similar to the trends observed for models close to the onset of convection and depicted in \sectionname{}~\ref{close to onset}. As a matter of fact, in weakly turbulent convection, the emergence of dynamo magnetic fields can retroactively lead to the laminarization of the flow \citep[e.g.,][]{Petitdemange2018}. Therefore, the properties of the convective shell are then expected to be very similar to those of a model with a slightly lower value of the Rayleigh number, that is, close to the linear onset of convection. For the models with $Pm=0.5$ and $Pm=2$ represented in \figurename{}~\ref{fig:V_transition_Np4b_main}, we can check this interpretation holds valid in a simple way by considering the model in \figurename{}~\ref{fig:onset_Np4a} obtained with the same parameters but with a slightly lower value of $Ra/Ra_{\rm c}$. Although the mean value of the Nusselt number remains naturally larger for the models with the largest value of $Ra$, as expected from simple scaling laws, we see that the shape of the profiles are very similar in both figures, even between the nonmagnetic and the magnetic models at $Pm=0.5$, hence supporting the laminarization effect. 

Our set of simulations comprises other examples of such inversion. A clear example can be found in \figurename{}~\ref{fig:Np3c} for $E=3\times10^{-5}$, $Pr=1$, $N_\rho=3$, and $Ra \approx 3.7~Ra_{\rm c}$. In this figure, the model with $Pm=1$ exhibits an hemispherical dynamo while the model with $Pm=3$ has a dominant dipole component. The laminarizing effect of the magnetic field is again suggested when comparing this figure with the models in \figurename{}~\ref{fig:Nu_onset} for $Ra\approx 1.7~Ra_{\rm c}$. Another example can be found in \figurename{}~\ref{fig:Nu_Np6_transition}, for a model with $E=3\times 10^{-4}$, $Pr=1$, $N_\rho=6$, and $Ra \approx 3~Ra_{\rm c}$, exhibiting an oscillatory dynamo. We note that this model with such a low value of $Ra/Ra_{\rm c}$ requires values of $Pm$ larger than about $6$ to trigger a magnetic dynamo.
We thus conclude that, whatever the magnetic topology , magnetic fields can globally induce an inversion of the pole-equator brightness contrast, going from a darker to a brighter equator as the magnetic intensity increases. This generally happens for models at the transition between the onset of convection and more turbulent flows.

\subsection{Turbulent models}

We finally consider turbulent models with values of $Ra$ much larger than the threshold value above which the equator becomes darker than the poles (i.e., typically $Ra \gtrsim 5~Ra_{\rm c}$). As an illustration, we plot in \figurename{}~\ref{fig:Nu_turb} the Nusselt profile of such models with $E=3\times 10^{-4}$, $Pr=1$, $N_\rho=6$, and $Ra \approx 8~Ra_{\rm c}$. We remind that the mean azimuthal velocity of these models is represented in \figurename{}~\ref{fig:zonal}. The associated latitudinal profiles at the surface are also plotted in \figurename{}~\ref{fig:V_Np6_8} for more details. As already shown in Paper I and in \sectionname{}~\ref{transition} for less turbulent models, we can perceive the strong correlation between the prograde jet and the dark band close to the equator in the hydrodynamic case.
Dynamo models are then compared in the same figures.
We can see in \figurename{}~\ref{fig:Nu_turb} that increasing $Pm$ and thus the magnetic field intensity tends to flatten the pole-equator brightness contrast. This is actually a general trend that is systematically observed in our turbulent models. This is summarized in \appendixname{}~\ref{app:Nu} through Figs.~\ref{fig:turb_Np4}, \ref{fig:turb_Np35}-
\ref{fig:turb_Np5e} and \ref{fig:turb_Np6a}-\ref{fig:turb_Np6d}, for values of $N_\rho$ between $3.5$ and $6$, and values of $Ra/Ra_{\rm c}$ between about $5$ and $30$. For such values of $Ra/Ra_{\rm c}$, the models generally exhibit a multipolar magnetic structure. 
This result suggests that far enough from the onset of convection, the magnetic fields globally tend to reduce the surface brightness contrast between the poles and the equator. This effect can be significant: if $Pm$ is large enough, a luminosity contrast between the equator and the pole of about 150\% such as depicted in Paper I in the hydrodynamic case can be totally smoothed out. 

Similarly to the transition models discussed in \sectionname{}~\ref{transition}, the global reduction of the pole-equator brightness contrast depicted in \figurename{}~\ref{fig:Nu_turb} is clearly correlated with the quenching of the surface differential rotation by the magnetic fields, as we can see \figurename{}~\ref{fig:V_Np6_8}. This is again a systematic trend observed in all the turbulent models of our sample. Several examples can be found for instance in Figs.~\ref{fig:V_turb_Np4c}, \ref{fig:V_turb_Np35}-\ref{fig:V_turb_Np5d} and \ref{fig:V_Np6_4}-\ref{fig:V_Np6_16}, for values of $N_\rho$ between $3.5$ and $6$, as well as values of $Ra/Ra_{\rm c}$ between about $5$ and $20$. For the models with no dynamo in these figures (solid black lines), the prograde equatorial jet reinfoces as $Ra$ increases, as noted before. In contrast, increasing $Pm$ in all these figures results in the reduction of the large scale differential rotation, which is directly responsible for the homogenization of the surface heat flux in this regime.

When increasing the value of $Ra/Ra_{\rm c}$ toward very large values (i.e., typically $Ra/Ra_{\rm c} \gtrsim 30$ in our sample), the effect of inertia on the flow then starts dominating the Coriolis force at some point, reducing ipso facto the angular momentum transfer from the small scales to the mean flow and hence, the  magnitude of the surface differential rotation. In order to conserve the total angular momentum, the system has no choice but to generate a retrograde equatorial zonal jet at the surface, with a sharp transition between both regimes \citep[e.g.,][]{Gastine2013}. In the hydrodynamic case, we have shown in Paper I that the mean surface brightness contrast tends to decouple from the differential rotation and homogenize as the surface convective envelopes become more and more turbulent. Exploring this kind of dynamo models in the present parametric study is difficult since they require long numerical integration. We census only eight models around the transition between a solar (prograde) and an antisolar (retrograde) differential rotation in our set of simulations. First, we have two models with $E=10^{-4}$, $Pr=1$, $N_\rho=5$, and $Ra=32~Ra_{\rm c}$, whose surface azimuthal velocity profiles are plotted in \figurename{}~\ref{fig:V_turb_Np5e}. A quick comparison with the nonmagnetic models at lower values of $Ra/Ra_{\rm c}$ in Figs.~\ref{fig:V_turb_Np5a}-\ref{fig:V_turb_Np5d} (black lines) suggests that these models are  very close to the transition: indeed, we can see in \figurename{}~\ref{fig:V_turb_Np5e} that the equatorial jet starts decreasing slightly around $Ra=32~Ra_{\rm c}$ in the hydrodynamical case (black line). We see in the same figure that the dynamo magnetic fields also quench the differential rotation in this case (blue line), without any clear differences with less turbulent models. Second, we have six other models with $E=10^{-4}$, $Pr=1$, $N_\rho=4$, and $Ra \approx 29~Ra_{\rm c}$, and with $E=3\times 10^{-4}$, $Pr=1$, $N_\rho=6$, and $Ra \approx 32~Ra_{\rm c}$, the surface azimuthal velocity profiles of which are plotted in Figs.~\ref{fig:V_turb_Np4_3} and \ref{fig:V_Np6_32}, respectively. In both figures, we can see that the models without magnetic fields (solid black lines) exhibit a solar differential rotation. Regarding the dynamo models computed with the same parameters, we see either that the zonal equatorial jet is quenched or that a strong antisolar jet develops when adding magnetic fields; we nevertheless do not depict any clear dependence on the value of $Pm$. Actually, the transition between a solar and an antisolar differential rotation is known to be hysteretic in the hydrodynamic case and thus depends on the initial conditions \citep[e.g.,][]{Gastine2014.mnras}. In contrast, \cite{Karak2015} gave some evidence that dynamo magnetic fields may remove the flow bistability in this regime. In this context, it is thus difficult to explain our results through a simple trend as a function of $Pm$. In any case, the surface Nusselt profiles of these eight models displayed in Figs.~\ref{fig:turb_Np43}, \ref{fig:turb_Np5e} and \ref{fig:turb_Np6d} demonstrate that the surface brightness contrast is always reduced compared to the hydrodynamic case. This indicates that the main trend observed for slightly less turbulent models with a solar differential rotation regime also holds true for more turbulent models with an antisolar differential rotation. We nonetheless acknowledge that the transition regime toward an antisolar differential rotation deserves a dedicated study \citep[e.g.,][]{Simitev2015,Noraz2022a,Noraz2022b}, which is beyond the scope of this paper.


\section{Discussion}

\label{discussion}

In this section, we discuss the results of this work in view of the past studies and observational context.

\subsection{Magnetic brightness flattening: applicability and limitations}
\label{applicability}

\begin{figure}
   \centering
   \includegraphics[width=\hsize,trim=0.5cm 0.cm 1.5cm 1.5cm,clip]{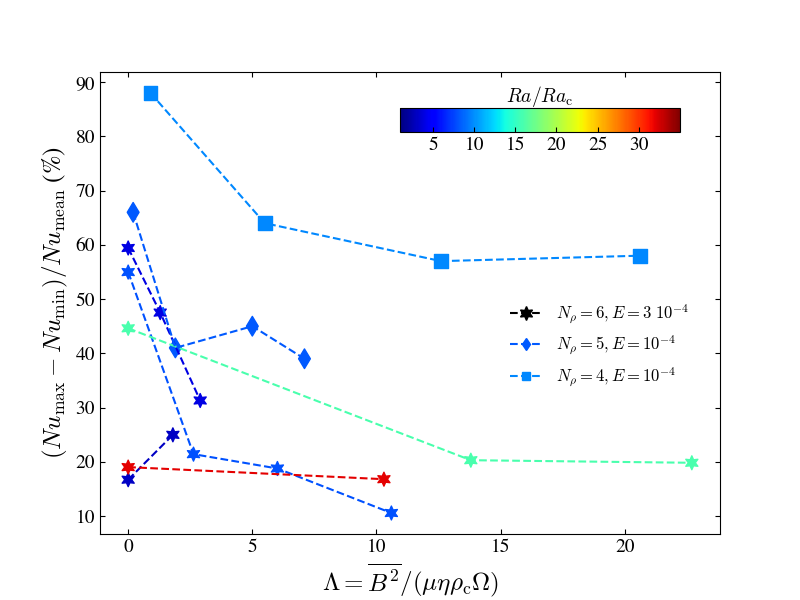}
      \caption{Maximum brightness contrast as a function of the mean Elsasser number, i.e., a proxy of the mean magnitude of the magnetic field, computed for models with $E=3\times10^{-4}$, $N_\rho=6$ (stars), $E=10^{-4}$, $N_\rho=5$ (diamonds), and $E=10^{-4}$, $N_\rho=4$ (squares); the color codes the Rayleigh number. For all these models, $Pr=1$.
              }
         \label{DNu_Np6}
   \end{figure}

Our parametric study demonstrates that magnetic fields and the mean brightness contrast between the poles and equator in rotating cool stars (i.e., with relatively thick turbulent convective envelopes) are tightly related. This was shown to result from the systematic quenching of the large-scale differential rotation by the magnetic fields (see \sectionname{}~\ref{quenching}), which in turn affects the convective transport of energy, as previously expected from \cite{Yadav2016} and Paper I. The larger the magnetic intensity, the lower the differential rotation, and the lower the surface heat flux variations at the surface. Moreover, this turns out to hold true at leading order whatever the magnetic topology. This is illustrated in \figurename{}~\ref{DNu_Np6} for some of the most stratified models of our sample with $Pr=1$, $N_\rho$ between $4$ and $6$, and different values of $Ra/Ra_{\rm c}$; the mean surface Nusselt and zonal flow of these models are plotted in Figs.~\ref{fig:Nu_Np6} and \ref{fig:V_Np6} for $N_\rho=6$, in Figs.~\ref{fig:turb_Np5c} and \ref{fig:V_turb_Np5c} for $N_\rho=5$, and Figs.~\ref{fig:turb_Np42} and \ref{fig:V_turb_Np4_2} for $N_\rho=4$. In \figurename{}~\ref{DNu_Np6}, the relative difference between the maximum and minimum values of the surface Nusselt number (normalized by its mean value) is plotted as a function of the global Elsasser number~$\Lambda = \overline{B^2}/(\Omega \rho_{\rm c}\mu \eta)$.
Since the surface Nusselt profile can sometimes exhibit localized oscillating variations around the mean trend, its maximum and minimum values are taken by computing a rolling average over a latitude interval of 10\textdegree~beforehand.

First, Fig.~\ref{DNu_Np6} shows that the brightness contrast tends to increase as $N_\rho$ decreases whatever the value of $\Lambda$, as expected from Paper I. Indeed, as $N_\rho$ increases, the surface convective Rossby number tends to increase so that the turbulent standardization of the surface heat flux is more efficient (e.g., see Figs.~5 and 7 in Paper I). The trend observed in Paper I in the hydrodynamical case is thus shown to hold valid when magnetic fields set in.
Second, focusing on $N_\rho=6$ models only, we can observe a nonmonotonic behavior as a function of $Ra/Ra_{\rm c}$ in the nonmagnetic case (i.e., $\Lambda=0$); the brightness contrast first increases from $Ra/Ra_{\rm c}=3$ to $Ra/Ra_{\rm c}=4$, before decreasing for larger values of $Ra/Ra_{\rm c}$. This change of regime has also already been observed in Paper I. It has been shown to result initially from the progressive reinforcement of the prograde equatorial jet and, then, the progressive homogenization of the heat transport as turbulence is enhanced. We have checked that the surface Rossby number of these models increases from about~$0.1$ to about $2$ as $Ra/Ra_{\rm c}$ increases from $3$ to $32$, that is, covering the above-mentioned transition from the onset of convection toward the start of the convective homogenization  (e.g., see Fig. 7 in Paper I). Besides, in the dynamo case, we do distinguish two regimes already depicted in \sectionname{}~\ref{brightness}:
\begin{itemize}
    \item On the one hand, the model with $N_\rho=6$ and $Ra/Ra_{\rm c}=3$ exhibits a typical reversal of the pole-equator brightness contrast in presence of magnetic fields, resulting in an increase in its value as $\Lambda$ increases. Such a case is associated with models close to the linear onset of convection and, although being of interest from a theoretical point of view, it is likely not to be applicable to very turbulent stellar convective envelopes.\\
    \item  On the other hand, for more turbulent models, we clearly see the systematic decrease of the surface brightness contrast as $\Lambda$ increases. This relation between the magnetic field intensity and the global brightness contrast, although counterintuitive at first, has been demonstrated to be physically grounded and could have implications for interpreting the combination of photometric and spectropolarimetric data.
\end{itemize}
From a quantitative point of view, the surface heat flux contrast predicted by our models in \figurename{}~\ref{DNu_Np6} can reach an upper limit of about 100\% of the mean value in the hydrodynamic case (see also Paper I). In order to interpret such differences, it is worth translating them in terms of temperature variations. By construction and for sake of simplicity, a constant temperature (or entropy) is imposed at the surface of our models. A simple way to circumvent this limitation is to use the Stefan-Boltzmann's law at the outer sphere to link the relative variations in the Nusselt number with those in temperature, that is, \smash{$\Delta Nu / Nu \sim 4 ~\Delta T / T  $.} The surface temperature deviations are therefore expected to reach about 25\% at most in our simulations. As an illustration, we know that the temperature contrast between the quiet photosphere and cool nonaxisymmetric stellar spots in G, K, and M dwarfs can reach up to about 20\% \citep[e.g.,][]{Solanki2003,Afram2015,Garcia2022}. In absence of additional constraints on the mean axisymmetric brightness distribution in thick stellar convective envelopes (see \sectionname{}~\ref{test}), this comparison, although simplistic, does not rule out the somehow large surface temperature and brightness contrasts predicted by the present set of simulations; the large relative Nusselt deviations predicted and represented in \figurename{}~\ref{DNu_Np6} thus do not appear untenable.

Nevertheless, before going further in the comparison, it is necessary to discuss in more detail the applicability and limitations of these findings in an astrophysical context.Indeed, the parameter space covered by our simulations is ineluctably limited and far from stellar regimes for numerical reasons. As a first attempt to compare simulations and actual stars, we usually rely on
the predominant force balance. In all our models, the geostrophic balance (i.e., between the pressure gradient and the Coriolis acceleration) is globally met at large scales. Regarding the convective fluctuations around this global state, we have seen that the local Rossby number, $Ro_{\rm surf}$, lies between $0.1$ and about unity at the surface, and thus is even smaller in deeper layers where it decreases (e.g., see Fig.~5 in Paper I). Therefore, the Coriolis acceleration is mainly responsible for the dynamical properties of convection, even if
buoyancy starts acting in the surface layers for the most turbulent
models. Using such considerations, we can attempt to define a first applicability domain: cool stars with relatively thick convective
envelopes and $Ro_{\rm surf} \lesssim 1$. This actually corresponds to the range of values where the contrast in
the brightness distribution is expected to be the largest in the
hydrodynamic case, with predicted contrasts reaching up to 150\% (e.g., see Fig. 7 of Paper I).
According to observations and stellar modeling, this low Rossby number regime should be met in most of the K and M dwarfs \citep[e.g.,][]{Wright2011,Wright2018} as well as pre-main sequence stars \citep[e.g.,][]{Landin2023}. However, this is not met for the other usual type of cool stars, namely evolved red giant stars. Indeed, using asteroseismic data, we know for such stars that the surface granulation turnover timescale $\tau_{\rm conv}\lesssim 1$ day and that the mean rotation period, $P \gtrsim 10$ days \citep[e.g.,][]{Samadi2013,Peralta2018,Gehan2018}, which permits us to estimate a lower limit for the surface local Rossby number, that is, $Ro_{\rm surf} \sim P / \tau_{\rm conv} \gtrsim 10$. Such values are out of the parameter space considered in the present paper. Besides, it is also worth mentioning RS CVn close binary stars in which the red-giant component can rotate very rapidly. However, the presence of strong tides adds important dynamical effects that go beyond the framework of this work. More detailed investigations are needed in order to tackle the regime of large Rossby number, which will demand to study the transition from a solar to an antisolar differential rotation and very turbulent flows. According to Paper I and this work, we expect a rather uniform mean brightness distribution as the turbulence increases at the surface and the heat flux decouples from the differential rotation. We recall that this agrees with the Sun whose surface Rossby number is quite large and photosphere is uniform at leading order, regardless of the strong prograde differential rotation. Of course, another current question is to understand the formation and persistence of such strong prograde differential rotation in such regimes, but this is out of the scope of this paper. As a consequence, to tackle the regime of large Rossby numbers, it will be necessary to go beyond the large-scale and mean diagnostic used in this work and to account for other higher-order effects such as the impact of small-scale and time-variant surface magnetic features.

\subsection{Observational tests and potential implications}
\label{test}

In order to test our theoretical results in a direct way, we need observational targets for which we are able to measure both the surface magnetic fields and brightness distribution in addition to their fundamental properties (i.e., mean density, rotation rate, luminosity). Zeeman-Doppler imaging techniques permit to map magnetic fields, but it can only give insights on nonaxisymmetric brightness structures \citep[e.g.,][]{Kochukhov2016}. In that context, interferometric imaging seems the most adequate way to map directly the mean axisymmetric brightness distribution of stars. Recent instruments such as GRAVITY \citep[e.g.,][]{Gravity2017} can provide an angular resolution of about 1 mas on order of magnitude. While such
precision enables us to estimate the apparent radii of the closest K or M dwarfs \citep[e.g.,][]{Kervella2017}, it does not allow to map the surface of these stars directly. Indeed, assuming typical radii of about $0.5~R_\odot$ \citep[e.g.,][]{Chabrier1997}, resolving for instance a fifth of the stellar disk at least with this precision requires parallaxes larger than about 1". For pre-main sequence stars, assuming typical radii of about $2~R_\odot$ \citep[e.g.,][]{Antona2017}, this would still require parallaxes larger than about 300 mas. Both estimated parallaxes are unrealistic, but
simple checks in the available astrophysical database (e.g., the Gaia database) show that increasing the angular resolution by a factor of three would permit to have access to an interesting sample of potential targets. This sample is composed of one pre-main sequence star of about $20$~Myr (i.e., V* AP Col), and of about twenty K and M dwarfs, which is quite promising for forthcoming projects. We can note especially the recent instrument CHARA/SPICA at Mount Wilson that is expected to improve the accessible angular resolution by a factor of two at least in a very near future \citep[e.g.,][]{Mourard2017,Rigi2023}.

\begin{figure}
   \centering
   \includegraphics[width=\hsize]{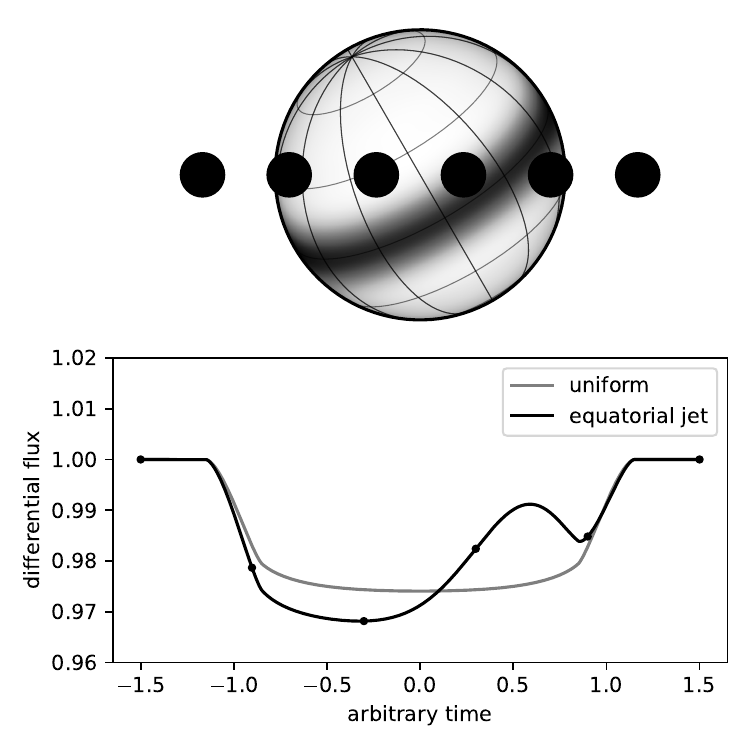}
      \caption{Transit light curve of an exoplanet occulting a star with an equatorial jet dimmer than the rest of its photosphere. This simulation made use of \texttt{starry} \citep{Luger2019}.
      It features a limb-darkened star with an adimensional unit radius and an occulting spherical body of relative radius 0.15. The equatorial band has a contrast of 1 (for illustration purposes) and an angular radius of 0.2 radians. The upper plot shows the system at scale for six different times. The bottom plot shows the occultation light curve (black line), the differential flux measured for the six times shown above (points), as well as the transit signal that would be observed for a uniform limb-darkened star (gray).}
        \label{jet_occultation}
   \end{figure}

While direct tests through interferometry remain currently challenging, indirect measurements could also help in probing the (magnetic) Coriolis darkening effect. One of this indirect method is to compare the spectral energy distribution of stars with a template assuming a uniform surface brightness. Indeed, the differences in temperature in the dark and bright surface bands should induce local differences in the spectral lines and thus deviations in the total integrated spectrum. Such method has already been used to probe the gravity darkening in fastly-rotating intermediate-mass stars \citep[e.g.,][]{Girardi2019}. While the effect is certainly subtle and the result depends on the model of darkening considered, analyzing a large sample of cool stars of similar mass with different inclination angle can be envisaged to lift degeneracies \citep[e.g.,][]{Zorec2023}. Another indirect method to test the Coriolis darkening consists in observing a companion body passing in front of the star, and to use the photometric and spectroscopic measurements of this event to infer the presence and properties of the star’s surface axisymmetric structure. Indeed, assuming an equator much darker than the poles at the stellar surface, as expected from our simulations, the transit depth in the observed light curve is expected to vary with the transit phase, with a minimum occurring when the companion obscures the dim stellar equatorial band. For that purpose, transiting close-in exoplanets are excellent candidates \citep[e.g.,][]{Morris2017} as these bodies are dark compared to the star and have a small relative size (in opposition to stellar companions), enabling the discrete occultation of finer details on the host star photosphere. Nonetheless, the detection of such transit events will only be possible under stringent orbital configurations: it requires the orbital plane of the transiting exoplanet to be sufficiently misaligned with the equatorial plane of the star to produce a change in the apparent brightness that differs from the change expected during the transit of a simple limb-darkened body. Such a configuration, and its resulting light curve, are illustrated in \figurename{}~\ref{jet_occultation}. In order for the detection of this structure to be significant, it will be interesting to combine photometric measurements with time-series spectroscopic measurements that are able to encapsulate Doppler effects due to the surface differential rotation of the star.
While this could be achievable through the modeling of the Rossiter–McLaughlin effect \citep{Rossiter1924, Ohta2005}, the extent to which a proper model comparison can be made in order to decipher an equatorial jet from a simple stellar spot occultation is under investigation, and will be featured in a future study.

In case the spin-orbit angle
is close to 90\textdegree, that is when the transit chord of the planet lies within the equatorial band, it is worth mentioning that a significant pole-equator brightness contrast as predicted in this work may also have indirect implications on the interpretation of exoplanets transmission spectroscopy measurements.
During a transit, the light coming from the star is transmitted through the planet's atmosphere. 
Owing to the wavelength dependence of its atmosphere opacity, the apparent radius of the planet inferred via the analysis of a transit can vary as a function of the observed wavelength, leading to a transmission spectrum that contains precious information about the planet composition \citep[e.g.,][]{Kreidberg2014}.
In practice, this technique often assumes that the transited chord is representative of the rest of the stellar photosphere. However, any difference between this chord and the unocculted stellar disk results
in the contamination of the planet's transmission spectrum by this unocculted part.
The low effective temperatures of cool stars lead to stellar atmospheres containing complex molecules (and clouds for the latest types) that are hard to separate from the features of cool exoplanets atmospheres \citep[e.g.,][]{Rackham2018M}. For this reason, stellar contamination is a major concern for the study of exoplanets atmospheres around late M dwarfs, and a great obstacle to the characterization of rocky exoplanets envisioned for the next decades \citep{Rackham2023}. Given the large-scale structures resulting from the magnetic Coriolis darkening presented in this work, this effect results in a greater level of transmission spectra stellar contamination, with the worst-case scenario being when
the spin-orbit angle is equal to 90\textdegree.
In this case, the transit chord is everywhere different from the rest of the stellar disk, with no observable capable of identifying the presence of an equatorial band. A quantitative analysis of the impact of such structures on the interpretation of exoplanets transmission spectra is out of the scope of this paper, but certainly deserves further investigation.



\subsection{On the effect of the magnetic morphology}
\label{disc:morph}

 According to our work, the magnetic morphology does not affect the leading-order magnetic flattening of the surface brightness distribution. Nevertheless, going beyond this leading-order trend, there are clues that the magnetic field structure may impact this distribution. A first example at the transition between the onset of convection and more turbulent models can be found in \figurename{}~\ref{fig:Np3c} for $N_\rho=3$ and $Ra=3.7~Ra_{\rm c}$. In this figure, the dynamo for $Pm=1$ is hemispherical with a more magnetized south pole than the north one, while the dynamo is dipolar for $Pm=3$. We can see that while the heat flux is symmetric with respect to the equator for the dipolar model, it becomes somehow asymmetric in the hemispherical model with a brighter north hemisphere. Another striking example is given in \figurename{}~\ref{fig:turb_Np6a} for $N_\rho=6$ and $Ra/Ra_{\rm c}=4$. These dynamo models are also hemispherical with a more magnetized south hemisphere, as we can see in Figs.~\ref{fig:hemi} and \ref{fig:but_hemi} for $Pm=2$. Inspecting the surface Nusselt profile, we can see that although magnetic fields reduce the global brightness contrast, the north pole also becomes brighter than the south pole. The concentration of the magnetic energy in the south seems to inhibit the heat flux in this hemisphere. It is also important to note that this south-north deviation is not directly correlated with the differential rotation (see \figurename{}~\ref{fig:V_Np6_4}), supporting the magnetic field topology as the responsible of this trend. The magnetic impingement of the brightness contrast still holds from a global point of view, but it is clear that these higher-order effects remain to be addressed properly in the future. This point also encompasses the impact of potential polar spots, which are already known to be able to play a significant role \citep[e.g.,][]{Yadav2015}.

 In the same spirit as in the previous section, it is also worth discussing the way to test the relation between the magnetic topology and the surface heat flux distribution. To do so, a relevant indicator is necessary to compare the predictions of dynamo simulations and observations. The simplest diagnostic consists in classifying the magnetic topology into the two main families, that is, large-scale dipoles or small-scale multipoles. For this purpose, we usually rely on the mean dipole field strength indicator, which is the time average of $f_{\rm dip}^{\ell_{\rm cut}}$ defined in \eq{f_dip} \citep[e.g.,][]{Christensen2006}. Above a value of $0.5$, we merely consider that the topology is dipolar; otherwise, it is multipolar. The value of this indicator directly depends on the value of the cut-off degree, $\ell_{\rm cut}$. For sake of the consistency, $\ell_{\rm cut}$ is generally set to a value corresponding to the maximum spatial resolution reachable with the considered observational techniques, that is, between about $5$ and $12$ for Zeeman-Doppler reconstruction \citep[e.g.,][]{Yadav2015b,Zaire2022}. As seen in Figs.~\ref{f_dip plot} and \ref{f_dipax}, changing
 $\ell_{\rm cut}$ from $5$ to $12$ can decrease the value of $f_{\rm dip}^{\ell_{\rm cut}}$ by about 10\%. For the dipole model in \figurename{}~\ref{f_dip plot}, the diagnostic does not depend on the value of $\ell_{\rm cut}$ globally, except close to polarity reversal events where $f_{\rm dip}^{\ell_{\rm cut}}$ can be either greater or smaller than $0.5$. Without the knowledge of the long-term evolution of $f_{\rm dip}^{\ell_{\rm cut}}$, the diagnostic can thus remain ambiguous, especially if the measurement is performed close to a polarity reversal \cite[e.g.,][]{Petit2009}. As another example, the equatorial dipole considered in \figurename{}~\ref{f_dipax} exhibits $f_{\rm dip}^{\ell_{\rm cut}}$ very close to the threshold $0.5$, because of the substantial contribution of the small-scale dynamics superimposed to the dipole component, especially for highly stratified models with a well-developed turbulence in the 
 near-surface layers (because of the sharp density drop). When confronting magnetic morphologies from dynamo simulations and observations of stars, the knowledge of the time evolution of the magnetic field as well as the contribution of the intermediate- and small-scale dynamics therefore appears crucial to make a relevant diagnostic of the magnetic morphology in such stratified systems.


\section{Conclusion}

\label{conclu}

In \citet[][Paper I]{Raynaud2018}, the effect of the Coriolis acceleration on the mean latitudinal surface brightness distribution of cool stars (i.e., with a relatively thick convective envelope) has been studied through direct numerical simulations. This latter study has highlighted the existence of the so-called Coriolis darkening phenomenon, that is, the emergence of a darker surface equatorial band in fastly-rotating convective shells, and whose nature differs from the usual gravity darkening effect expected in the oblate radiative envelopes of hotter highly-rotating stars. In this paper, we extend this latter work taking into account the additional effect of magnetic fields generated by a self-sustained convective dynamo. To do so, we perform a parametric study based on a large set of direct numerical simulations of convection in rotating  shells with large density stratification. We focus on stars with a convective envelope thickness of about 65\% of the total radius, which is representative of $\sim 0.35~M_\odot$ cool M dwarfs, stars at the beginning of the ascent of the red giant branch and stars on the pre-main sequence.

Based on the computation of the azimuthally  and time averaged latitudinal Nusselt profile, we distinguish three different leading-order regimes, depending on how far we are from the onset of convection. First, close to convection threshold, while the mean heat flux is maximum at the equator for nonmagnetic models, dynamo magnetic fields tend in contrast to increase it globally inside the tangent cylinder (i.e., close to the rotation poles). Second, for weakly turbulent models, that is, for models in which the emergence of a mean zonal surface jet starts inhibiting the heat flux close to the equator in the hydrodynamic case, the presence of dynamo magnetic fields can inverse the pole-equator brightness contrast, hence generating a brighter equator on average. Owing to the laminarization effect of the magnetic fields on the convective flow, we retrieve at some point, when the magnetic fields are strong enough, the same global behavior as that observed close to the onset of convection; the shape of the Nusselt profile is similar in both cases but the mean value consistently remains larger for larger values of the Rayleigh number. Third, for well-developed turbulent flows (i.e., for large values of the Rayleigh number $Ra \gtrsim 5~Ra_{\rm c}$), magnetic fields tend to smooth out the mean latitudinal brightness distribution. In all these regimes, we show that the magnetic fields systematically reduce the surface large-scale differential rotation level. Whereas the mean azimuthal velocity exhibits a cylindrical symmetry in the nonmagnetic case according to the Taylor-Proudman constraint, magnetic fields tend to rearrange the flow into a conical-like symmetry while progressively quenching the latitudinal surface variations as its magnitude increases. The lowering of the mean equatorial jet turns out to be directly correlated with the increase of the heat flux close to the equator in the case of either a pole-equator brightness contrast inversion in the second regime or of its global reduction in the third turbulent regime. At least, this generally happens for models with a surface convective Rossby number, $Ro_{\rm surf}$, lower than unity. Besides, for the most turbulent models (i.e., $Ro_{\rm surf} \gtrsim 1$), the surface heat flux is expected to decouple from the differential rotation and to become homogeneous; this happens close to the transition from a solar to an antisolar surface differential rotation. However, the same mean behavior is observed: magnetic fields tend to reduce independently the global surface brightness contrast during this transition.

These predictions obtained about the global magnetic Coriolis darkening now have to be tested through observations. We expect low-mass nearby M dwarfs to be one of the most interesting targets in this context. Direct mapping of the surface brightness distribution of such stars remains nonetheless challenging as it would require very-high-resolution interferometric imaging, still unreachable with the current instruments. Indirect measurements exploiting the variations in the light curve of M-dwarf stars induced by the transits of close-in exoplanets can be helpful too to scan the surface features, although the whole problem is somehow degenerate with the unknown planet parameters. The potential of such indirect constraints, as well as the natural repercussions of the Coriolis darkening effect on the estimate of the exoplanet properties, certainly deserve a dedicated study in a near future.
As the next step in the investigation, the systematic computation of light curves based on the mean brightness distribution profile derived from MHD simulations \citep[e.g.,][]{Yadav2015} will be necessary to make the most relevant possible comparison with observational constraints and their intrinsic limitations (e.g., spatial and time resolution). This will also provide practical models of the stellar disk light that could be used for applications in interferometry observations, Zeeman-Doppler imaging or exoplanet transit analyses. In addition, this will open the door to the study of the temporal dimension of the problem by means of global dynamo simulations, which has been intentionally put aside in this work for a first-step study. Focusing on the leading-order average latitudinal brightness distribution, we indeed did not take into account the impact of rotating nonaxisymmetric structures or of dynamo cycles. Going further in this direction, we will question the notion of stellar spots, their emergence, their persistence, or the relation between the surface magnetic structures and the emitted light. Regarded as higher-order effects compared to the mean behavior in this study, we can expect them to dominate the Coriolis darkening in case the mean brightness distribution tends to become uniform, especially for slowly-rotating cool stars, as for instance the Sun. Over a longer term, such a project will certainly need to account for a relevant description of the radiative transfer in the stellar atmosphere. While numerical tools already exist to tackle this issue locally \citep[e.g.,][]{Panja2020,Witske2022,Ludwig2023}, with imposed initial equilibrium MHD properties (e.g., magnetic field flux intensity, bottom velocity boundary conditions), a strong effort will be needed in the future to couple such local atmosphere models with the global dynamo simulations performed in this study.

\begin{acknowledgements}
We are indebted to C. Catala for very fruitful discussions on the paper. During this work, C. P. was funded by postdoctoral grants from Sorbonne Université (\'Emergence project) and the Research Federation PLAS@PAR. CP and MR acknowledge financial support from Centre National d'\'Etudes Spatiales (CNES). MR also acknowledges support from the European Research Council (ERC) under the Horizon Europe programme (Synergy Grant agreement N$^\circ$101071505: 4D-STAR).  While funded by the European Union, views and opinions expressed are however those of the authors only and do not necessarily reflect those of the European Union or the European Research Council. Neither the European Union nor the granting authority can be held responsible for them.

This study was granted access to the HPC resources of
MesoPSL financed by the Région Île-de-France and the
project Equip@Meso (reference ANR-10-EQPX-29-01) of
the programme Investissements d’Avenir supervised by the
Agence Nationale pour la Recherche.
This work was also supported by the "Programme National de Physique Stellaire" (PNPS) and the "Programme National des Hautes \'Energies" (PNHE) of CNRS/INSU co-funded by CEA and CNES, by a grant from Labex OSUG@2020 (Investissements
d'avenir - ANR10 LABX56), and by  l’Agence Nationale de la Recherche (ANR), project ANR-22-CE31-0020.
It has made use of data from the European Space Agency (ESA) mission
{\it Gaia} (\url{https://www.cosmos.esa.int/gaia}), processed by the {\it Gaia}
Data Processing and Analysis Consortium (DPAC,
\url{https://www.cosmos.esa.int/web/gaia/dpac/consortium}), as well as of the SIMBAD database, operated at CDS, Strasbourg, France.

\end{acknowledgements}

\bibliographystyle{aa}
\bibliography{bib.bib}

\appendix

\section{MHD anelastic equations}\label{app:eq}

Following \cite{Jones2011}, we use the LBR formulation of the anelastic approximation \citep{Lantz1999,Braginsky1995}, in which the closure relation for the heat flux is expressed as a diffusive law of the entropy, that is, $-\kappa \rho c_p T \vec{\nabla} S $ with $c_p$ the specific heat at constant pressure, $\rho$ the mean density and $S$ the entropy. The reference state is assumed to be close to the adiabatic hydrostatic equilibrium, that is, following the polytropic structure for the pressure, density, and temperature
\algn{
P=P_{\rm c} \varw^{n+1}\; ,~~~~\rho = \rho_{\rm c} \varw^n\; ,~~~~ T = T_{\rm c} \varw \; ,
}
with
\algn{
&\varw=c_0+\frac{c_1d }{r}\; ,~~~~c_0=\frac{2\varw_{\rm o}-\chi-1}{1-\chi}\; ,\\
&c_1=\frac{(1+\chi)(1-\varw_{\rm o})}{(1-\chi)^2} \; , ~~~~\varw_{\rm o} =\frac{\chi+1}{\chi \exp(N_\rho/n)+1} \; .
}
In the above expressions, $n$ is the polytropic index and \smash{$N_\rho = \ln [\rho(r_{\rm i}) /\rho(r_{\rm o}) ]$}is the number of density scale heights considered in the simulation domain. The values $\rho_{\rm c}$, $T_{\rm c}$, and $P_{\rm c}$ are the reference-state density, temperature, and pressure midway between the inner and outer
boundaries. A small deviation from an adiabatic state is nevertheless needed if we want convection to start. Convection is thus triggered by imposing an entropy drop $\Delta S$ between $r_{\rm i}$ and $r_{\rm o}$. By scaling length by the shell width $d$, time by the magnetic diffusion timescale $d^2 /\eta$, entropy by $\Delta S$, pressure by $\Omega \rho_{\rm c} \eta$, density by $\rho_{\rm c}$, temperature by $T_{\rm c}$, and magnetic field by $\sqrt{\Omega\rho_{\rm c} \mu \eta}$, where $\mu$ is the magnetic permeability, the equations governing the anelastic dynamics in the frame co-rotating with the reference shell are given by
\algn{
\derivp{\vec{\varv}}{t} + \vec{\varv}\cdot \vec{\nabla} \vec{\varv} =&  - \frac{Pm}{E} \vec{\nabla}\left( \frac{P^\prime}{\varw^n}\right) + \frac{Ra Pm^2}{Pr} \frac{S}{r^2} \vec{e}_r -\frac{2Pm}{E} \vec{e}_z \wedge \vec{\varv} \nonumber \\
&+ \frac{Pm}{E\varw^n} (\vec{\nabla}\wedge \vec{B})\wedge \vec{B}+Pm\vec{F}_\nu \; , \label{moment}\\
\derivp{S}{t} + \vec{\varv} \cdot \vec{\nabla} S =&\frac{Pm}{\varw^{n+1} Pr} \vec{\nabla} \cdot \left(\varw^{n+1} \vec{\nabla} S \right) \nonumber \\
&+ \frac{Di}{\varw} \left[\frac{(\vec{\nabla}\wedge \vec{B})^2}{E \varw^{n}} +Q_\nu \right]\label{chaleur}\\
\derivp{\vec{B}}{t} =& \vec{\nabla}\wedge\left(\vec{\varv}\wedge\vec{B} \right)+ \vec{\nabla}^2 \vec{B} \label{induction}\\
\vec{\nabla}\cdot \vec{B} =&0 \label{flux B}\\
\vec{\nabla} \cdot \left(\varw^n \vec{\varv} \right) =&0 \label{flux m}\; .
}
with $P^\prime$ the pressure perturbation compared to the reference state. The viscous force $\vec{F}_\nu$ is equal to $\vec{F}_\nu= \varw^{-n} \vec{\nabla} \mathsf{S}$, where $\mathsf{S}$ is the rate of strain tensor provided by
\algn{
\mathsf{S}_{ij}=2\varw^{n} \left( e_{ij}-\frac{1}{3} \delta_{ij} \vec{\nabla} \cdot \varv \right)\; , ~~~~e_{ij}=\frac{1}{2} \left(\derivp{\varv_i}{x_j}+\derivp{\varv_j}{x_i} \right) \; .
}
Moreover, the dissipation parameter and the viscous heating are respectively equal to
\algn{
Di=\frac{c_1 Pr}{Pm Ra}\; , ~~~~Q_{\nu} = 2 \left[e_{ij}e_{ij}-\frac{1}{3} (\vec{\nabla}\cdot\vec{\varv})^2 \right] \; .
}

\section{Simulation parameters}
The critical values of the Rayleigh number for the set of parameters used in this work are provided in \tablename{}~\ref{table:Rac}. An overview of the parameter space covered by our simulations is given in \tablename{}~\ref{param}. Finally, the set of simulations considered for the paper is presented in more detail in \tablename{}~\ref{param 2}.

\begin{table}
\caption{Critical Rayleigh numbers, $Ra_{\rm c}$, at the linear onset of convection for different values of the Ekman number, Prandtl number, and density stratification. The critical azimuthal order of the first unstable mode is given by $m_{\rm c}$.}             
\label{table:Rac}      
\centering                          
\begin{tabular}{c c c c c}        
\hline\hline                 
$E$ & $Pr$ & $N_\rho$ & $Ra_{\rm c}$  & $m_{\rm c}$ \\    
\hline                        
   $3\times 10^{-5}$& 0.3& $3.0$ & $6.12\times 10^6$ &  31\\
   $3\times 10^{-5} $& 0.3 & $4.0$ & $8.31\times 10^6$ &  46\\
   $3\times 10^{-5}$ & 1 & $3.0$ & $1.47\times 10^7$ &  43\\
   $10^{-4}$ & 1 & $3.5$ & $3.62\times10^6$ &  37\\
   $10^{-4}$ & 1 & $4.0$ & $4.09\times10^6$ &  43\\
   $10^{-4}$ & 1 & $5.0$ & $4.82~\times0^6$ &  51\\
   $3\times 10^{-4}$ & 1 & $6.0$ & $1.63\times10^6$ &  37\\

\hline                                   
\end{tabular}
\end{table}
\begin{table}
\caption{Overview of the range of values considered for the dimensionless parameters in this work.}             
\label{param}      
\centering                          
\begin{tabular}{c c c c c}        
\hline\hline                 
$N_\rho$ & $Pr$ & $E$ & $Ra/Ra_{\rm c}$  & $Pm$ \\    
\hline                        
   $3$& 0.3 -- 1& $3\times 10^{-5}$ & $1.7-4$ &  $0.25-4$\\
   $3.5$& 0.3 -- 1 & $10^{-4}$ & $2.7-5.5$ &  $0.5-5$\\
   $4$ & 1 & $3\times 10^{-5}-10^{-4}$ & $2.5-30$ &  $0.1-3$\\
   $5$ & 1 & $10^{-4}$ & $4-32$ &  $0.25-5$\\
   $6$ & 1 & $3\times 10^{-4}$ & $3-32$ &  $1-3$\\
\hline                                   
\end{tabular}
\end{table}

\newpage
\tablefirsthead{%
    \hline
    \hline
        $N_\rho$ & $Pr$ & $E$ & $Ra/Ra_{\rm c}$  & $Pm$ & $Ro$ & $Re$ & $Nu$ \\\hline}
\tablehead{%
    \hline
    \hline
        $N_\rho$ & $Pr$ & $E$ & $Ra/Ra_{\rm c}$  & $Pm$ & $Ro$ & $Re$ & $Nu$ \\\hline}
\topcaption{Set of simulations used in this work. The third last columns provide the time- and volume-averaged global Rossby number \smash{$Ro=u_{\rm rms} / \Omega d$,} Reynolds number $Re=Ro/E$, and Nusselt number $Nu$. Here, $u_{\rm rms}$ is the time-and volume-averaged root mean squared total velocity. The ($/$) symbol for $Pm$ indicates hydrodynamic runs.}
\begin{supertabular}{cccccccc}
\label{param 2}
 $3$ & 1 & $3\times 10^{-5}$ & $1.7$ & 0.5 &  0.0030 & 101   &  1.71 \\
  $3$ & 1 & $3\times 10^{-5}$ & $1.7$ &  1   &  0.0029& 97  &  1.83 \\
  $3$ & 1 & $3\times 10^{-5}$ & $1.7$ &  2   &  0.0031& 104 & 1.73 \\
  $3$ & 1 & $3\times 10^{-5}$ & $1.7$ &  3   &  0.0029& 98 & 1.84 \\
   $3$ & 1 & $3\times 10^{-5}$ & $2.5$ & 0.5 &  0.0062& 207 & 2.81 \\
   $3$ & 1 & $3\times 10^{-5}$ & $2.5$ &  1  &  0.0055& 183 & 2.82 \\
   $3$ & 1 & $3\times 10^{-5}$ & $2.5$ &  2  &  0.0057& 189 & 2.95 \\
   $3$ & 1 & $3\times 10^{-5}$ & $2.5$ &  3  &  0.0053& 178 & 2.85 \\
   $3$ & 1 & $3\times 10^{-5}$ & $2.5$ &  4  &  0.0045& 149 & 2.73 \\
   $3$ & 1 & $3\times 10^{-5}$ & $3.7$ &  1  &  0.0095& 318 & 4.85 \\
  $3$ & 1 & $3\times 10^{-5}$ & $3.7$ &   3  &  0.0090 & 301 & 5.30 \\
   \hline
   $3$ & 0.3 & $3\times 10^{-5}$ & $2.8$ & 0.5 & 0.0075 & 251 & 1.49 \\
  $3$ & 0.3 & $3\times 10^{-5}$ & $2.8$ &  1   & 0.0079 & 262 & 1.70 \\
  $3$ & 0.3 & $3\times 10^{-5}$ & $2.8$ &  2   & 0.0072 & 240 & 1.66\\
   $3$ & 0.3 & $3\times 10^{-5}$ & $3.4$ & 0.5 & 0.010  & 338 & 1.86 \\
   $3$ & 0.3 & $3\times 10^{-5}$ & $3.4$ &  1  & 0.0090  & 300 & 1.94\\
   $3$ & 0.3 & $3\times 10^{-5}$ & $3.4$ &  2  & 0.0082 & 274 & 2.10 \\
   $3$ & 0.3 & $3\times 10^{-5}$ & $4.1$ & 0.5 & 0.015  & 496 & 2.35 \\
   $3$ & 0.3 & $3\times 10^{-5}$ & $4.1$ &  1  & 0.014  & 463 & 2.90 \\
   $3$ & 0.3 & $3\times 10^{-5}$ & $4.1$ &   2  & 0.010 & 341 & 2.50 \\
   \hline
   $3.5$ & 1 & $10^{-4}$ & 2.8 & 0.5 & 0.014 & 142 & 2.86  \\
   $3.5$ & 1 & $10^{-4}$ & 2.8 & 1 & 0.013 & 132 & 2.80   \\
   $3.5$ & 1 & $10^{-4}$ & 2.8 & 5 & 0.013 & 134 & 3.16  \\
   $3.5$ & 1 & $10^{-4}$ & 5.5 & 0.25 &  0.031 & 310 & 5.46 \\
   $3.5$ & 1 & $10^{-4}$ & 5.5 & 0.5  &  0.032 & 316 & 5.70 \\
   $3.5$ & 1 & $10^{-4}$ & 5.5 & 1    &  0.030 & 299 & 5.80 \\
   $3.5$ & 1 & $10^{-4}$ & 5.5 & 5    &  0.024 & 241 & 5.65  \\
 \hline
    $4$ & 0.3 & $3\times 10^{-5}$ & $2.5$ & 0.5 &  0.0071 & 235 & 1.25 \\
  $4$ & 0.3 & $3\times 10^{-5}$ & $2.5$ &  2   &   0.0058 & 193 & 1.51 \\
   $4$ & 0.3 & $3\times 10^{-5}$ & $3 $ & 0.3 &  0.015 & 491  & 1.12 \\
   $4$ & 0.3 & $3\times 10^{-5}$ & $3 $ & 0.5 &  0.0091 & 303  & 1.50 \\
   $4$ & 0.3 & $3\times 10^{-5}$ & $3 $ &  2  &  0.0070 & 235  & 1.52 \\
   $4$ & 0.3 & $3\times 10^{-5}$ & $6.3$ & 0.1 & 0.026 & 854 & 3.25 \\
   $4$ & 0.3 & $3\times 10^{-5}$ & $6.3$ & 0.15 & 0.025 & 828 & 3.63 \\
   $4$ & 0.3 & $3\times 10^{-5}$ & $6.3$ & 0.3 &  0.024 & 805 & 3.72 \\
   $4$ & 0.3 & $3\times 10^{-5}$ & $6.3$ & 0.5  & 0.023 & 775 & 3.90 \\
   $4$ & 0.3 & $3\times 10^{-5}$ & $6.3$ & 1  & 0.021 & 707 & 4.00 \\
   $4$ & 0.3 & $3\times 10^{-5}$ & $6.3$ &   2  & 0.022 & 735 & 3.80 \\
\hline
    $4$ & 0.3 & $3\times 10^{-5}$ & $2.5$ & 0.5 &  0.0071 & 235 & 1.25 \\
  $4$ & 0.3 & $3\times 10^{-5}$ & $2.5$ &  2   &   0.0058 & 193 & 1.51 \\
   $4$ & 0.3 & $3\times 10^{-5}$ & $3 $ & 0.3 &  0.015 & 491  & 1.12 \\
   $4$ & 0.3 & $3\times 10^{-5}$ & $3 $ & 0.5 &  0.0091 & 303  & 1.50 \\
   $4$ & 0.3 & $3\times 10^{-5}$ & $3 $ &  2  &  0.0070 & 235  & 1.52 \\
   $4$ & 1 & $10^{-4}$ & 9.8  & 0.15 &  0.062 & 622 & 7.29 \\
   $4$ & 1 & $10^{-4}$ & 9.8  & 0.25 &  0.049 & 490 & 7.75  \\
   $4$ & 1 & $10^{-4}$ & 9.8  &  1   &  0.045 & 452 & 7.75  \\
   $4$ & 1 & $10^{-4}$ & 9.8  &  2   &  0.040 & 396 & 7.76  \\
   $4$ & 1 & $10^{-4}$ & 9.8  &  3   &  0.036 & 363 & 7.78 \\
   $4$& 1 & $10^{-4}$ & 29   &  \slash   &  0.23  & 2300 & 12.4 \\
   $4$ & 1 & $10^{-4}$ & 29  &  0.15  & 0.23  & 2275 & 17.2 \\
   $4$ & 1 & $10^{-4}$ & 29  &  0.25  & 0.33  & 3300 & 16.0 \\
   $4$ & 1 & $10^{-4}$ & 29  &  1     & 0.042 & 420  & 14.5 \\
   \hline
   $5$ & 1 & $10^{-4}$ & 4.2 & 0.25 & 0.022 & 221  &  2.35 \\
  $5$ & 1 & $10^{-4}$ & 4.2 & 1 &  0.016 & 163 & 2.65 \\
  $5$ & 1 & $10^{-4}$ & 4.2 & 5 &  0.014 & 138 & 2.82 \\
    $5$ & 1 & $10^{-4}$ & 6.2 & 0.25 & 0.026 &  264 &  3.48  \\
  $5$ & 1 & $10^{-4}$ & 6.2 & 0.4 &  0.024 & 242 & 3.90\\
  $5$ & 1 & $10^{-4}$ & 6.2 & 1 &  0.024  &  238 & 3.75 \\
  $5$ & 1 & $10^{-4}$ & 6.2 & 2 &  0.024 &  235 & 4.05 \\
  $5$ & 1 & $10^{-4}$ & 6.2 & 3 &  0.023  & 226 & 4.00 \\
  $5$ & 1 & $10^{-4}$ & 6.2 & 5 &  0.019  & 190 & 3.80 \\
 $5$ & 1 & $10^{-4}$ & 8.3 & 0.25 & 0.032 & 319  & 4.32  \\
  $5$ & 1 & $10^{-4}$ & 8.3 & 0.4 & 0.028 & 282 & 5.03 \\
  $5$ & 1 & $10^{-4}$ & 8.3 & 1 &  0.026 & 263 &  5.00 \\
  $5$ & 1 & $10^{-4}$ & 8.3 & 2 &  0.027 & 270 & 4.80 \\
  $5$ & 1 & $10^{-4}$ & 8.3 & 3 &  0.028 & 282 & 4.90 \\
  $5$ & 1 & $10^{-4}$ & 16 & 0.25 & 0.16 & 1586  &  5.60 \\
  $5$ & 1 & $10^{-4}$ & 16 & 1 &  0.048 & 483  & 6.90 \\
  $5$ & 1 & $10^{-4}$ & 16 & 2 &   0.044 & 439 & 7.05 \\
  $5$ & 1 & $10^{-4}$ & 32 & 0.25 & 0.46 & 4550 & 20.5 \\
  $5$ & 1 & $10^{-4}$ & 32 & 1    & 0.11 & 1056 & 16.1 \\
\hline  
   $6$ & 1 & $3\times 10^{-4}$ & $3$ &  \slash &$0.021$&$70$&$1.39$\\
   $6$ & 1 & $3\times 10^{-4}$ & $3$ &  $6$ &$0.017$&$57$&$1.43$\\
   $6$ & 1 & $3\times 10^{-4}$ & $4$ &  $1$ &$0.035$&$117$&$1.70$\\
   $6$ & 1 & $3\times 10^{-4}$ & $4$ &  $2$ &$0.028$&$94$&$1.75$\\
   $6$ & 1 & $3\times 10^{-4}$ & $4$ &  $3$ &$0.024$&$81$&$1.81$\\
   $6$ & 1 & $3\times 10^{-4}$ & $8$ &  \slash &$0.087$&$288$&$2.42$\\
   $6$ & 1 & $3\times 10^{-4}$ & $8$ &  $1$ &$0.047$&$155$&$2.62$\\
   $6$ & 1 & $3\times 10^{-4}$ & $8$ &  $2$ &$0.043$&$142$&$2.73$\\
   $6$ & 1 & $3\times 10^{-4}$ & $8$ &  $3$ &$0.042$&$139$&$2.79$\\
   $6$ & 1 & $3\times 10^{-4}$ & $16$ &  \slash &$0.15$&$505$&$3.43$\\
   $6$ & 1 & $3\times 10^{-4}$ & $16$ &  $1$ &$0.15$&$501$&$3.32$\\
   $6$ & 1 & $3\times 10^{-4}$ & $16$ &  $2$ &$0.066$&$221$&$3.52$\\
   $6$ & 1 & $3\times 10^{-4}$ & $16$ &  $3$ &$0.065$&$218$&$3.55$\\
   $6$ & 1 & $3\times 10^{-4}$ & $32$ &  \slash &$0.18$&$611$&$5.51$\\
   $6$ & 1 & $3\times 10^{-4}$ & $32$ &  $2$ &$0.20$&$673$&$4.88$\\
\hline                                   
\end{supertabular}
%

\section{Mean surface Nusselt profiles}
\captionsetup[subfigure]{justification=centering,singlelinecheck=false}
\label{app:Nu}


    \begin{figure*}
    \centering
    \textbf{Nusselt profile: $E=3\times 10^{-5},~Pr = 1$}\par\medskip
    \rotatebox[origin=c]{90}{$N_\rho=3$} \quad
    \begin{subfigure}[c]{0.31\hsize}
    \centering
    \includegraphics[trim=0cm 0cm 2cm 1cm, clip,width=\hsize]{Np3/Nu_3e-05_1.0_1.70.png}
    \caption{$Ra=1.7~Ra_{\rm c}$}
    \label{fig:Np3a}
    \end{subfigure}
    \begin{subfigure}[c]{0.31\hsize}
    \centering
    \includegraphics[trim=0cm 0cm 2cm 1cm, clip,width=\hsize]{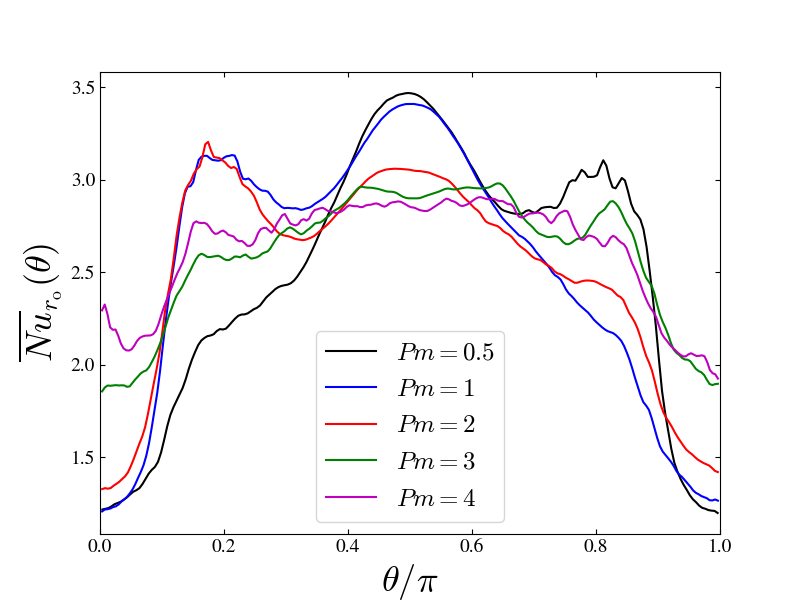}
    \caption{$Ra=2.5~Ra_{\rm c}$}
    \label{fig:Np3b}
    \end{subfigure}
    \begin{subfigure}[c]{0.31\hsize}
    \centering
    \includegraphics[trim=0cm 0cm 2cm 1cm, clip, width=\hsize]{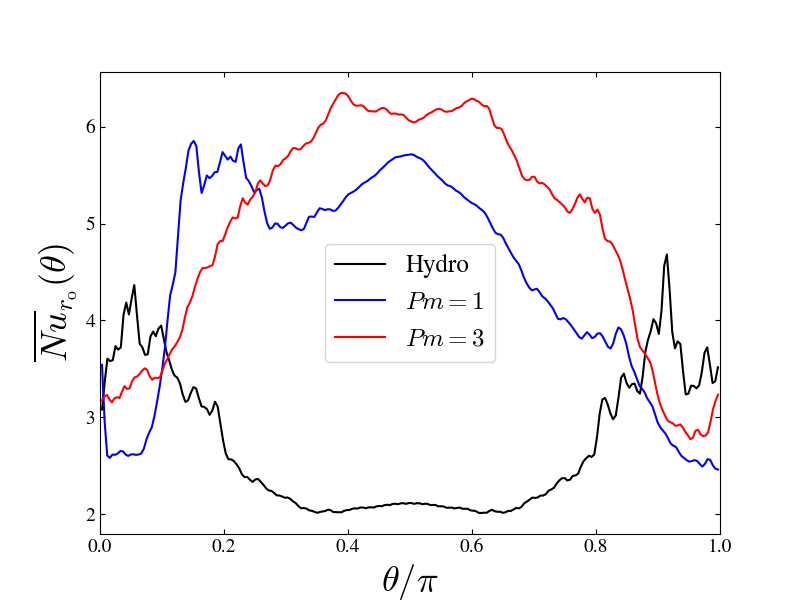}
    \caption{$Ra=3.7~Ra_{\rm c}$}
    \label{fig:Np3c}
    \end{subfigure}
    \caption{Mean latitudinal profile of the Nusselt number defined in \eq{Nu}, for models with $N_\rho=3$, $E=3\times10^{-5}$, and $Pr=1$. The value of the Rayleigh number $Ra$ increases from the left to the right and the magnetic Prandtl number $Pm$ is varied. Solid black lines represent either pure hydrodynamic models or MHD models without magnetic dynamo (i.e., too low values of $Pm$). Note that panel (a) is similar to \figurename{}~\ref{fig:Nu_onset} but is added here for the sake of completeness and comparison.}
    \label{fig:onset_Np3}
    \end{figure*}
    
    
    \begin{figure*}
    \centering
    \textbf{Nusselt profile: $E=3\times 10^{-5},~ Pr = 0.3$}\par\medskip
    \rotatebox[origin=c]{90}{$N_\rho=3$} \quad
    \begin{subfigure}[c]{0.31\hsize}
    \centering
    \includegraphics[trim=0cm 0cm 2cm 1cm, clip, width=\hsize]{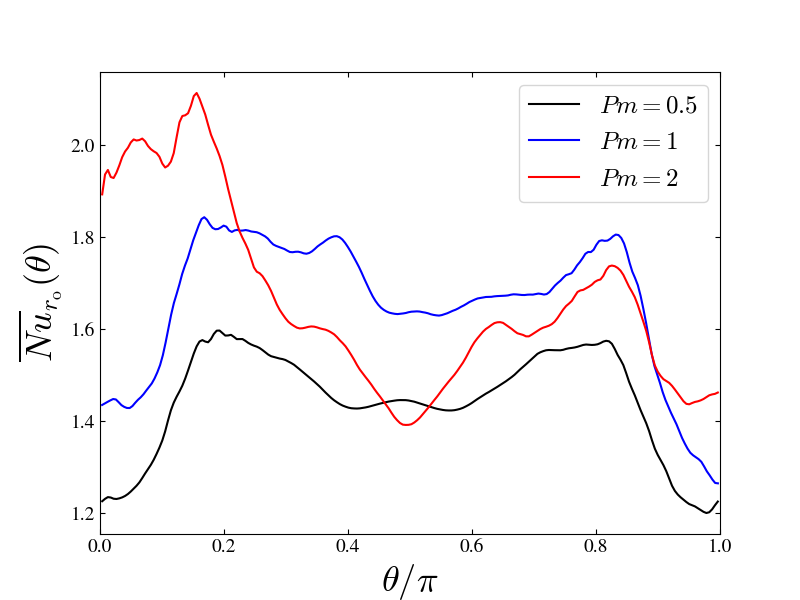}
    \caption{$Ra=2.8~Ra_{\rm c}$}
    \label{fig:onset_Np3a}
    \end{subfigure}
    \begin{subfigure}[c]{0.31\hsize}
    \centering
    \includegraphics[trim=0cm 0cm 2cm 1cm, clip, width=\hsize]{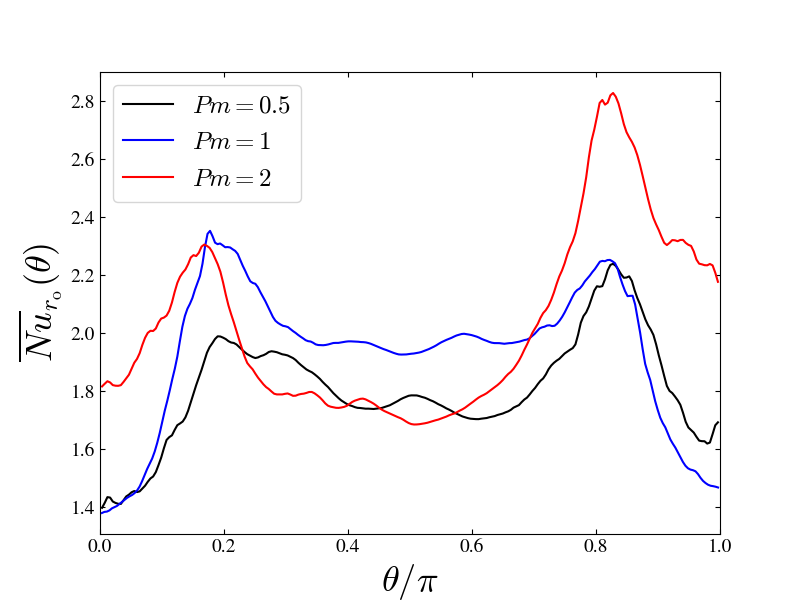}
    \caption{$Ra=3.4~Ra_{\rm c}$}
    \label{fig:onset_Np3b}
    \end{subfigure}
    \begin{subfigure}[c]{0.31\hsize}
    \centering
    \includegraphics[trim=0cm 0cm 2cm 1cm, clip, width=\hsize]{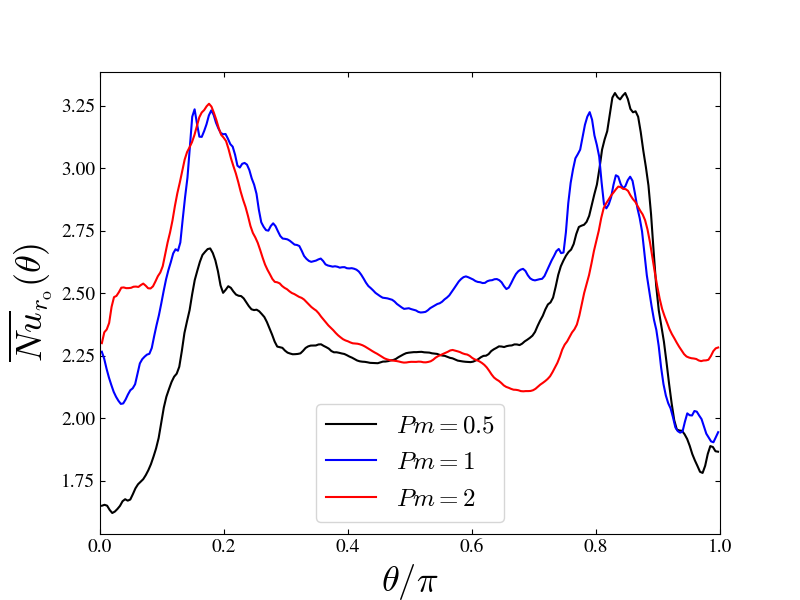}
    \caption{$Ra=4.1~Ra_{\rm c}$}
    \label{fig:onset_Np3c}
    \end{subfigure}
    \rotatebox[origin=c]{90}{$N_\rho=4$} \quad
    \begin{subfigure}[c]{0.31\hsize}
    \centering
    \includegraphics[trim=0cm 0cm 2cm 1cm, clip, width=\hsize]{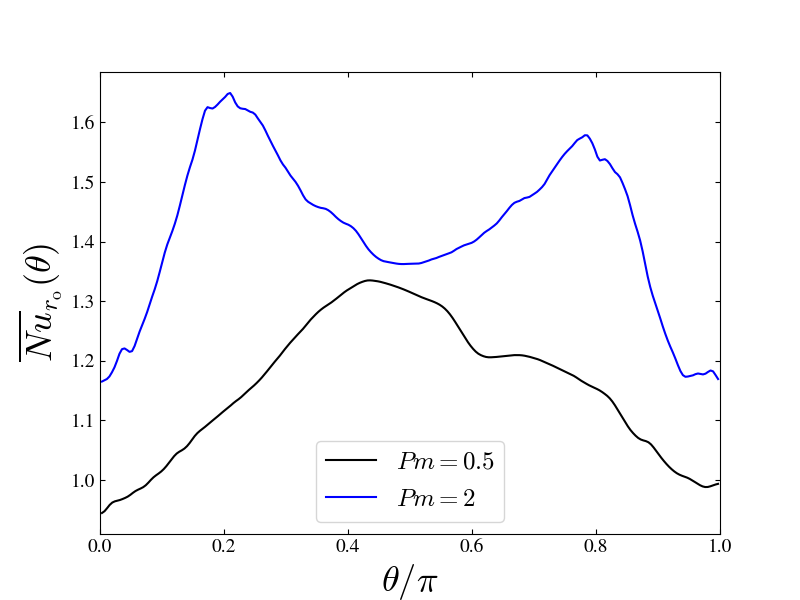}
    \caption{$Ra=2.5~Ra_{\rm c}$}
    \label{fig:onset_Np4a}
    \end{subfigure}
    \begin{subfigure}[c]{0.31\hsize}
    \centering
    \includegraphics[trim=0cm 0cm 2cm 1cm, clip, width=\hsize]{Np4/Nu_3e-05_0.3_3.01.png}
    \caption{$Ra=3.0~Ra_{\rm c}$}
    \label{fig:trans_Np4}
    \end{subfigure}
    \begin{subfigure}[c]{0.31\hsize}
    \centering
    \includegraphics[trim=0cm 0cm 2cm 1cm, clip, width=0.95\hsize]{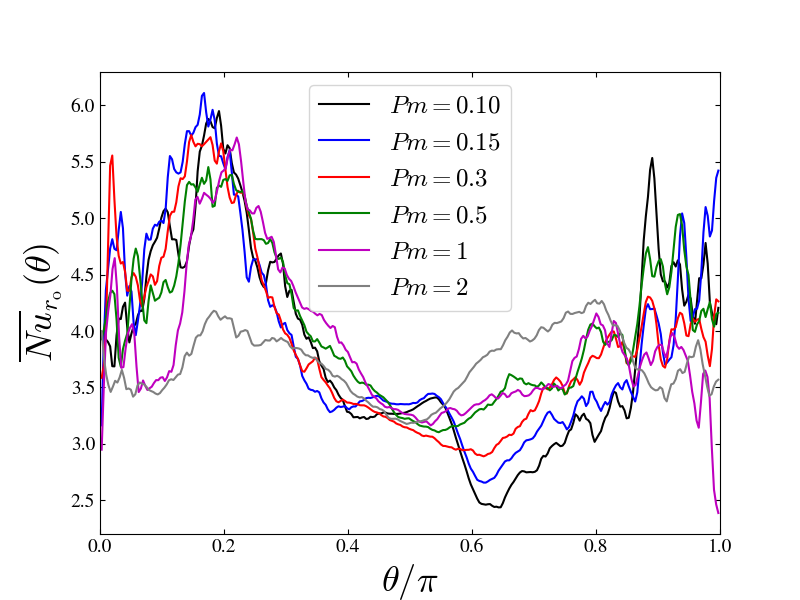}
    \caption{$Ra=6.3~Ra_{\rm c}$}
    \label{fig:turb_Np4}
    \end{subfigure}
    \caption{Same as in \figurename{}~\ref{fig:onset_Np3} but for models with $E=3\times 10^{-5}$, $Pr=0.3$, and moderate values of the Rayleigh number. The first and second rows correspond to $N_\rho=3$ and $4$, respectively. Note that panel (e) is similar to \figurename{}~\ref{fig:Nu_transition} but is added here for the sake of completeness and comparison.}
    \label{fig:Np34}
    \end{figure*}


    \begin{figure*}
    \centering
    \textbf{Nusselt profile: $E= 10^{-4},~ Pr = 1$}\par\medskip
    \flushleft
    \rotatebox[origin=c]{90}{$N_\rho=3.5$} \quad
    \begin{subfigure}[c]{0.31\hsize}
    \includegraphics[trim=0cm 0cm 2cm 1cm, clip, width=\hsize]{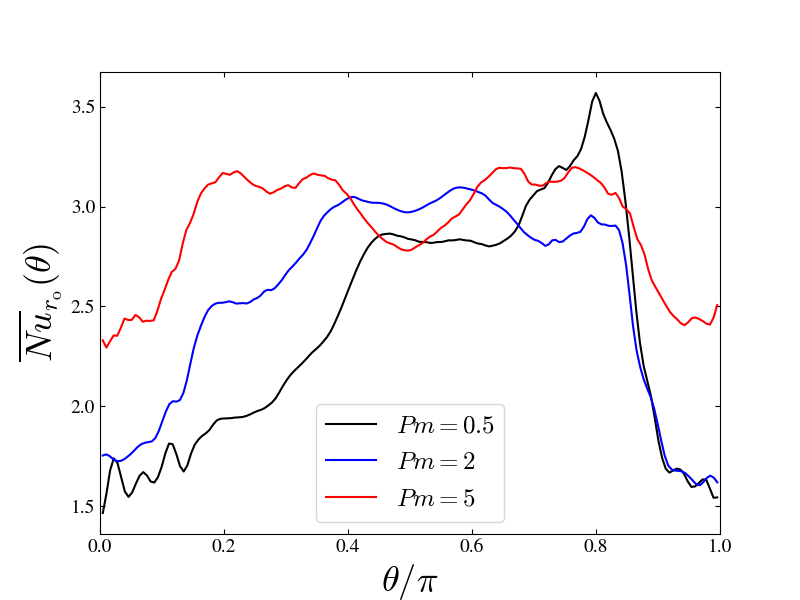}
    \caption{$Ra=2.8~Ra_{\rm c}$}
    \label{fig:onset_Np35}
    \end{subfigure}
    \begin{subfigure}[c]{0.31\hsize}
    \centering
    \includegraphics[trim=0cm 0cm 2cm 1cm, clip, width=\hsize]{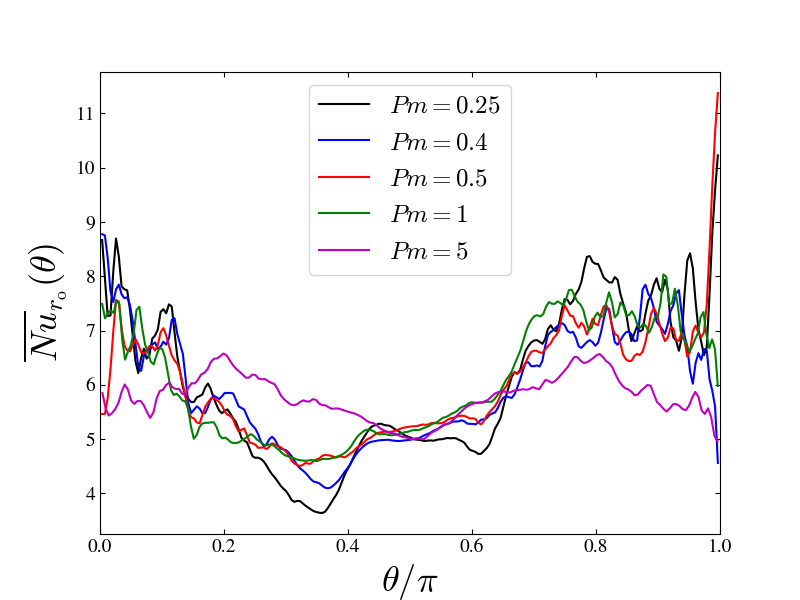}
    \caption{$Ra=5.5~Ra_{\rm c}$}
    \label{fig:turb_Np35}
    \end{subfigure}
    
    \rotatebox[origin=c]{90}{$N_\rho=4$} \quad
    \begin{subfigure}[c]{0.31\hsize}
    \centering
    \includegraphics[trim=0cm 0cm 2cm 1cm, clip, width=\hsize]{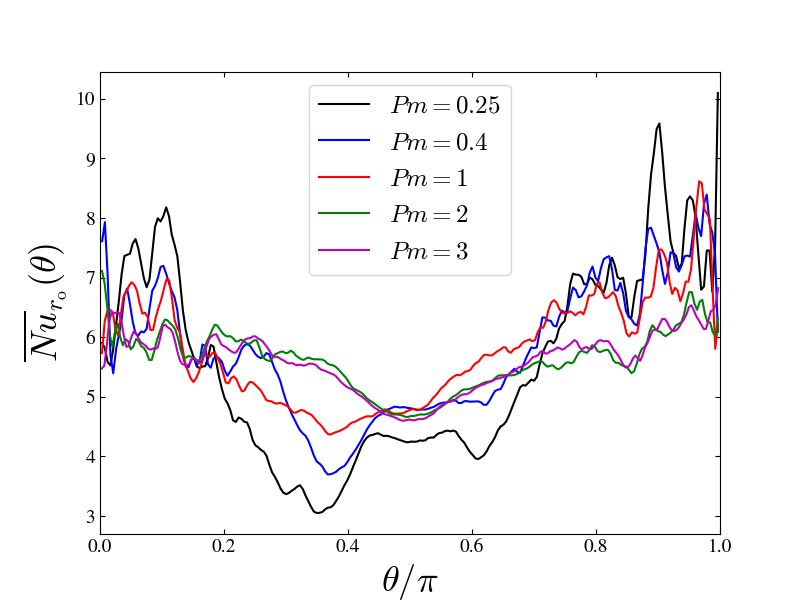}
    \caption{$Ra=6.1~Ra_{\rm c}$}
    \label{fig:turb_Np41}
    \end{subfigure}
    \begin{subfigure}[c]{0.31\hsize}
    \centering
    \includegraphics[trim=0cm 0cm 2cm 1cm, clip, width=\hsize]{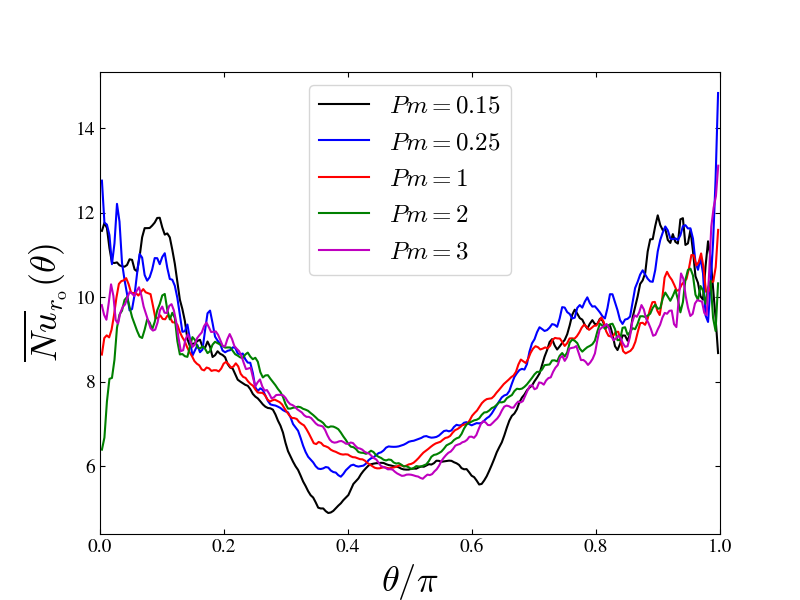}
    \caption{$Ra=9.8~Ra_{\rm c}$}
    \label{fig:turb_Np42}
    \end{subfigure}
    \begin{subfigure}[c]{0.31\hsize}
    \centering
    \includegraphics[trim=0cm 0cm 2cm 1cm, clip, width=\hsize]{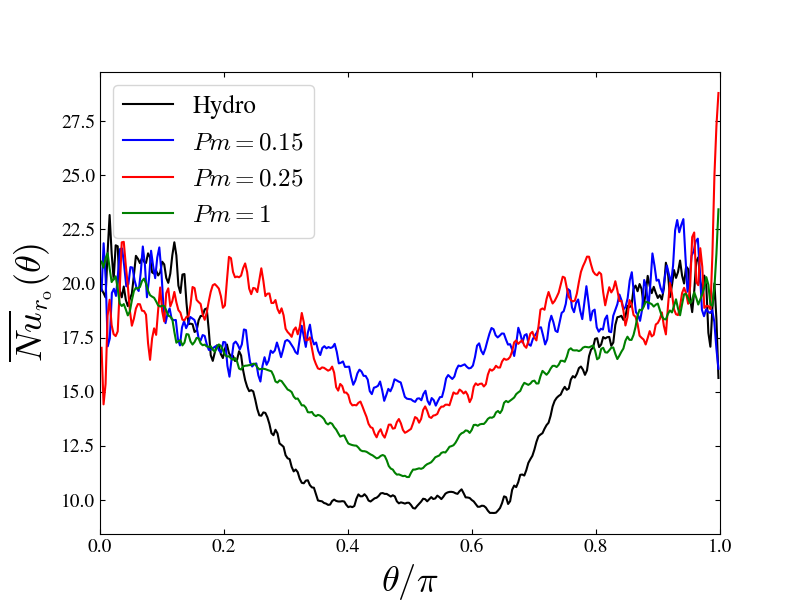}
    \caption{$Ra=29~Ra_{\rm c}$}
    \label{fig:turb_Np43}
    \end{subfigure}
    
    \rotatebox[origin=c]{90}{$N_\rho=5$} \quad
    \begin{subfigure}[c]{0.31\hsize}
    \centering
    \includegraphics[trim=0cm 0cm 2cm 1cm, clip, width=\hsize]{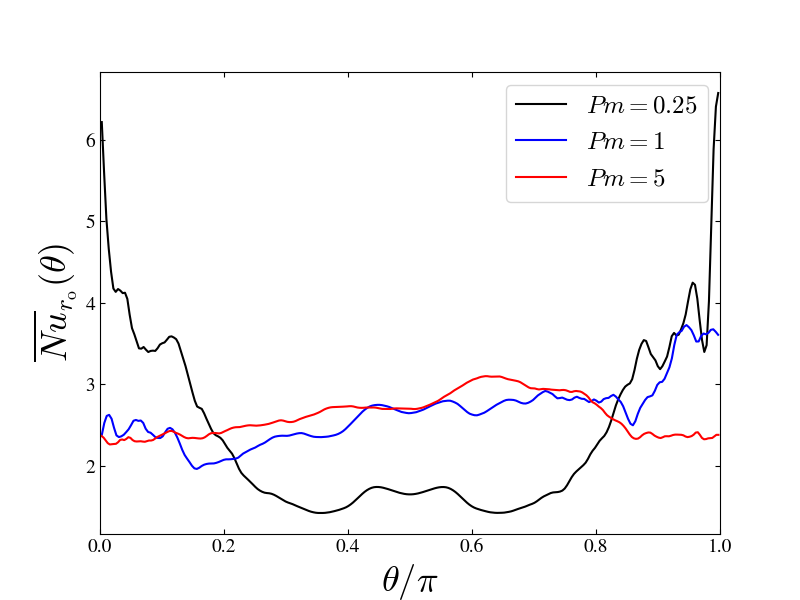}
    \caption{$Ra=4.2~Ra_{\rm c}$}
    \label{fig:turb_Np5a}
    \end{subfigure}
    \begin{subfigure}[c]{0.31\hsize}
    \centering
    \includegraphics[trim=0cm 0cm 2cm 1cm, clip, width=\hsize]{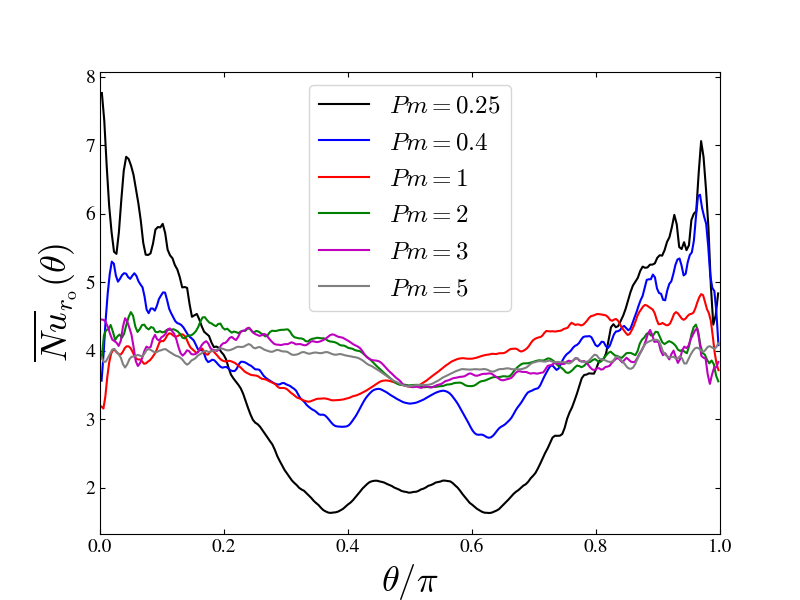}
    \caption{$Ra=6.2~Ra_{\rm c}$}
    \label{fig:turb_Np5b}
    \end{subfigure}
    \begin{subfigure}[c]{0.31\hsize}
    \centering
    \includegraphics[trim=0cm 0cm 2cm 1cm, clip, width=\hsize]{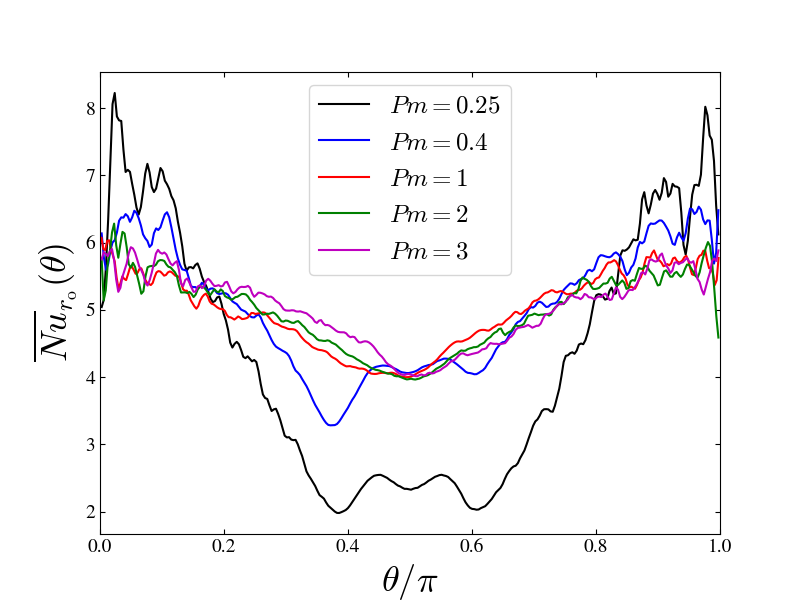}
    \caption{$Ra=8.3~Ra_{\rm c}$}
    \label{fig:turb_Np5c}
    \end{subfigure}
    \rotatebox[origin=c]{90}{$N_\rho=5$} \quad
    \begin{subfigure}[c]{0.31\hsize}
    \centering
    \includegraphics[trim=0cm 0cm 2cm 1cm, clip, width=\hsize]{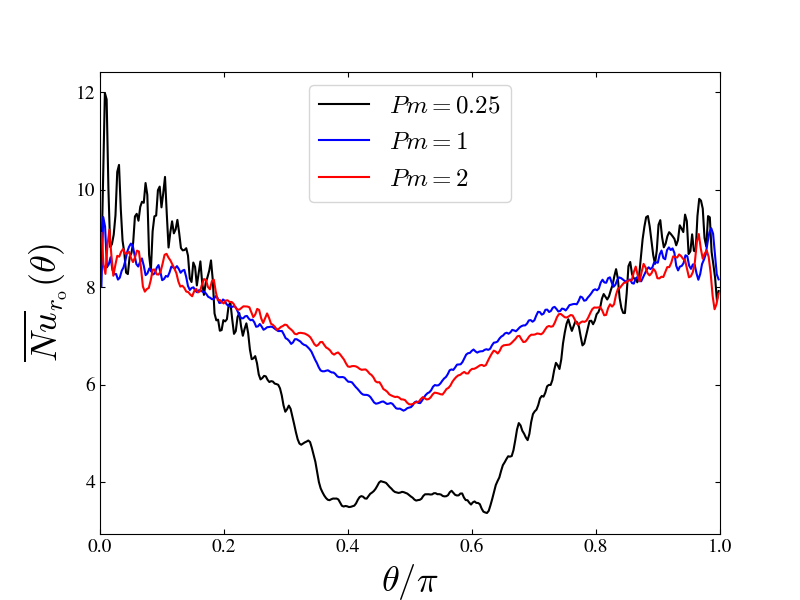}
    \caption{$Ra=16~Ra_{\rm c}$}
    \label{fig:turb_Np5d}
    \end{subfigure}
    \begin{subfigure}[c]{0.31\hsize}
    \centering
    \includegraphics[trim=0cm 0cm 2cm 1cm, clip, width=\hsize]{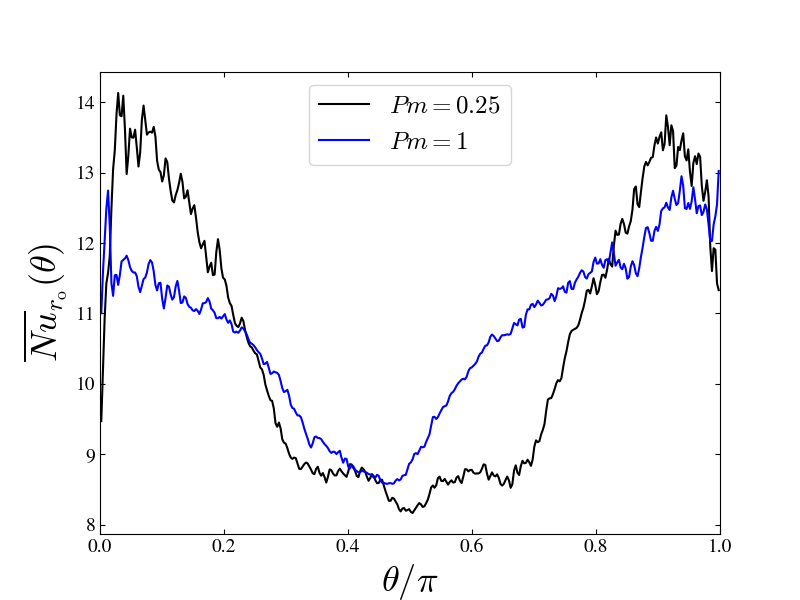}
    \caption{$Ra=32~Ra_{\rm c}$}
    \label{fig:turb_Np5e}
    \end{subfigure}
    \caption{Same as in \figurename{}~\ref{fig:Np34}, but for models with $E=10^{-4}$ and $Pr=1$. The first and second rows correspond to $N_\rho=3.5$ and $4$, respectively, while the last rows correspond to $N_\rho=5$.}
    \label{fig:Nu_Np35}
    \end{figure*}

    
    \begin{figure*}
    \centering
    \textbf{Nusselt profile: $E= 3\times10^{-4},~ Pr = 1$}\par\medskip
    \rotatebox[origin=c]{90}{$N_\rho=6$} \quad
    \begin{subfigure}[c]{0.31\hsize}
    \centering
    \includegraphics[trim=0cm 0cm 2cm 1cm, clip, width=\hsize]{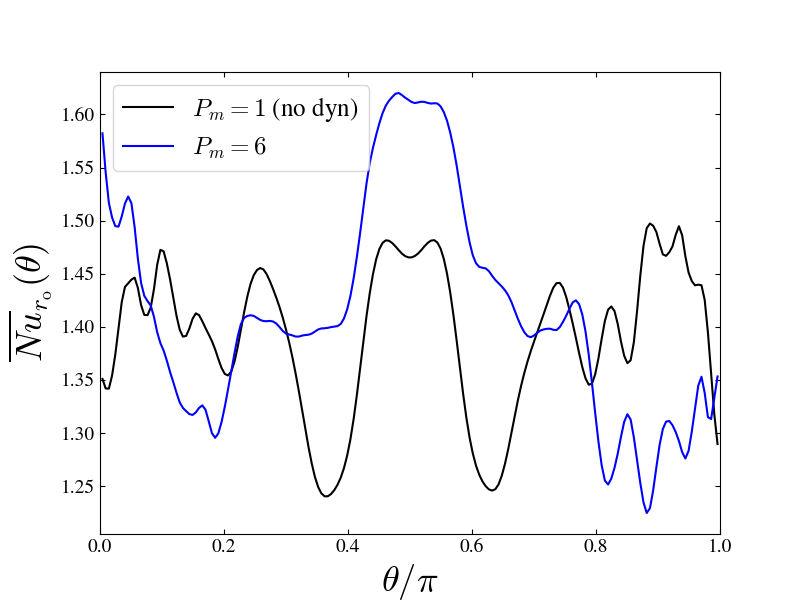}
    \caption{$Ra=3.0~Ra_{\rm c}$}
    \label{fig:Nu_Np6_transition}
    \end{subfigure}
    \begin{subfigure}[c]{0.31\hsize}
    \centering
    \includegraphics[trim=0cm 0cm 2cm 1cm, clip, width=\hsize]{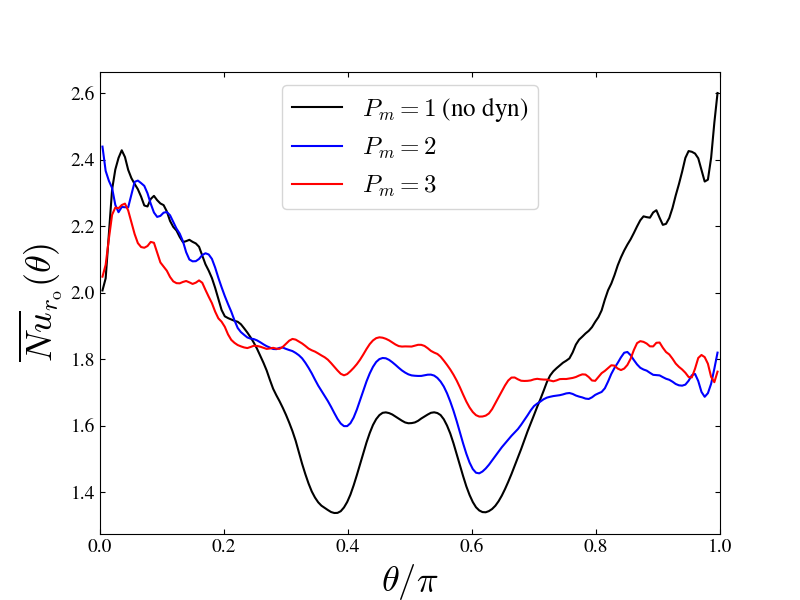}
    \caption{$Ra=4.0~Ra_{\rm c}$}
    \label{fig:turb_Np6a}
    \end{subfigure}
    \begin{subfigure}[c]{0.31\hsize}
    \centering
    \includegraphics[trim=0cm 0cm 2cm 1cm, clip, width=\hsize]{Np6/Nu_8.png}
    \caption{$Ra=8.0~Ra_{\rm c}$}
    \label{fig:turb_Np6b}
    \end{subfigure}   
    
    \flushleft
    \rotatebox[origin=c]{90}{$N_\rho=6$} \quad
    \begin{subfigure}[c]{0.31\hsize}
    \centering
    \includegraphics[trim=0cm 0cm 2cm 1cm, clip, width=\hsize]{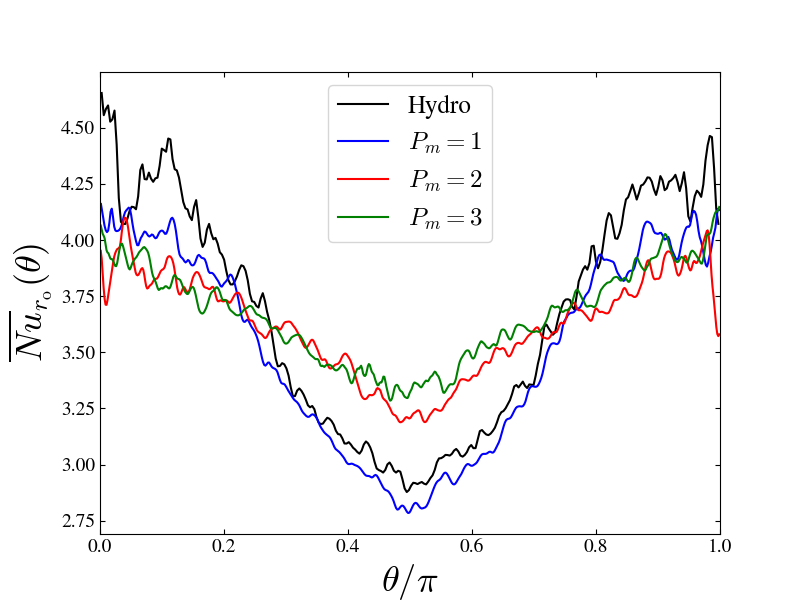}
    \caption{$Ra=16~Ra_{\rm c}$}
    \label{fig:turb_Np6c}
    \end{subfigure}
    \begin{subfigure}[c]{0.31\hsize}
    \centering
    \includegraphics[trim=0cm 0cm 2cm 1cm, clip, width=\hsize]{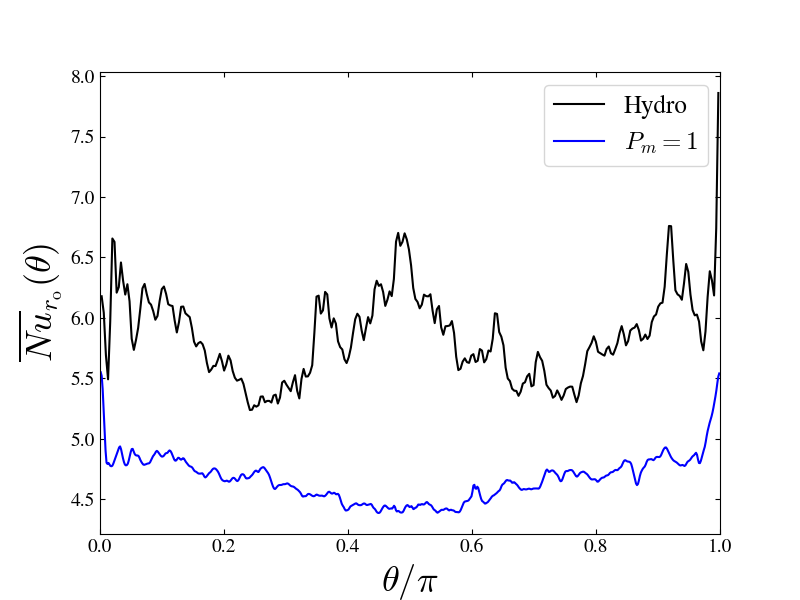}
    \caption{$Ra=32~Ra_{\rm c}$}
    \label{fig:turb_Np6d}
    \end{subfigure}
    \caption{Same as in \figurename{}~\ref{fig:Nu_Np35}, but for models with $E=3\times 10^{-4}$, $Pr=1$, and $N_\rho=6$. The model with $Ra=32~Ra_{\rm c}$ requires a very high resolution and may be not fully resolved or relaxed. We checked this has no effect on the shape of the differential rotation; we nevertheless have to be cautious when considering the absolute value of the surface Nusselt. Note that panel (c) is similar to \figurename{}~\ref{fig:Nu_turb} but is added here for the sake of completeness and comparison.}
    \label{fig:Nu_Np6}
    \end{figure*}

\section{Surface azimuthal velocity profiles}
\label{app:V}

    \clearpage
    \begin{figure*}
    \centering
    \textbf{Azimuthal velocity profile: $E= 3\times10^{-5},~ Pr = 0.3$}\par\medskip
    \rotatebox[origin=c]{90}{$N_\rho=4$} \quad
    \begin{subfigure}[c]{0.31\hsize}
    \centering
    \includegraphics[trim=0cm 0cm 2cm 1cm, clip, width=\hsize]{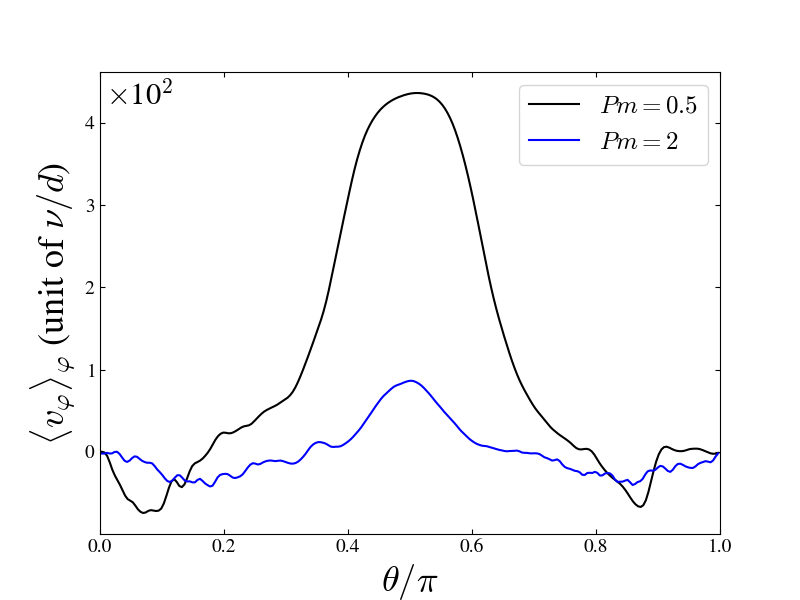}
    \caption{$Ra=2.5~Ra_{\rm c}$}
    \label{fig:V_onset_Np4a}
    \end{subfigure}
    \begin{subfigure}[c]{0.31\hsize}
    \centering
    \includegraphics[trim=0cm 0cm 2cm 1cm, clip, width=\hsize]{Np4/V_3e-05_0.3_3.01.png}
    \caption{$Ra=3.0~Ra_{\rm c}$}
    \label{fig:V_transition_Np4b}
    \end{subfigure}
    \begin{subfigure}[c]{0.31\hsize}
    \centering
    \includegraphics[trim=0cm 0cm 2cm 1cm, clip, width=\hsize]{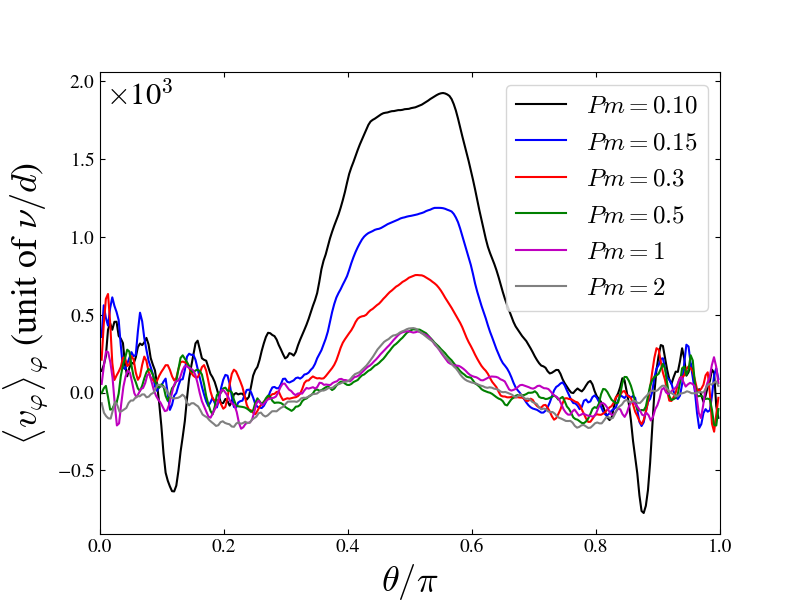}
    \caption{$Ra=6.3~Ra_{\rm c}$}
    \label{fig:V_turb_Np4c}
    \end{subfigure}
    \caption{Same as in the second row of \figurename{}~\ref{fig:Np34}, but for the mean azimuthal velocity profile as a function of the colatitude. Note that panel (b) is similar to \figurename{}~\ref{fig:V_transition_Np4b_main} but is added here for the sake of completeness and comparison.}
    \label{fig:V_Np4}
    \end{figure*}

    \begin{figure*}
    \centering
    \textbf{Azimuthal velocity profile: $E= 10^{-4},~ Pr = 1$}\par\medskip
    \flushleft
    \rotatebox[origin=c]{90}{$N_\rho=3.5$} \quad
    \begin{subfigure}[c]{0.31\hsize}
    \centering
    \includegraphics[trim=0cm 0cm 2cm 1cm, clip, width=\hsize]{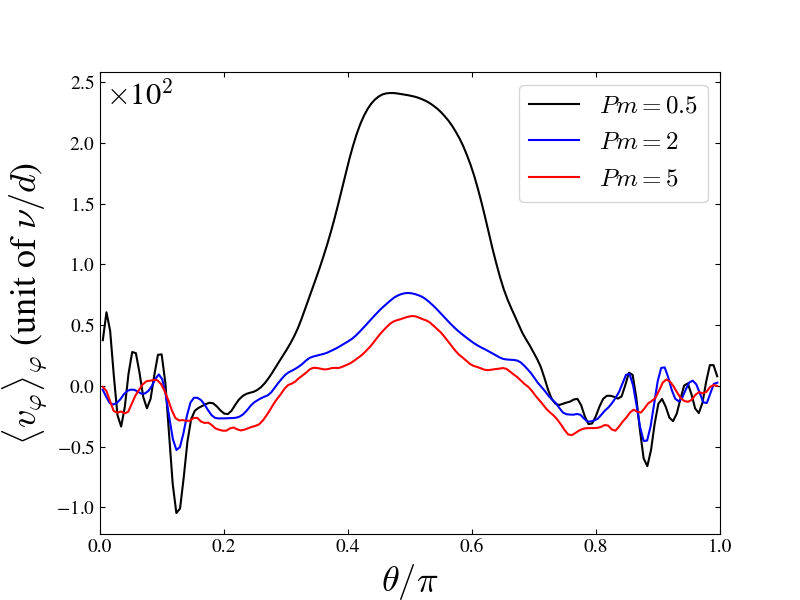}
    \caption{$Ra=2.8~Ra_{\rm c}$}
    \label{fig:V_onset_Np35}
    \end{subfigure}
    \begin{subfigure}[c]{0.31\hsize}
    \centering
    \includegraphics[trim=0cm 0cm 2cm 1cm, clip, width=\hsize]{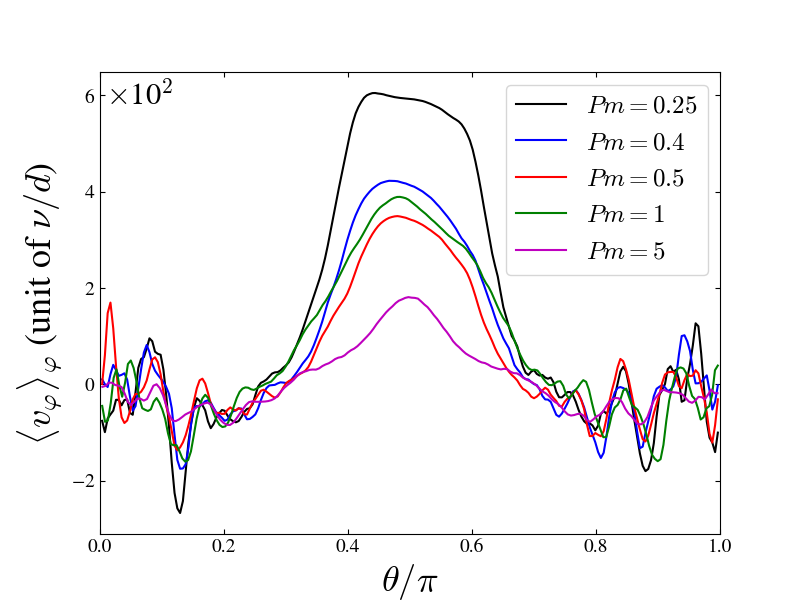}
    \caption{$Ra=5.5~Ra_{\rm c}$}
    \label{fig:V_turb_Np35}
    \end{subfigure}

    \rotatebox[origin=c]{90}{$N_\rho=4$} \quad
    \begin{subfigure}[c]{0.31\hsize}
    \includegraphics[trim=0cm 0cm 2cm 1cm, clip, width=\hsize]{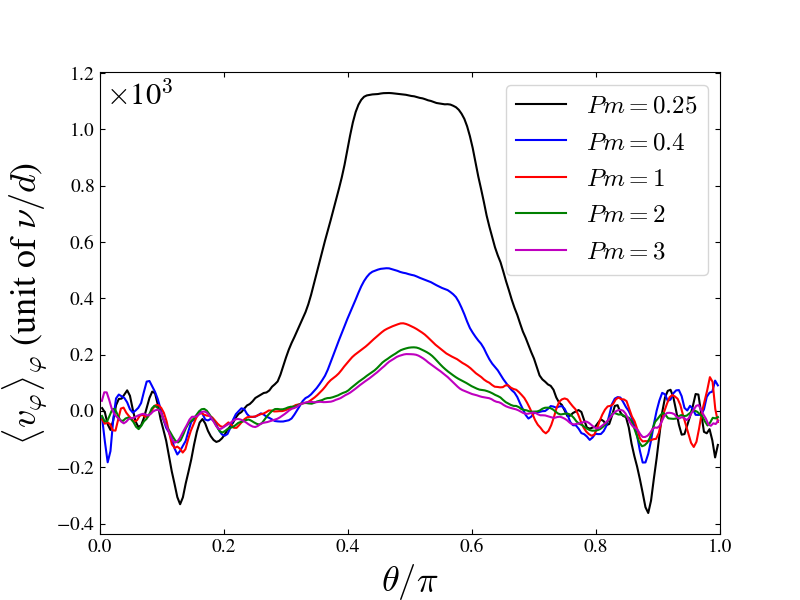}
    \caption{$Ra=6.1~Ra_{\rm c}$}
    \label{fig:V_turb_Np4_1}
    \end{subfigure}
    \begin{subfigure}[c]{0.31\hsize}
    \centering
    \includegraphics[trim=0cm 0cm 2cm 1cm, clip, width=\hsize]{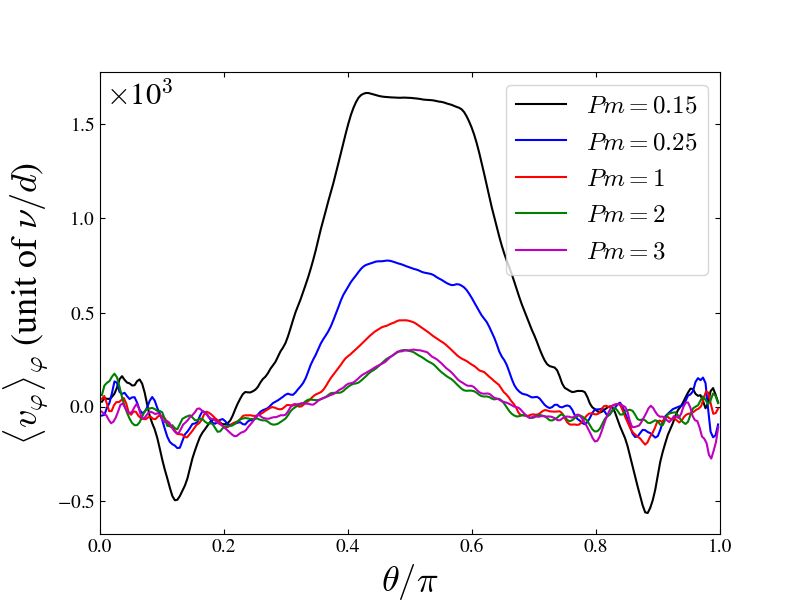}
    \caption{$Ra=9.8~Ra_{\rm c}$}
    \label{fig:V_turb_Np4_2}
    \end{subfigure}
    \begin{subfigure}[c]{0.31\hsize}
    \centering
    \includegraphics[trim=0cm 0cm 2cm 1cm, clip, width=\hsize]{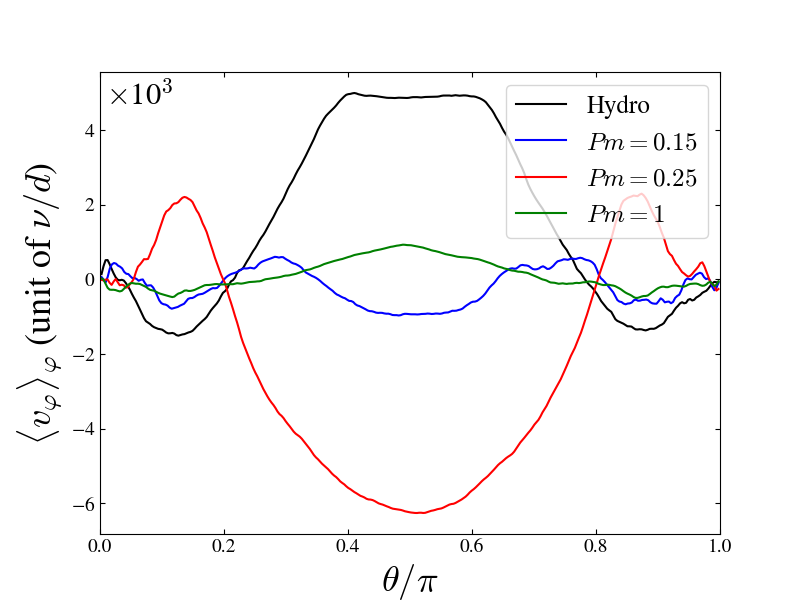}
    \caption{$Ra=29~Ra_{\rm c}$}
    \label{fig:V_turb_Np4_3}
    \end{subfigure}
    
    \rotatebox[origin=c]{90}{$N_\rho=5$} \quad
    \begin{subfigure}[c]{0.31\hsize}
    \centering
    \includegraphics[trim=0cm 0cm 2cm 1cm, clip, width=\hsize]{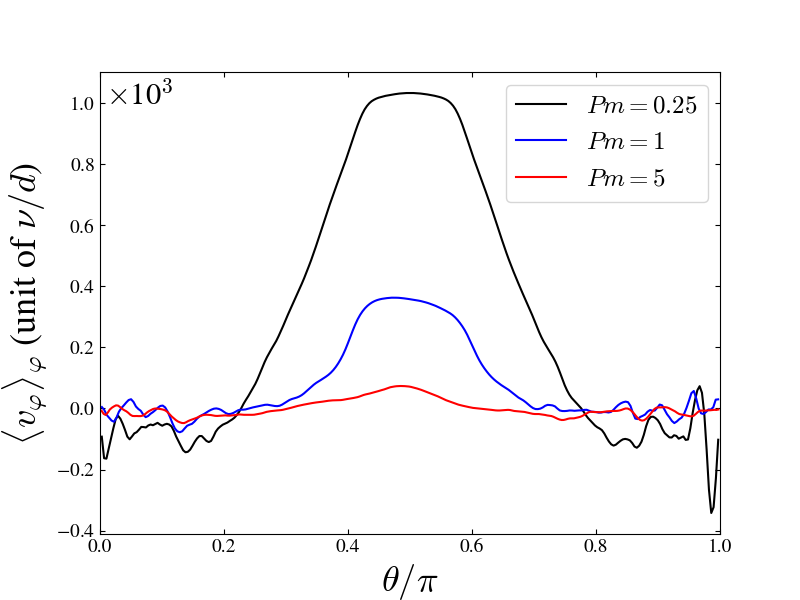}
    \caption{$Ra=4.2~Ra_{\rm c}$}
    \label{fig:V_turb_Np5a}
    \end{subfigure}
    \begin{subfigure}[c]{0.31\hsize}
    \centering
    \includegraphics[trim=0cm 0cm 2cm 1cm, clip, width=\hsize]{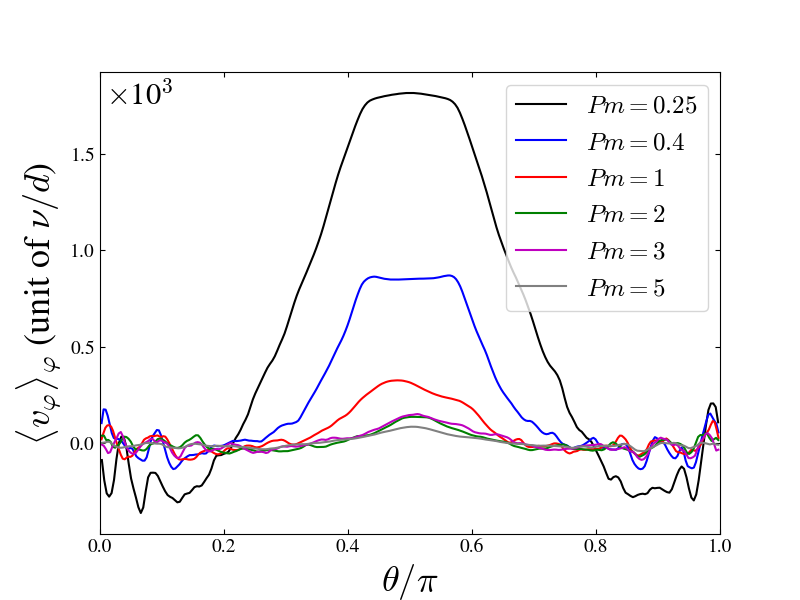}
    \caption{$Ra=6.2~Ra_{\rm c}$}
     \label{fig:V_turb_Np5b}
     \end{subfigure}
    \begin{subfigure}[c]{0.31\hsize}
    \centering
    \includegraphics[trim=0cm 0cm 2cm 1cm, clip, width=\hsize]{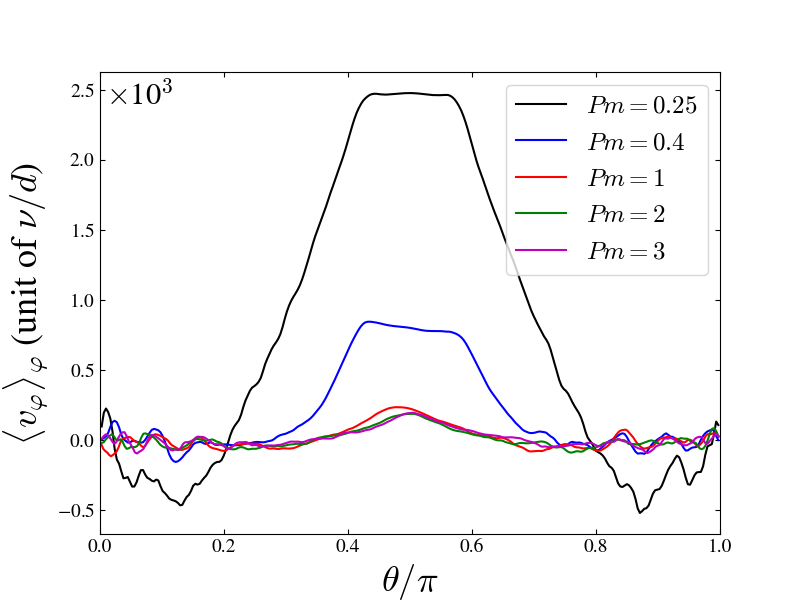}
    \caption{$Ra=8.3~Ra_{\rm c}$}
    \label{fig:V_turb_Np5c}
    \end{subfigure}
    \rotatebox[origin=c]{90}{$N_\rho=5$} \quad
    \begin{subfigure}[c]{0.31\hsize}
    \centering
    \includegraphics[trim=0cm 0cm 2cm 1cm, clip, width=\hsize]{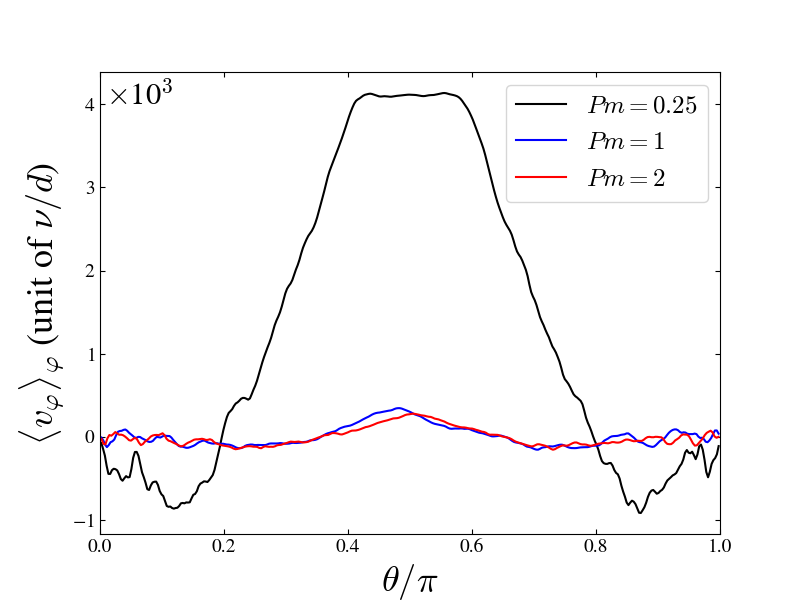}
    \caption{$Ra=16~Ra_{\rm c}$}
    \label{fig:V_turb_Np5d}
    \end{subfigure}
    \begin{subfigure}[c]{0.31\hsize}
    \centering
    \includegraphics[trim=0cm 0cm 2cm 1cm, clip, width=\hsize]{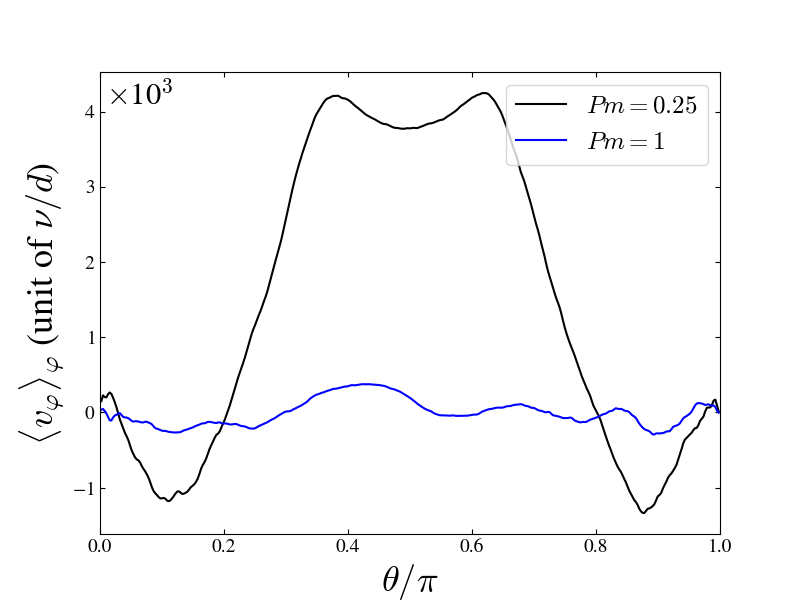}
    \caption{$Ra=32~Ra_{\rm c}$}
    \label{fig:V_turb_Np5e}
    \end{subfigure}
    \caption{Same as in \figurename{}~\ref{fig:Nu_Np35}, but for the mean azimuthal velocity profile as a function of the colatitude.}
    \label{fig:V_Np35}
    \end{figure*}

    \begin{figure*}
    \centering
    \textbf{Azimuthal velocity profile: $E= 3 \times 10^{-4},~ Pr = 1$}\par\medskip
    \flushleft
    \rotatebox[origin=c]{90}{$N_\rho=6$} \quad
    \begin{subfigure}[c]{0.31\hsize}
    \centering
    \includegraphics[trim=0cm 0cm 2cm 1cm, clip, width=\hsize]{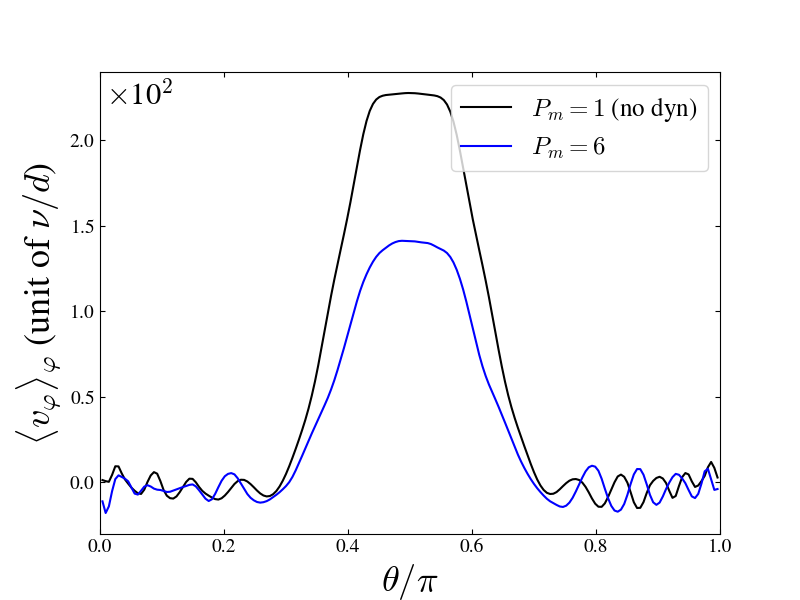}
    \caption{$Ra=3.0~Ra_{\rm c}$}
    \label{fig:V_Np6_3}
    \end{subfigure}
    \begin{subfigure}[c]{0.31\hsize}
    \centering
    \includegraphics[trim=0cm 0cm 2cm 1cm, clip, width=\hsize]{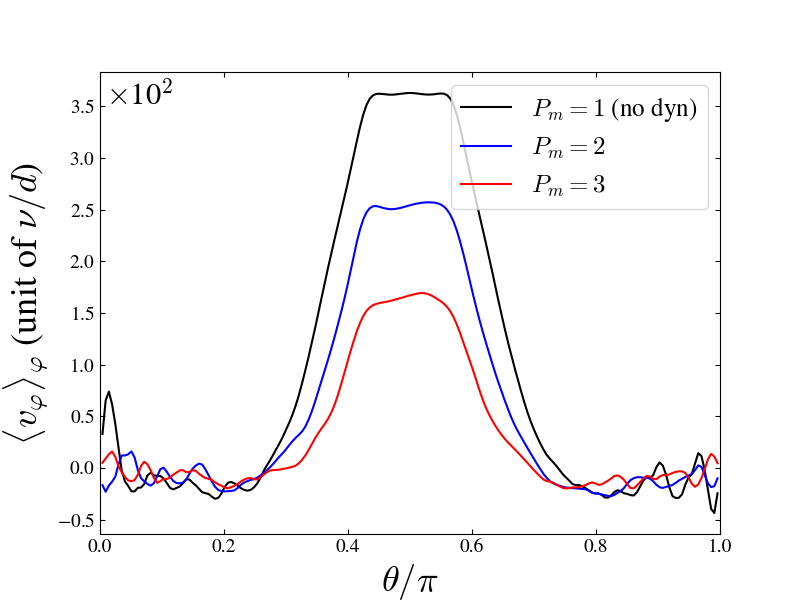}
    \caption{$Ra=4.0~Ra_{\rm c}$}
    \label{fig:V_Np6_4}
    \end{subfigure}
    \begin{subfigure}[c]{0.31\hsize}
    \centering
    \includegraphics[trim=0cm 0cm 2cm 1cm, clip, width=\hsize]{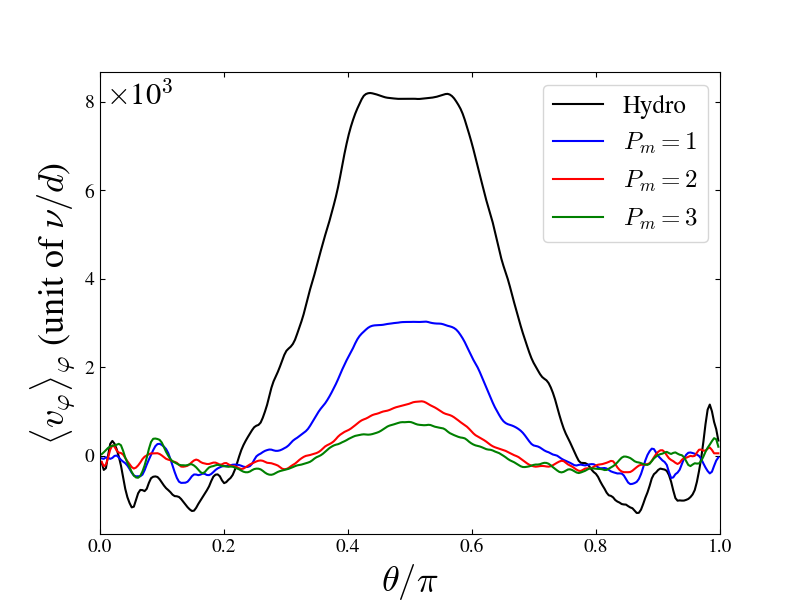}
    \caption{$Ra=8~Ra_{\rm c}$}
    \label{fig:V_Np6_8}
    \end{subfigure}
    
    \flushleft
    \rotatebox[origin=c]{90}{$N_\rho=6$} \quad
    \begin{subfigure}[c]{0.31\hsize}
    \centering
    \includegraphics[trim=0cm 0cm 2cm 1cm, clip, width=\hsize]{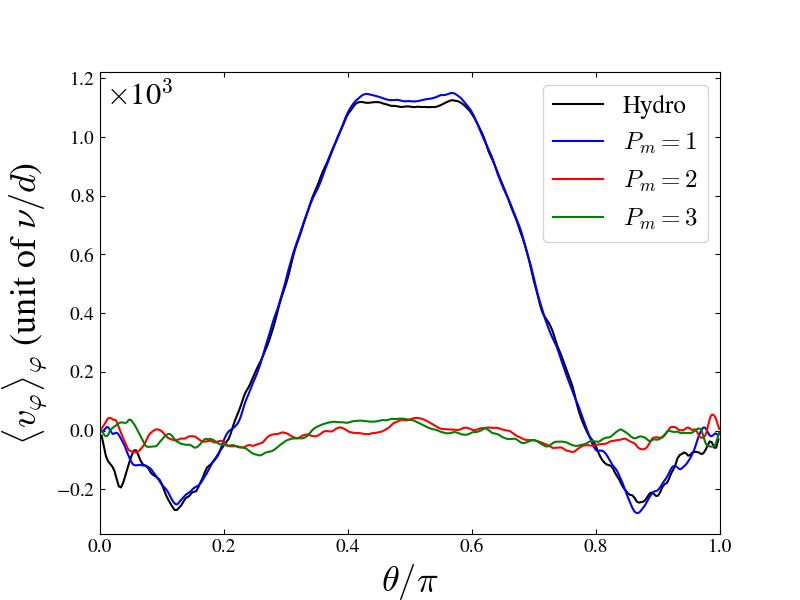}
    \caption{$Ra=16~Ra_{\rm c}$}
    \label{fig:V_Np6_16}
    \end{subfigure}
    \begin{subfigure}[c]{0.31\hsize}
    \centering
    \includegraphics[trim=0cm 0cm 2cm 1cm, clip, width=\hsize]{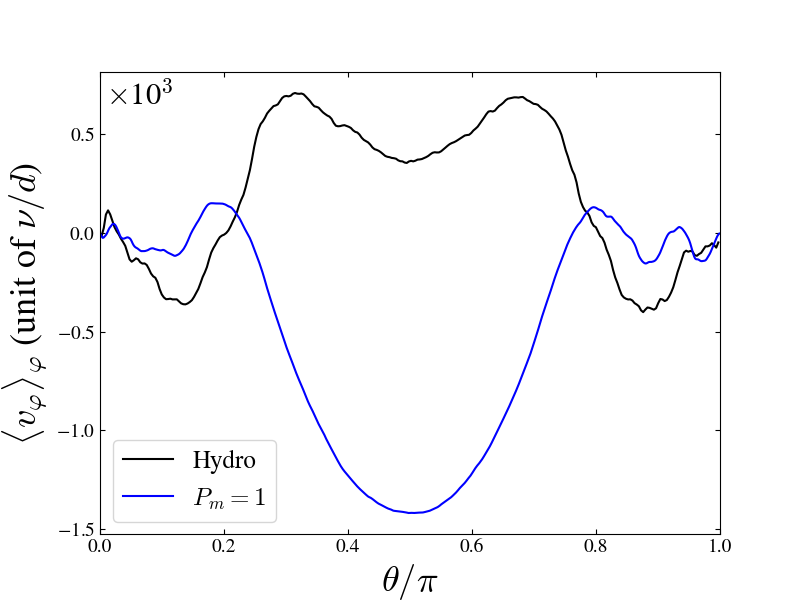}
    \caption{$Ra=32~Ra_{\rm c}$}
    \label{fig:V_Np6_32}
    \end{subfigure}
    \caption{Same as in \figurename{}~\ref{fig:Nu_Np6}, but for the mean azimuthal velocity profile as a function of the colatitude.}
    \label{fig:V_Np6}
    \end{figure*}
   
\end{document}